\documentclass[a4paper,11pt]{article}

\pdfoutput=1
\usepackage{comment}
\usepackage{fixltx2e}
\usepackage[table,xcdraw]{xcolor}
\usepackage{jcappub} 
\usepackage{float} 
\usepackage{bm}
\usepackage{stmaryrd}
\usepackage[normalem]{ulem}
\usepackage{amsmath}
\usepackage{bbm}
\usepackage{amssymb}
\usepackage{geometry}
\usepackage{adjustbox}

\def\be{\begin{equation}}
\def\ee{\end{equation}}
\def\bea{\begin{eqnarray}}
\def\eea{\end{eqnarray}}
\newcommand{\vs}{\nonumber\\}
\def\ba#1\ea{\begin{align}#1\end{align}}
\def\bg#1\eg{\begin{gather}#1\end{gather}}

\newcommand{\refeq}[1]{Eq.~(\ref{eq:#1})}

\newcommand{\reffig}[1]{Fig.~\ref{fig:#1}}          
          
\newcommand{\reftab}[1]{Tab.~\ref{tab:#1}}          
\newcommand{\refsec}[1]{Sec.~\ref{sec:#1}}          
\newcommand{\refapp}[1]{App.~\ref{app:#1}}

\def\Plin{P_{\rm L}}

\def\vphi{\varphi}

\renewcommand{\v}[1]{\bm{#1}}

\renewcommand{\emph}[1]{\textit{#1}}

\newcommand{\vx}{\v{x}}
\newcommand{\vy}{\v{y}}
\newcommand{\vz}{\v{z}}

\newcommand{\vk}{\v{k}}
\newcommand{\vdisp}{\v{s}}
\newcommand{\vq}{\v{q}}
\newcommand{\vp}{\v{p}}

\newcommand{\vK}{\bm{k}_{\rm coll}}

\renewcommand{\d}{\delta}
\newcommand{\D}{\Delta}

\newcommand{\fnl}{f_{\rm NL}}

\def\L{\Lambda}

\def\dirac{\delta_{\rm D}}
\def\diracpi{\hat{\delta}_{\rm D}}

\def\dlin{\delta_{\rm L}}

\def\L{\Lambda}

\def\O{\mathcal{O}}

\newcommand{\perm}[1]{ \expandafter\ifstrempty\expandafter{#1} {\mbox{perm.}} {\mbox{$#1$ perm.}} }

\makeatletter
\newlength{\apb@width}
\newcommand{\autoparbox}[2][c]{\settowidth{\apb@width}{#2}\parbox[#1]{\apb@width}{#2}}
\newcommand{\includegraphicsbox}[2][]{\autoparbox{\includegraphics[#1]{#2}}}
\makeatother

\def\dirac{\delta_{\rm D}}
\def\diracpi{\hat{\delta}_{\rm D}}

\renewcommand{\comment}[1]{}

\usepackage[T1]{fontenc}
\usepackage{graphbox}
\usepackage{subfig}
\usepackage{booktabs}
\title{Galaxy Bias and Soft Limits from Inflation}
\usepackage{amsmath}
\author[a]{Charalampos Nikolis}
\author[b,c,d]{and Henrique Rubira}
\affiliation[a]{The Oskar Klein Centre, Department of Physics,
Stockholm University, Stockholm 106 91, Sweden}
\affiliation[b]{University Observatory, Faculty of Physics, Ludwig-Maximilians-Universit\"at, Scheinerstr. 1, D-81679 München, Germany}
\affiliation[c]{Kavli Institute for Cosmology Cambridge, Madingley Road, Cambridge CB3 0HA, UK}
\affiliation[d]{Centre for Theoretical Cosmology, Department of Applied Mathematics and Theoretical Physics
University of Cambridge, Wilberforce Road, Cambridge, CB3 0WA, UK}

\emailAdd{charalampos.nikolis@fysik.su.se}
\emailAdd{henrique.rubira@lmu.de}

\abstract{
Primordial non-Gaussianity (PNG) leaves an imprint on large-scale structure through a scale-dependent galaxy bias, and properly incorporating the effects of higher order PNG into the bias expansion requires a consistent renormalization framework that extends beyond the squeezed limit of the bispectrum. We develop the multi-loop renormalization of bias operators in the presence of non-vanishing primordial n-point functions of arbitrary order, including stochastic contributions. We show that galaxy bias renormalization is determined by the multi-soft limits of primordial correlators and bias kernels. We provide explicit computations for one- and two-loop renormalization of deterministic operators up to fifth order in the bias expansion. Finally, we derive the renormalization group equations for the bias parameters at one and two loops.
}


\begin{document}

\maketitle
\flushbottom

\section{Introduction}\label{sec:intro}

The imprint of primordial non-Gaussianity (PNG) on large-scale structure (LSS) provides a unique window into the physics of inflation \cite{Guth:1980zm, Linde:1981mu,Guth:1982ec, PhysRevD.42.3936,maldacena:2003,Cheung:2007st}. A broad class of primordial interactions can generate observable signatures in LSS, with their characteristic momentum dependence encoded in the soft limits of primordial correlators. At leading order in PNG, the squeezed limit of the primordial bispectrum induces non-Gaussian contributions of order $f_{\rm NL}$ to the two-point statistics of biased tracers \cite{2008PhRvD..77l3514D,2008ApJ...677L..77M} (see \cite{Giannantonio:2009ak, Yokoyama:2012az,K-Smith} for cases including local $g_{\rm NL}$ and \cite{2013JCAP...05..001B, Gleyzes:2016tdh,MoradinezhadDizgah:2017szk} for scenarios including different shapes and fields, and \cite{2015JCAP...10..032S} for the effect of PNG on galaxy shapes), requiring new classes of non-Gaussian operators in the bias expansion \cite{McDonald:2008sc,Schmidt:2010gw,Assassi:2015EFT,Assassi:2015fma}, and models with resonant and oscillatory features \cite{2018JCAP...09..003C}. We refer the reader to \cite{Desjacques:2016bnm} and references therein for a review of the bias expansion, and the modifications needed in the presence of PNG.

The need for these PNG operators can also be understood by requiring the bias basis to remain closed under renormalization \cite{Assassi:2015fma}. Extending this renormalization to higher-loop orders and higher-order non-Gaussian initial conditions ($f_{\rm NL},\, g_{\rm NL},\, \dots$) remains an open problem, which we address in this paper. Furthermore, we develop the framework for the renormalization of stochastic contributions \cite{Rubira:2024tea} in the presence of PNG.

Moreover, we introduced in Paper I \cite{Nikolis:2024kbx} the renormalization-group (RG) approach to biased tracers in the presence of $f_{\rm NL}$ interactions, generalizing the results of \cite{Rubira:2023vzw, Rubira:2024tea, Bakx:2025cvu, Bakx:2026rmd} valid for Gaussian initial conditions. In this work, we extend this framework beyond the leading $f_{\rm NL}$ contributions and to higher loops in the bias expansion, relating the resulting RG flow directly to the soft limits of primordial correlators and bias kernels. We explicitly renormalize the bias operators in the presence of local-type PNG at one and two loops and solve the corresponding RG equations.
We show that soft limits provide an organizing principle for the renormalization of biased tracers. Our results make explicit how bias operators probe squeezed configurations of the initial conditions\footnote{Terms beyond the squeezed limit contribute at the same order as higher-derivative contributions \cite{Schmidt:2013nsa}, and are therefore subleading.}, connecting with consistency relations developed for squeezed-limit bispectra and higher-point functions \cite{maldacena:2003,Paolo_Creminelli_2004,Assassi_2012,Joyce_2015,Mirbabayi_2016}.

We find that the two-loop renormalization of the extended operator basis in the presence of local-type PNG involves two distinct topologies, in contrast to the Gaussian case \cite{Bakx:2025cvu}. Interestingly, all generalized bias kernels up to fifth order in the bias expansion reduce to \textit{just} two functions, one for each topology, which correspond to the unsymmetrized versions of the functions previously identified in \cite{Bakx:2025cvu} for the renormalization of deterministic and stochastic operators. Specific kinematic limits of these functions allow us to reconstruct certain two-loop results entirely in terms of sequential one-loop contributions.

This paper is organized as follows. In \refsec{soft_limits}, we develop the soft limits of primordial correlators. We present the bias basis and operator kernels in \refsec{biasbasis}. In \refsec{renorm}, we derive the renormalization of deterministic bias operators at one, two, and arbitrary loop order, while \refsec{stoch} extends the analysis to stochastic contributions. In \refsec{running}, we obtain the corresponding one- and two-loop RG equations. We conclude in \refsec{disc}. \refapp{disp} considers the displacement kernels for non-Gaussian operators, and \refapp{tables} collects mixing-matrix coefficients.

Throughout this paper, we denote hard and integrated momenta by $\vp$ (with subscripts $\vp_1, \vp_2, \dots$) and soft momenta by $\vk$ (with subscripts $\vk_1, \vk_2, \dots$). Bold variables denote vectors, and their magnitudes are written in normal typeface. For brevity, we use the shorthand $\vp_{1 \dots n} = \vp_{1} + \dots + \vp_{n}$ to indicate the sum of vectors. Momentum-space integrals are expressed as
\be
\int_{\vp_1,\dots,\vp_n} = \int \frac{d^3p_1}{(2\pi)^3} \dots \int \frac{d^3p_n}{(2\pi)^3}\,.
\ee
We use prime notation in $n$-point functions $\langle \dots \rangle'$ to indicate that the momentum-conservation factor $\diracpi=(2\pi)^3\dirac$, the Dirac delta function, has been omitted.

\newpage

\section{Soft limits of primordial \texorpdfstring{$n$}{n}-point functions}
\label{sec:soft_limits}

As we will see in detail in \refsec{renorm}, the renormalization of bias coefficients in the presence of primordial non-Gaussianity is controlled by soft limits of primordial correlators, together with the corresponding soft limits of the bias kernels.
We denote the primordial $N$-point functions of curvature perturbations by $G^{(N)}$ and the $\ell$th-order perturbative kernel of a bias operator by $K^{(\ell)}$.
At one loop, a primordial $N$-point function enters through its $(N-2)$-soft limit, in which two momenta are hard compared with the remaining $N-2$ momenta. Schematically, for a bias operator correlated with $n$ external linear density fields, the one-loop contribution takes the form
\ba \label{eq:loop1sec2}
\boxed{{\rm 1-loop:}\;\sim \int_p
K^{(n+2-s)}
\!\left(\vp,-\vp-\vk_{1\dots s},\dots\right)
G^{(s+2)}
\!\left(\vp,-\vp-\vk_{1\dots s},\dots\right)}\,.
\ea
The external momenta $\vk_i$ are taken soft relative to the loop momentum $\vp$, with $\vk_{1\dots s}\equiv\sum_{i=1}^{s}\vk_i$.
Here $s$ external fields enter the primordial correlator, while the remaining $n-s$ attach directly to the bias kernel, with ellipses denoting the corresponding soft momenta. As we discuss in \refsec{renorm}, at two loops there are contributions involving either one primordial correlator with three hard legs (topology $[3]$) or two primordial correlators with two hard legs each (topology $[22]$). Using the same schematic notation, the two-loop contributions are
\ba \label{eq:connectedsec2}
\boxed{
\begin{aligned}
&{\rm 2-loop}\,[3]:\;\sim
\int_{p_1,p_2}
K^{(n+3-s)}
\!\left(
\vp_1,\vp_2,-\vp_1-\vp_2-\vk_{1\dots s},\dots
\right)
\\
&\qquad
\times G^{(s+3)}
\!\left(
\vp_1,\vp_2,-\vp_1-\vp_2-\vk_{1\dots s},\dots
\right)\,,
\\[1ex]
&{\rm 2-loop}\,[22]:\;\sim
\int_{p_1,p_2}
K^{(n+4-s)}
\!\left(
\vp_1,-\vp_1-\vk_{1\dots s_1},
\vp_2,-\vp_2-\vk_{s_1+1\dots s},\dots
\right)
\\
&\qquad
\times G^{(s_1+2)}
\!\left(
\vp_1,-\vp_1-\vk_{1\dots s_1},\dots
\right)
G^{(s_2+2)}
\!\left(
\vp_2,-\vp_2-\vk_{s_1+1\dots s},\dots
\right)\,,
\end{aligned}
}
\ea
where $s=s_1+s_2$ in the $[22]$ topology.
The $[3]$ topology therefore probes the $(N-3)$-soft limit of a primordial $N$-point function, whereas the $[22]$ topology involves two soft limits of the same type as at one loop.
We emphasize that {\it these renormalization structures are independent of the inflationary model}.

At any loop order, the renormalization combines multi-soft limits of primordial correlators, determined by the inflationary model, with the corresponding soft limits of the bias kernels. For the case of higher loops, we probe the full kinematic structure of the bispectrum and higher $n$-point functions. Computing these higher-point functions is challenging and has motivated substantial progress in both analytical methods (see, e.g.~\cite{baumann2022snowmass}) and numerical algorithms \cite{Werth:2024aui}.
The soft limits of the bias kernels have been systematically computed through their projection onto operators of lower order, at one and two loops and at both leading and next-to-leading order in the gradient expansion \cite{Bakx:2025cvu,Bakx:2026rmd} (see \cite{Philcox:2026myi} for the case of redshift space). We discuss in \refsec{renorm} the extension of the bias basis in the presence of PNG (see also \cite{Assassi:2015fma}), while here we focus on the soft limits of primordial correlators.

Soft limits of inflationary $n$-point functions have been extensively studied in the context of consistency conditions for different models of inflation \cite{maldacena:2003, Paolo_Creminelli_2004,Assassi_2012,Joyce_2015, Mirbabayi_2016} and their momentum dependence probes the particle content during inflation. One example is the exchange of a massive scalar coupled to the inflaton, which can produce distinct power-law or oscillatory signatures in the soft limit, depending on the mass of the scalar \cite{Chen_2010,Baumann_2012,arkanihamed2015cosmological,Lee_2016}. The exchange of particles with nonzero spin introduces additional angular dependence, described by Legendre polynomials \cite{arkanihamed2015cosmological,Lee_2016,arkanihamed2019cosmological}.
In general, due to statistical homogeneity, the single-soft (or squeezed) limit of the bispectrum can only depend on $k,\,p$ and $\vk\cdot\vp$ when $k\ll p$. As a result, that limit can be written as an expansion in Legendre polynomials of even order
\cite{Assassi:2015EFT,Assassi:2015fma}
\bea
B(k,p,|\vp+\vk|)
\xrightarrow{k\ll p}
\fnl \sum_{L,i}
a_{L,i}\left(\frac{k}{p}\right)^{\Delta_i} \mathcal{P}_L(\mathbf{\hat{k}}\cdot\mathbf{\hat{p}})
P(k)P(p)\;,
\eea
where $\mathcal{P}_L$ is a Legendre polynomial of order $L$, and $i$ labels distinct scaling contributions with exponents $\Delta_i\in\mathbb{C}$, amplitudes $a_{L,i}$ and $P$ is the power spectrum. Only even $L$ contribute at leading order in the squeezed limit \cite{Assassi:2015EFT}, with the $L=0$ term being isotropic and capturing scalar exchange, while higher even multipoles are associated with particles of nonzero spin. In the following, we restrict our analysis to the spin-zero case.

The exponent $\Delta$ controls the momentum scaling of the squeezed-limit signal. The case $\Delta=0$ reproduces the squeezed-limit scaling of local-type PNG, parametrized by $f_{\rm NL}$ and commonly generated in multifield inflation. For scalar exchange, $0<\Delta<3/2$ corresponds to massive fields in the complementary series. For heavier fields in the principal series, the scaling exponent becomes complex, leading to oscillations in the momentum ratio $k/p$, characteristic of cosmological-collider signals \cite{arkanihamed2015cosmological}. 

As a case study for our analysis, we consider primordial non-Gaussianity of the local type ($\Delta = 0$), generated by an effectively single source. This allows us to focus on the structures that arise when extending renormalization to higher loop orders and higher orders in primordial non-Gaussianity. We highlight once more that the organization of these contributions in terms of soft limits as in \refeq{loop1sec2} and \refeq{connectedsec2} is general, and the framework can be extended to other inflationary models by including the appropriate bias operators to ensure closure of the basis under renormalization.

In this section, we introduce the notation for fields and correlators used throughout the paper and then discuss the multi-soft limits of primordial $n$-point functions for single-source local PNG.

\subsection{Fields and \texorpdfstring{$n$}{n}-point functions} 

We denote the primordial gravitational potential by $\vphi$ and the comoving curvature perturbation by $\zeta$. While inflationary soft theorems and consistency relations are typically formulated in terms of $\zeta$, $\vphi$ provides a convenient description of the initial conditions for large-scale structure and the two fields are related by a constant rescaling on superhorizon scales,
\be
\label{eq:vphi-zeta}
\vphi(\vk) \;=\; \frac{3}{5}\,\zeta(\vk)\,.
\ee
For bias renormalization, we work directly with the primordial $n$-point functions of $\vphi$, defined as
\bea
\label{eq:non-vanishingB}
G^{(n)}_{\vphi}\left(\vk_1,\dots,\vk_n\right)
\equiv
\left\langle
\vphi(\vk_1)\cdots\vphi(\vk_n)
\right\rangle' \,,
\eea
such that $G^{(n)}_\vphi$ includes both connected and disconnected contributions. Therefore, the soft-limit structure of $G^{(n)}_\vphi$ and $G^{(n)}_\zeta$ coincides up to an overall normalization, where $G^{(n)}_\zeta$ are the $n$-point functions of $\zeta$.

We also work with the linear matter overdensity $\dlin$, with power spectrum $\Plin$ and variance
\ba \label{eq:var}
\sigma^2_{\Lambda}
=
\int_{\vp}^\Lambda
\left\langle
\dlin(\vp)\dlin(-\vp)
\right\rangle'
=
\int_{\vp}^\Lambda \Plin(p)\,.
\ea
Yet, bias renormalization forces us to introduce another field $\Psi$ \cite{Assassi:2015fma,Nikolis:2024kbx}
\be
\label{eq:Psi-def}
\Psi(\vk) \;\equiv\; k^{\D} \frac{\dlin(\vk)}{M(k)}
\;=\; k^{\D}\!\varphi(\vk)\,,
\ee
such that each $\Psi$ leg contributes a factor $k^\D$ relative to a $\vphi$ leg. Since we restrict ourselves to the local PNG shape $\D=0$, $\Psi = \vphi$.
We use
\be
\dlin(\vk,\tau)=M(k,\tau)\varphi(\vk)\,,
\ee
coming from Poisson's equation, with
\be
M(k,\tau)=\frac{2}{3}\frac{k^2T(k)D(\tau)}{H_0^2\Omega_{m,0}}\,,
\label{eq:Mdef}
\ee
where $H_0$ is the Hubble constant, $\Omega_{m,0}$ the matter fraction, $D$ the growth factor, and $T(k)$ is the {\it transfer function}, which describes the linear evolution through horizon crossing and is normalized such that $T(k)\to 1$ for $k\ll k_{\rm eq}$, with $k_{\rm eq}$ the equality scale. For the following, we omit the time dependence in $\dlin$ and $M$.
Therefore, explicitly, for $r$ legs of $\Psi$ and $n-r$ legs of $\dlin$,
\bea
\label{eq:phi-to-Psi-deltaL}
\left\langle
\Psi(\vk_1)\cdots\Psi(\vk_r)\,
\dlin(\vk_{r+1})\cdots\dlin(\vk_n)
\right\rangle'
\;=\;
G^{(n)}_\varphi(\vk_1,\dots,\vk_n)
\prod_{j=r+1}^{n}M(k_j)\,,
\eea
where we abbreviate $\dlin$ by $\d$ in field subscripts. Moreover,
\be \label{eq:Ppsidelta}
P_{\Psi\d}(k)\equiv\langle\Psi\dlin\rangle'
=
\frac{\Plin(k)}{M(k)}\,,
\quad \textrm{and} \quad
P_{\Psi\Psi}(k)\equiv\langle\Psi\Psi\rangle'
=
\frac{\Plin(k)}{M^2(k)}\,.
\ee
For $X_1,X_2\in\{\Psi,\dlin\}$, the corresponding power spectra are
\bea
P_{X_1X_2}(k)=
\begin{cases}
    \Plin(k), &
    X_1=X_2=\dlin,\\
    \dfrac{\Plin(k)}{M(k)}, &
    (X_1,X_2)\in\{(\dlin,\Psi),(\Psi,\dlin)\},\\
    \dfrac{\Plin(k)}{M^2(k)}, &
    X_1=X_2=\Psi.
\end{cases}
\eea
Throughout this paper, whenever an $n$-point function carries only a single field subscript, such as $P_\varphi$, $B_\varphi$, and $T_\varphi$ for the power spectrum, bispectrum, and trispectrum, or more generally $G^{(n)}_\varphi$, it denotes the correlator of $n$ copies of the same field $\varphi$.

\subsection{Soft limits for local PNG}

We consider primordial non-Gaussianity of the local type, generated by a single Gaussian source and parametrized by \cite{Komatsu:2001rj}
\begin{equation}
    \varphi
    =
    \varphi_{\rm G}
    +\fnl\varphi_{\rm G}^2
    +g_{\rm NL}\varphi_{\rm G}^3
    +h_{\rm NL}\varphi_{\rm G}^4
    +\cdots\;,
\end{equation}
where $\varphi_{\rm G}$ is Gaussian and homogeneous mean subtractions are understood.
Moreover, the primordial correlators below are evaluated at tree level, independently of the loop order considered in bias renormalization.

For one-loop renormalization, sketched in \refeq{loop1sec2}, we need the $(n-2)$-soft limit of a primordial $n$-point function, for which the simplest example is the single-soft (squeezed) limit of the bispectrum,
\bea
B_\varphi(\vk,\vp,-\vp-\vk)
\xrightarrow{k\ll p}
4\fnl P_\varphi(k)P_\varphi(p)\;.
\eea
This factorization requires the inclusion of a (full non-Gaussian) $\Psi$ operator in the bias basis, as we see in \refsec{renorm}.
For the trispectrum, the corresponding double-soft limit is
\ba \label{eq:softT}
T_\varphi(\vk_1,\vk_2,\vp,-\vp-\vk_{12})
&\xrightarrow{k_i\ll p}
4\fnl B_\varphi(\vk_1,\vk_2,-\vk_{12})P_\varphi(p) 
+12g_{\rm NL}P_\varphi(k_1)P_\varphi(k_2)P_\varphi(p)\;.
\ea
The first term contains a fully soft bispectrum and is already reproduced by the non-Gaussian contraction $\langle\Psi\dlin\dlin\rangle$ of the operator introduced above, but the second term has a new soft structure and requires an operator proportional to $\Psi^2$ to close the basis under renormalization.
More generally, the $(n-2)$-soft limit contains terms involving fully soft correlators that are already reproduced by lower powers of $\Psi$.
After accounting for these terms, the remaining contribution takes the form
\ba
\label{eq:multi-soft}
\boxed{G_\varphi^{(n)}
\!\left(
\vk_1,\dots,\vk_{n-2},
\vp,-\vp-\vk_{1\dots n-2}
\right)\Big|^{{}^{\rm soft}_{\rm 2pt}}
\xrightarrow{k_i\ll p}
\lambda_{\rm NL}^{(n)}
P_\varphi(k_1)\cdots P_\varphi(k_{n-2})P_\varphi(p)\;.}
\ea
Here the restriction $\Big|^{{}^{\rm soft}_{\rm 2pt}}$ selects the connected contribution remaining after subtracting the terms reproduced by operators with lower powers of $\Psi$, these being the term in which soft modes appear only through two-point functions as the last term of \refeq{softT}. This term requires an operator proportional to $\Psi^{n-2}$ in the bias expansion to ensure closure under renormalization. The first $\lambda_{\rm NL}^{(n)}$ coefficients are 
\begin{alignat}{2} \label{eq:lambdas}
\lambda_{\rm NL}^{(2)}&=1\,,
&\qquad \lambda_{\rm NL}^{(3)}&=4\fnl\,,
\\
\lambda_{\rm NL}^{(4)}&=12g_{\rm NL}\,,
&\qquad \lambda_{\rm NL}^{(5)}&=48h_{\rm NL}-24\fnl g_{\rm NL}\;. \nonumber
\end{alignat}
This factorization indicates that we need to include operators of the form $\Psi^m$ which scale with $\lambda_{\rm NL}^{(m+2)}$. For example, $\Psi^2$ scales with $\lambda_{\rm NL}^{(4)}=12g_{\rm NL}$, while the $f_{\rm NL}^2$ contribution is already accounted from non-Gaussian contractions of the linear $\Psi$.

The connected two-loop topology $[3]$, sketched in \refeq{connectedsec2}, probes the $(n-3)$-soft limit of a primordial $n$-point function.
The first relevant example is the single-soft limit of the trispectrum,
\bea
T_\varphi(\vk,\vp,\vq,-\vp-\vq-\vk)
\xrightarrow{k\ll p,q,|\vp+\vq|}
P_\varphi(k) \mathcal{B}_{\varphi,4}(\vp,\vq,-\vp-\vq)\,,
\eea
factorizing into a soft power spectrum and a hard bispectrum response $\mathcal{B}_{\varphi,4}$, with the label 4 standing for the order of the original $n$-point function. Its soft dependence is therefore absorbed by a counterterm proportional to the same field $\Psi$ that appears in the squeezed-bispectrum case. For a general primordial $n$-point function, the $(n-3)$-soft limit also contains terms involving fully soft correlators that are already reproduced by lower powers of $\Psi$. After accounting for these terms, the remaining contribution generalizes \refeq{multi-soft} to
\begin{equation}
\label{eq:Gthreehard}
\boxed{
\begin{aligned}
&
G_\varphi^{(n)}
\!\left(
\vk_1,\dots,\vk_{n-3},
\vp,\vq,-\vp-\vq-\vk_{1\dots n-3}
\right)\Big|^{{}^{\rm soft}_{\rm 2pt}}
\\
&\hspace{3cm}
\xrightarrow{k_i\ll p,q,|\vp+\vq|}
P_\varphi(k_1)\cdots P_\varphi(k_{n-3})
\mathcal{B}_{\varphi,n}(\vp,\vq,-\vp-\vq)\,,
\end{aligned}
}
\end{equation}
with a new bispectrum response $\mathcal{B}_{\varphi,n}$, and a soft structure absorbed by a counterterm proportional to $\Psi^{n-3}$. The first two response functions for the local PNG case are given by
\bea
&&
\mathcal{B}_{\varphi,3}(\vp_1,\vp_2,\vp_3)
=
2\fnl
\left[P_\varphi(p_1)P_\varphi(p_2)+2\ {\rm perms.}\right],
\vs
&&
\mathcal{B}_{\varphi,4}(\vp_1,\vp_2,\vp_3)
=
\left(8\fnl^2+6g_{\rm NL}\right)
\left[P_\varphi(p_1)P_\varphi(p_2)+2\ {\rm perms.}\right]\,.
\eea
We also define the response with three hard linear-density legs by
\bea
\label{eq:bispec-response}
\mathcal{B}_{n}(\vp_1,\vp_2,\vp_3)
=
M(p_1)M(p_2)M(p_3)
\mathcal{B}_{\varphi,n}(\vp_1,\vp_2,\vp_3)\,.
\eea
For mixed hard legs $X_i\in\{\Psi,\dlin\}$, the corresponding response is
\bea \label{eq:BXXX}
\mathcal{B}_{X_1X_2X_3,n}(\vp_1,\vp_2,\vp_3)
=
\frac{
\mathcal{B}_{n}(\vp_1,\vp_2,\vp_3)
}{
M^{q_1}(p_1)M^{q_2}(p_2)M^{q_3}(p_3)
}\,,
\eea
where $q_i=1$ if $X_i=\Psi$ and $q_i=0$ if $X_i=\dlin$, using $\Psi=\varphi$ for local PNG.

Following the same procedure, the new contribution in a general $(n-h)$-soft limit factorizes into $n-h$ soft power spectra and a response of the hard $h$-point function
\begin{equation} \label{eq:multihard}
\boxed{
\begin{aligned}
&
G_\varphi^{(n)}
\!\left(
\vk_1,\dots,\vk_{n-h},
\vp_1,\dots,\vp_{h-1},
\vp_h-\vk_{1\dots n-h}
\right)\Big|^{{}^{\rm soft}_{\rm 2pt}}
\\
&\qquad
\xrightarrow{k_i\ll p_a}
\left[\prod_{i=1}^{n-h}P_\varphi(k_i)\right]
\mathcal{G}_{\varphi,n}^{(h)}
(\vp_1,\dots,\vp_h)\,,
\end{aligned}
}
\end{equation}
with $\vp_h=-\vp_{1\dots h-1}$ and where the hard response $\mathcal{G}_{\varphi,n}^{(h)}$ can be further decomposed into powers of the linear power spectrum for the case of local PNG. In summary, as we see in detail in \refsec{renorm}, at $L$ loops, there is a new topology probed by renormalization whose contribution involves a single connected primordial correlator containing $L+1$ hard legs. Thus, one loop probes a hard power spectrum, two loops a hard bispectrum, three loops a hard trispectrum, and so on.
It is also convenient to define the response with $h$ hard linear-density legs by
\be
\mathcal{G}_{n}^{(h)}(\vp_1,\dots,\vp_h)
=
\left[\prod_{a=1}^{h}M(p_a)\right]
\mathcal{G}_{\varphi,n}^{(h)}(\vp_1,\dots,\vp_h)\,,
\ee
and for mixed hard legs $X_a\in\{\Psi,\dlin\}$ 
\be \label{eq:GXXXX}
\mathcal{G}_{X_1\cdots X_h,n}^{(h)}(\vp_1,\dots,\vp_h)
=
\frac{
\mathcal{G}_{n}^{(h)}(\vp_1,\dots,\vp_h)
}{
\displaystyle\prod_{a=1}^{h}M^{q_a}(p_a)
}\,,
\ee
where $q_a=1$ if $X_a=\Psi$ and $q_a=0$ if $X_a=\dlin$, using $\Psi=\varphi$ for local PNG.

Finally, we consider the collapsed limit. Whereas the multi-soft limits discussed above determine the renormalization of the deterministic bias parameters, starting at two loops the renormalization of the stochastic sector discussed in \refsec{stoch} also probes collapsed configurations. For the trispectrum, we define the collapsed channel by
\be
\label{eq:collapse-kinematics}
\vk\equiv\vp_{12}=-\vp_{34}\;\to\;0\,,
\qquad
k\ll p_i\,,
\ee
such that the two pairs of hard momenta become separately back-to-back, which can be understood as the correlation between two short-scale power spectra modulated by the same long-wavelength background. In that case, individual external momenta remain hard and only the momentum exchanged between the two pairs becomes soft.
For single-source local PNG, the leading collapsed contribution at tree level is
\be
\label{eq:collapse-trispectrum}
T_\varphi(\vp_1,\vp_2,\vp_3,\vp_4)
\Big|_{\mathrm{coll}(12|34)}
\;\xrightarrow{\,k\ll p_i\,}\;
16\fnl^2
P_\varphi(k)\,P_\varphi(p_1)\,P_\varphi(p_3)\,,
\ee
while for PNG generated by multiple primordial sources, the collapsed amplitude is conventionally parametrized by $\tau_{\rm NL}$.  At tree level, single-source local PNG satisfies $\tau_{\rm NL}=16\fnl^2$, whereas multiple sources can produce a parametrically larger collapsed amplitude \cite{PhysRevLett.107.191301,2013JCAP...05..001B}.

In multi-source scenarios, the long-wavelength modulation of small-scale statistics need not be fully correlated with $\varphi$ and describing the uncorrelated component generally requires additional PNG fields in the bias expansion \cite{Assassi:2015fma}, such as a field $\hat{\psi}$ satisfying $\langle\hat{\psi}\varphi\rangle=0$. 
For the remainder of the paper, however, we restrict our analysis to single-source local PNG with $\D=0$, for which no extra field beyond $\Psi$ is necessary.

\newpage

\section{The bias expansion} \label{sec:biasbasis}

In this section, we introduce the bias operator basis in \refsec{basis}, including its extension to primordial non-Gaussianity. We then construct the corresponding displaced non-Gaussian kernels in \refsec{kernels}.

\subsection{The basis} \label{sec:basis}

The deterministic part of the bias expansion at a fixed time is written as \cite{Desjacques:2016bnm}
\ba
\label{eq:bias-expansion}
    \d_{g,\rm det}(\vk)&= b_A \O_A(\vk)\,,
\ea
where repeated indices are summed. Here, both the bias coefficients and the operators are bare and therefore cutoff dependent.
We consider a Gaussian set of operators extended to include
\be
{\cal O}_{a_m} = {\cal O}_a\,\Psi^m\,,
\qquad m=0,1,2,\ldots\,,
\ee
where $m$ counts the number of powers of $\Psi$.
In this work, we restrict ourselves to operators at leading order in derivatives (see \cite{Bakx:2026rmd} for the next-to-leading-order Gaussian case) and to local spin-0 PNG. In this notation, $m=0$ denotes the purely Gaussian sector.  For convenience, we drop the index $m=0$ and write ${\cal O}_{a}\equiv{\cal O}_{a_0}$ for Gaussian operators.  We use capital indices $A,B,\ldots$ to label generic operators ${\cal O}_{A}$ in the extended basis. Moreover, we absorb the PNG amplitudes $f_{\rm NL},\,g_{\rm NL},\,\dots$ into the bias parameters $b_A$, in contrast to the usual convention of keeping these amplitudes explicit, also used in \cite{Nikolis:2024kbx}. 

We begin with the Gaussian operator basis through fifth order: 
\bea
\label{eq:bias_basis_G}
{\rm (1)} &&
{\rm tr}\big[\Pi^{[1]}\big],
\\
{\rm (2)} &&
{\rm tr}\big[\big(\Pi^{[1]}\big)^2\big],
~~\Big({\rm tr}\big[\Pi^{[1]}\big]\Big)^2,
\vs
{\rm (3)} &&
{\rm tr}\big[\big(\Pi^{[1]}\big)^3\big],
~~{\rm tr}\big[\big(\Pi^{[1]}\big)^2\big]
   {\rm tr}\big[\Pi^{[1]}\big],
~~\Big({\rm tr}\big[\Pi^{[1]}\big]\Big)^3,
~~{\rm tr}\big[\Pi^{[1]}\Pi^{[2]}\big],
\vs
{\rm (4)} &&
{\rm tr}\big[\big(\Pi^{[1]}\big)^2\big]
   \Big({\rm tr}\big[\Pi^{[1]}\big]\Big)^2,
~~{\rm tr}\big[\big(\Pi^{[1]}\big)^3\big]
   {\rm tr}\big[\Pi^{[1]}\big],
~~\Big({\rm tr}\big[\big(\Pi^{[1]}\big)^2\big]\Big)^2,
~~\Big({\rm tr}\big[\Pi^{[1]}\big]\Big)^4,
\vs
&&
{\rm tr}\big[\Pi^{[1]}\Pi^{[3]}\big],
~~{\rm tr}\big[\Pi^{[1]}\Pi^{[1]}\Pi^{[2]}\big],
~~{\rm tr}\big[\Pi^{[2]}\Pi^{[2]}\big],
~~{\rm tr}\big[\Pi^{[1]}\big]
   {\rm tr}\big[\Pi^{[1]}\Pi^{[2]}\big],
\vs
{\rm (5)} &&
{\rm tr}\big[\big(\Pi^{[1]}\big)^3\big]
   {\rm tr}\big[\big(\Pi^{[1]}\big)^2\big],
~~{\rm tr}\big[\big(\Pi^{[1]}\big)^3\big]
   \Big({\rm tr}\big[\Pi^{[1]}\big]\Big)^2,
~~{\rm tr}\big[\big(\Pi^{[1]}\big)^2\big]
   \Big({\rm tr}\big[\Pi^{[1]}\big]\Big)^3,
\vs
&&
{\rm tr}\big[\Pi^{[1]}\big]
   \Big({\rm tr}\big[\big(\Pi^{[1]}\big)^2\big]\Big)^2,
~~\Big({\rm tr}\big[\Pi^{[1]}\big]\Big)^5,
~~{\rm tr}\big[\Pi^{[1]}\big]
   {\rm tr}\big[\Pi^{[1]}\Pi^{[3]}\big],
~~{\rm tr}\big[\Pi^{[1]}\Pi^{[4]}\big],
\vs
&&
{\rm tr}\big[\Pi^{[1]}\Pi^{[1]}\Pi^{[3]}\big],
~~{\rm tr}\big[\Pi^{[1]}\Pi^{[2]}\Pi^{[2]}\big],
~~{\rm tr}\big[\Pi^{[2]}\Pi^{[3]}\big],
~~{\rm tr}\big[\Pi^{[1]}\big]
   {\rm tr}\big[\Pi^{[2]}\Pi^{[2]}\big],
\vs
&&
{\rm tr}\big[\Pi^{[1]}\big]
   {\rm tr}\big[\Pi^{[1]}\Pi^{[1]}\Pi^{[2]}\big],
~~{\rm tr}\big[\Pi^{[1]}\Pi^{[1]}\big]
   {\rm tr}\big[\Pi^{[1]}\Pi^{[2]}\big],
~~\Big({\rm tr}\big[\Pi^{[1]}\big]\Big)^2
   {\rm tr}\big[\Pi^{[1]}\Pi^{[2]}\big]\,,
\nonumber
\eea
where the $\Pi$ operators are defined as
\begin{align}
\label{eq:Pi-def}
\Pi^{[n]}_{ij}&=\frac{1}{(n-1)!}
\left[\frac{1}{aHf}\frac{\mathrm{D}}{\mathrm{D}\tau}
\Pi^{[n-1]}_{ij}-(n-1)\Pi^{[n-1]}_{ij}\right],
\qquad n\geq2\,,
\end{align}
following Refs.~\cite{MSZ,Desjacques:2016bnm} (see \cite{Bakx:2025cvu} for the completeness of this basis up to fifth order), with 
\be
\Pi^{[1]}_{ij}=\frac{\partial_i\partial_j}{\nabla^2}\d\,.
\ee
Here $\mathrm{D}/\mathrm{D}\tau=\partial_\tau+v^k\partial_k$ is the convective derivative and $f=\mathrm{d}\ln D/\mathrm{d}\ln a$, with $D$ the linear growth factor.
In the presence of non-Gaussian initial conditions, operators containing powers of $\Psi$ are necessary to ensure closure under renormalization \cite{2008PhRvD..77l3514D,Assassi:2015fma,Nikolis:2024kbx}, as we discuss in \refsec{renorm}. The corresponding non-Gaussian operator basis is given by
\bea
\notag
\label{eq:bias_basis}
{\rm (1)} &&
\Psi,
\\
{\rm (2)} &&
{\rm tr}\big[\Pi^{[1]}\big]\Psi,
\;\Psi^2,
\vs
{\rm (3)} &&
{\rm tr}\big[\big(\Pi^{[1]}\big)^2\big]\Psi,
~~\Big({\rm tr}\big[\Pi^{[1]}\big]\Big)^2\Psi,
\;{\rm tr}\big[\Pi^{[1]}\big]\Psi^2,
\;\Psi^3,
\vs
{\rm (4)} &&
{\rm tr}\big[\big(\Pi^{[1]}\big)^3\big]\Psi,
~~{\rm tr}\big[\big(\Pi^{[1]}\big)^2\big]
   {\rm tr}\big[\Pi^{[1]}\big]\Psi,
~~\Big({\rm tr}\big[\Pi^{[1]}\big]\Big)^3\Psi,
~~{\rm tr}\big[\Pi^{[1]}\Pi^{[2]}\big]\Psi,
\vs
&&
{\rm tr}\big[\big(\Pi^{[1]}\big)^2\big]\Psi^2,
~~\Big({\rm tr}\big[\Pi^{[1]}\big]\Big)^2\Psi^2,
\;{\rm tr}\big[\Pi^{[1]}\big]\Psi^3,
\;\Psi^4,
\vs
{\rm (5)} &&
{\rm tr}\big[\big(\Pi^{[1]}\big)^2\big]
   \Big({\rm tr}\big[\Pi^{[1]}\big]\Big)^2\Psi,
~~{\rm tr}\big[\big(\Pi^{[1]}\big)^3\big]
   {\rm tr}\big[\Pi^{[1]}\big]\Psi,
~~\Big({\rm tr}\big[\big(\Pi^{[1]}\big)^2\big]\Big)^2\Psi,
~~\Big({\rm tr}\big[\Pi^{[1]}\big]\Big)^4\Psi,
\vs
&&
{\rm tr}\big[\Pi^{[1]}\Pi^{[3]}\big]\Psi,
~~{\rm tr}\big[\Pi^{[1]}\Pi^{[1]}\Pi^{[2]}\big]\Psi,
~~{\rm tr}\big[\Pi^{[2]}\Pi^{[2]}\big]\Psi,
~~{\rm tr}\big[\Pi^{[1]}\big]
   {\rm tr}\big[\Pi^{[1]}\Pi^{[2]}\big]\Psi,
\vs
&&
{\rm tr}\big[\big(\Pi^{[1]}\big)^3\big]\Psi^2,
~~{\rm tr}\big[\big(\Pi^{[1]}\big)^2\big]
   {\rm tr}\big[\Pi^{[1]}\big]\Psi^2,
~~\Big({\rm tr}\big[\Pi^{[1]}\big]\Big)^3\Psi^2,
~~{\rm tr}\big[\Pi^{[1]}\Pi^{[2]}\big]\Psi^2,
\vs
&&
{\rm tr}\big[\big(\Pi^{[1]}\big)^2\big]\Psi^3,
~~\Big({\rm tr}\big[\Pi^{[1]}\big]\Big)^2\Psi^3,
\;{\rm tr}\big[\Pi^{[1]}\big]\Psi^4,
\;\Psi^5.
\eea
Therefore, the full basis has a double expansion, both in the order of gravitational evolution and in the order of primordial non-Gaussianity.

The above basis is closed by the inclusion of different powers of \textit{just one} additional field $\Psi$. This is a consequence of PNG being generated by a single source, as is the case for the local shape we consider. As briefly mentioned at the end of Sec.~\ref{sec:soft_limits}, when the non-Gaussianity of the curvature perturbations is sourced by multiple coupled fields (e.g. in the curvaton scenario \cite{PhysRevD.82.043531} or in quasi-single-field inflation \cite{Chen_2010}), we need to include new operators, collectively denoted as $\hat{\psi}$, which satisfy $\langle\hat{\psi}\varphi\rangle=0$, as already noted in \cite{Assassi:2015fma}. The inclusion of these operators is necessary to describe the parametrically larger collapsed-limit amplitude, which is typical in these kinds of models, as well as residual small-scale modulation that could appear in the multi-soft limit of $n$-point functions.


\subsection{Extension to displaced non-Gaussian bias kernels} \label{sec:kernels}

Before turning to the renormalization computation, we need to specify how the non-Gaussian operators are perturbatively expanded. 
The operator $\Psi(\vx)$ defined in \refeq{Psi-def} is linear in the initial conditions and can therefore be written as $\Psi(\vx) = \Psi^{(1)}(\vx)$.
However, since primordial non-Gaussianity is imprinted in the initial conditions, the field entering the bias expansion must be evaluated at the Lagrangian position $\vq = \vx-\vdisp(\vq)$ associated with the final Eulerian position $\vx$ \cite{Assassi:2015fma,Desjacques:2016bnm, Nikolis:2024kbx}
\be
  \Psi(\vq) = \Psi(\vx-\vdisp(\vq))\,,
\ee
where $\vdisp$ is the displacement field. Evaluating $\Psi$ directly at $\vx$ would generate terms that violate Galilean invariance, whereas the displacement expansion of $\Psi(\vq)$ cancels these contributions.

Since the displacement itself is evaluated at the Lagrangian position, it must also be expanded around $\vx$,
\be
\vdisp(\vq)=\vdisp\bigl(\vx-\vdisp(\vq)\bigr)
  =\vdisp(\vx)-\vdisp^j(\vq)\partial_j\vdisp(\vx)+\cdots\,,
\ee
so we perform a perturbative expansion of $\vdisp(\vq)=\sum_{i=1}^{\infty} \vdisp^{(i)}(\vq(\vx))$ and subsequently Taylor expand $\vdisp^{(i)}(\vq)$ around $\vx$ (see also Appendix B of \cite{Desjacques:2016bnm}).
The displaced field $\Psi(\vq)$ can then be expanded order by order in perturbation theory.
In this paper, we expand $\Psi$ up to fifth order, corresponding to fourth order in the displacement,
with the corresponding expressions for the displacement expansion given in \refapp{disp}.

We turn now to the bias kernels. The kernel expansion of a Gaussian bias operator in powers of $\dlin$ is given by:
\be
\O(\vk) = \sum_{\ell = {\rm ord}(\O)}^\infty
\int_{\vp_1, \dots, \vp_\ell}
\diracpi(\vk - \vp_{1 \dots \ell})
K_\O^{(\ell)}(\vp_1, \dots, \vp_\ell)
\dlin(\vp_1) \cdots \dlin(\vp_\ell) \,,
\label{eq:Ok}
\ee
where ${\rm ord}(\O)$ denotes the perturbative order at which the operator $\O$ first contributes.
These operators depend on the cutoff $\L$ introduced to regulate the momentum integrals or alternatively the linear density fields.
We also define
\be
K_\O \equiv \sum_{\ell={\rm ord}(\O)}^\infty K^{(\ell)}_\O\,,
\ee
and take each kernel $K_\O^{(\ell)}$ to be symmetrized with respect to its $\ell$ arguments.

For non-Gaussian operators, $\Psi$ is linear in the primordial initial conditions, whereas the displaced field $\Psi(\vq)=\Psi\bigl(\vx-\vdisp(\vq)\bigr)$ contains higher-order contributions generated by the displacement expansion.
We parametrize these contributions by kernels $\mathcal{S}^{(n)}$, defined by
\ba
\Psi^{(n)}(\vk)
=
\int_{\vp_1,\dots,\vp_n}
\diracpi(\vk-\vp_{12\dots n})
\mathcal{S}^{(n)}(\vp_1;\vp_2,\dots,\vp_n)
\Psi^{(1)}(\vp_1)\dlin(\vp_2)\cdots\dlin(\vp_n) \,,
\label{eq:SPsi_def}
\ea
with explicit expressions given in \refapp{disp}.
Here $\Psi^{(n)}$ denotes the contribution of total perturbative order $n$, counting $\Psi^{(1)}$ as first order.
Moreover, we take $\mathcal{S}^{(n)}$ to be symmetrized over its last $n-1$ arguments, which correspond to the $n-1$ $\dlin$ legs and we use a semicolon in $\mathcal{S}^{(n)}(\vp_1;\vp_2,\dots,\vp_n)$ to separate the first argument, associated with the $\Psi^{(1)}$ leg, from the remaining $\dlin$ arguments.
Using \refeq{Psi-def}, one could instead express $\Psi^{(1)}$ in terms of $\dlin$ and absorb the additional $M^{-1}(k)$ factor into the kernel but we keep the $\Psi^{(1)}$ leg explicit for convenience.

A non-Gaussian operator of the form ${\cal O}_a\Psi$ can then be written at total order $n$ as
\ba\label{eq:OKbar_onePsi}
{\cal O}_{a_1}^{(n)}(\vk)
=
({\cal O}_a\Psi)^{(n)}(\vk)
&=
\int_{\vp_1,\dots,\vp_n}
\diracpi\left(\vk-\vp_{1\dots n}\right)
K_{a_1}^{(n)}(\vp_1;\vp_2,\dots,\vp_n) \\
& \hspace{4cm}
\times 
\Psi^{(1)}(\vp_1)\dlin(\vp_2)\cdots\dlin(\vp_n)\,, \nonumber
\ea
where the arguments of $K_{a_1}^{(n)}$ are ordered such that the first one corresponds to the $\Psi^{(1)}$ leg, while the remaining $n-1$ correspond to $\dlin$ legs.
The kernel is therefore given by
\bea
K_{a_1}^{(n)}
=
\left\{K_a^{(n-1)}\mathcal{S}^{(1)}\right\}_{\rm sym.}
+
\left\{K_a^{(n-2)}\mathcal{S}^{(2)}\right\}_{\rm sym.}
+\cdots\,,
\label{eq:Kbar_onePsi}
\eea
where the symmetrization is taken over all $\dlin$ momenta, while the first argument of $\mathcal{S}$, corresponding to the $\Psi^{(1)}$ leg, is held fixed.
More generally, for an operator of the form ${\cal O}_a\Psi^m$, the order-$n$ contribution contains $m$ primordial $\Psi^{(1)}$ legs and $n-m$ $\dlin$ legs:
\begin{equation}
\label{eq:OKbar_mPsi}
\boxed{
\begin{aligned}
{\cal O}_{a_m}^{(n)}(\vk)
&=
({\cal O}_a\Psi^m)^{(n)}(\vk)
\\
&=
\int_{\vp_1,\dots,\vp_n}
\diracpi\left(\vk-\vp_{12\dots n}\right)
K_{a_m}^{(n)}(\vp_1,\dots,\vp_m;\vp_{m+1},\dots,\vp_n)
\\
&\hspace{1cm}\times
\Psi^{(1)}(\vp_1)\cdots\Psi^{(1)}(\vp_m)
\dlin(\vp_{m+1})\cdots\dlin(\vp_n)\,,
\end{aligned}
}
\end{equation}
with
\begin{equation}
\label{eq:Kbar_mPsi}
K_{a_m}^{(n)}
=
\sum_{\ell+n_1+\cdots+n_m=n}
\left\{
K_a^{(\ell)}\mathcal{S}^{(n_1)}
\cdots
\mathcal{S}^{(n_m)}
\right\}_{\rm sym.}\,.
\end{equation}
The kernel $K_{a_m}^{(n)}$ is symmetrized separately over the $m$ $\Psi$ legs and the $n-m$ $\dlin$ legs, but not between legs of different types.

\newpage
\section{Renormalization of deterministic terms} 
\label{sec:renorm}

We focus in this section on the renormalization of deterministic bias operators in the presence of primordial non-Gaussianity. 
We outline the renormalization conditions in \refsec{renormdet}.
The one- and two-loop results are given in \refsec{renorm_1L_det} and \refsec{renorm_2L_det}, respectively.
We then discuss the generalization to higher loop orders in \refsec{renorm_higherL_det}.

\subsection{Bias renormalization} \label{sec:renormdet}

We define renormalized operators by
\be\label{eq:biasren}
\boxed{  [{\cal O}_B]\equiv{\cal O}_B+  Z_{BA}{\cal O}_A\,,}
\ee
with a sum over $A$ implicit so that \refeq{bias-expansion} can be written as
\ba
\label{eq:bias-expansion-renorm}
    \d_{g,\rm det}(\vk)&= b_{[A]} [\O_{A}](\vk) \,,
\ea
with $b_{[A]}$ the renormalized bias parameters. Organized by the number of $\Psi$ insertions, the matrix $Z_{BA}$ has the block structure
\be
  Z_{BA}
  =
  \left(
  \begin{array}{cccc}
  Z_{b_0 a_0} & Z_{b_0 a_1} & Z_{b_0 a_2} & \cdots \\
  Z_{b_1 a_0} & Z_{b_1 a_1} & Z_{b_1 a_2} & \cdots \\
  Z_{b_2 a_0} & Z_{b_2 a_1} & Z_{b_2 a_2} & \cdots \\
  \vdots      & \vdots      & \vdots      & \ddots
  \end{array}
  \right)\,.
\ee

This matrix is, in principle, infinite-dimensional, since soft limits of higher primordial $n$-point functions can generate arbitrarily high powers of $\Psi$.
Moreover, as we show below, different $\Psi^m$ orders of the $Z_{BA}$ matrix mix, since non-Gaussian interactions raise $m$, whereas $\Psi$ legs integrated over in the loop lower $m$.
In practice, this matrix is truncated at the orders in primordial non-Gaussianity and perturbation theory relevant for a given calculation.
We truncate the bias expansion at fifth order. At one loop, we use three external linear density fields and include PNG up to order 3 (non-vanishing 5-point function). We thus keep terms with four linear power spectra (three external and one in the loop), and bias--PNG combinations up to $b_{a_3}$, $\lambda_{\rm NL}^{(3)}b_{a_2}$, $\lambda_{\rm NL}^{(4)}b_{a_1}$, and $\lambda_{\rm NL}^{(5)}b_{a_0}$. At two loops, we use one external linear density field and include PNG up to order two. This corresponds to three linear power spectra (one external and two in the loops), and combinations up to $b_{a_2}$, $\lambda_{\rm NL}^{(3)}b_{a_1}$, and $\lambda_{\rm NL}^{(4)}b_{a_0}$.

The matrix $Z_{BA}$ is fixed by imposing the following renormalization condition on connected correlators:
\be\label{eq:rencond}
\boxed{  \langle [{\cal O}_A] \delta_L({\bm k}_1)\cdots\delta_L({\bm k}_n)\rangle\big|_{{\bm k}_i\to 0}
  =
  \langle {\cal O}_A \delta_L({\bm k}_1)\cdots\delta_L({\bm k}_n)\rangle_\text{tree}\big|_{{\bm k}_i\to 0}\,,}
\ee
following \cite{Assassi2014, McDonald:2006mx} (see \cite{Assassi:2015EFT,Assassi:2015fma} for the non-Gaussian case and \cite{Bakx:2026rmd} for a generalization of this condition that includes higher-derivative counterterms).
This condition removes the cutoff-dependent contributions when the external momenta are taken soft, ensuring that the renormalized operators have the same large-scale correlators as their tree-level expressions.
Notably, the renormalized bias defined according to \refeq{rencond} matches the $\Lambda$-dependent bias parameters in the limit $\Lambda \to 0$ \cite{Rubira:2023vzw}, as confirmed by lattice measurements \cite{Huang:2026zvi}.
An important difference from the Gaussian case is that \refeq{rencond} also includes contributions involving connected primordial correlators of order three and higher.
The renormalization must therefore consistently remove the cutoff dependence arising from {\it all} possible contraction patterns.
We return to this point below.

The mixing matrix can be expanded loop by loop,
\be \label{eq:Zloops}
  Z_{BA} = Z_{BA}^\text{1L}+Z_{BA}^\text{2L}+\dots\,.
\ee
In the following, we compute the one- and two-loop contributions to $Z_{BA}$ in the deterministic sector. 

\subsection{One-loop}
\label{sec:renorm_1L_det}

\subsubsection{One-loop renormalization}
We now turn to the one-loop renormalization of deterministic operators in the presence of PNG.
The results of this section generalize those of \cite{Assassi:2015fma} by including interactions beyond the leading $\fnl$ terms and therefore probing not only the squeezed limit of the bispectrum, but also multi-soft limits of higher-point functions.
Substituting \refeq{biasren} into \refeq{rencond} and truncating at one-loop order, the renormalization condition becomes
\be\label{eq:rencond1L}
0 =
\langle {\cal O}_{b_{m'}}\delta_L({\bm k}_1)\cdots\delta_L({\bm k}_n)\rangle_\text{1L}
\big|_{{\bm k}_i\to 0}
+
Z_{b_{m'} a_m}^\text{1L}\,
\langle {\cal O}_{a_m}\delta_L({\bm k}_1)\cdots\delta_L({\bm k}_n)\rangle_\text{tree}
\big|_{{\bm k}_i\to 0}\,,
\ee
where a sum over the repeated extended index $a_m$ is understood. 
Taking $n={\rm ord}({\cal O}_{a_m})$ isolates the contribution in which ${\cal O}_{a_m}^{(n)}$ contracts with the external legs only through two-point functions, the tree-level $(n+1)$-point function appearing in the second term of \refeq{rencond1L} is then
\ba\label{eq:treelevel1L}
\langle
{\cal O}_{a_m}^{(n)}\delta_L({\bm k}_1)\cdots\delta_L({\bm k}_n)
\rangle_\text{tree}'
=
m!(n-m)!\,
&K_{a_m}^{(n)}({\bm k}_1,\dots,{\bm k}_m;{\bm k}_{m+1},\dots,{\bm k}_n)\, \vs
&\quad \prod_{i=1}^{m} P_{\Psi\d}(k_i)
\prod_{j=m+1}^{n} P_{\d\d}(k_j)
+{\rm Sym.}\,,
\ea
with $P_{\Psi\d}$ defined in \refeq{Ppsidelta}.
Here and below, ${\rm Sym.}$ denotes the inequivalent assignments of the external momenta to the $\Psi$ and $\dlin$ blocks. The displayed term uses the assignment in which the first $m$ external momenta contract with the $\Psi$ legs. 

The one-loop contribution from ${\cal O}_{b_{m'}}$ is generated by contractions among its $\dlin$ and $\Psi$ entries and the external $\dlin$ fields.
Two legs of ${\cal O}_{b_{m'}}$ are tied together through the loop momentum ${\bm p}$, while the remaining legs either contract directly with external soft fields or enter the soft arguments of a primordial correlator.
We denote by $s$ the number of external soft momenta entering the primordial correlator.
The relevant object is then an $(s+2)$-point function,
\be
G^{(s+2)}_{\d\cdots\d\,X_1X_2}
(\vk_1,\dots,\vk_s,\vp,-\vp-\vk_{1\dots s})\,,
\ee
where the first $s$ entries are of $\d$ type, since they correspond to external $\dlin$ legs, while $X_1,X_2\in\{\d,\Psi\}$ denote the types of the two loop legs of ${\cal O}_{b_{m'}}$.
We define
\be \label{eq:q1l}
q \equiv \#\{\Psi\;{\rm among}\;X_1,X_2\}\,,
\qquad q=0,1,2\,,
\ee
to keep track of how many of the two integrated loop legs are $\Psi$-type.
In the following, the sum over $X_1$ and $X_2$ runs once over the three inequivalent pairs $(\d,\d)$, $(\Psi,\d)$, and $(\Psi,\Psi)$. 
Since the operator ${\cal O}_{b_{m'}}$ contains $m'$ $\Psi$ legs in total, the directly contracted legs must contain $m'-q$ $\Psi$ legs. The legs of ${\cal O}_{b_{m'}}$ that contract externally with  $\delta_L(\vk_i)$ can be either $\Psi$ or $\d$ legs, thus if $Y_i\in\{\d,\Psi\}$ denotes the type of the leg of ${\cal O}_{b_{m'}}$ that contracts directly with $\delta_L(\vk_i)$, the allowed assignments satisfy $\#\{\Psi\;{\rm among}\;Y_{s+1},\dots,Y_n\}=m'-q$. We order these directly contracted legs such that the $m'-q$ $\Psi$ legs come first.
Moreover, the allowed values of $q$ obey $q_{\rm min}(s)\equiv \max(0,m'-n+s)$ and $q_{\rm max}\equiv \min(2,m')$ and the range of $s$ is also restricted by $s_{\rm max} \equiv \min\left[n,\, n+2-{\rm ord}({\cal O}_{b_{m'}}) \right]$.
The one-loop contribution can then be written as
\ba\label{eq:1Lcorr_raw}
&\langle{\cal O}_{b_{m'}}\delta_L({\bm k}_1)\cdots\delta_L({\bm k}_n)\rangle_{\text{1L}}'
\Big|^{{}^{\rm soft}_{\rm 2pt}}
=
\vs
&\sum_{s=0}^{s_{\rm max}}\,\,
\sum_{\substack{X_1,X_2\\
q_{\rm min}(s)\le q\le q_{\rm max}}}
\,
\frac{m'!(n+2-s-m')!}{q!(2-q)!}
\left[
\prod_{i=s+1}^{s+m'-q} P_{\Psi\d}(k_i)
\right]
\left[
\prod_{j=s+m'-q+1}^{n} P_{\d\d}(k_j)
\right]\times
\vs
&
\int_{p<\Lambda}
K_{b_{m'}}^{(n+2-s)}
\!\left(\vp,-\vp-\vk_{1\dots s},\vk_{s+1},\dots,\vk_n\right)
G^{(s+2)}_{\d\cdots\d\,X_1X_2}
(\vk_1,\dots,\vk_s,\vp,-\vp-\vk_{1\dots s})
+{\rm Sym.}\,,
\ea
with the integral loop being represented by the diagram
\ba
\raisebox{-0.0cm}{\includegraphicsbox[scale=0.7]{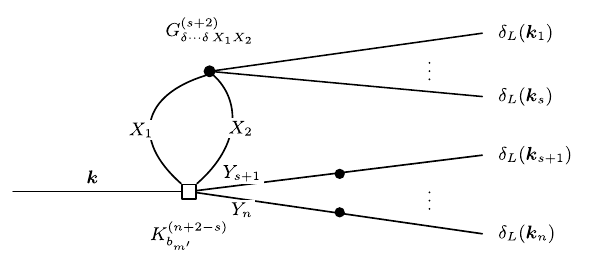}}\,,
\ea
and where for compactness, we have temporarily suppressed the ordering of the arguments of $K_{b_{m'}}^{(n+2-s)}$.
In \refeq{1Lcorr_raw}, we have restricted $\langle{\cal O}_{b_{m'}}\delta_L({\bm k}_1)\cdots\delta_L({\bm k}_n)\rangle_{\text{1L}}'\big|^{{}^{\rm soft}_{\rm 2pt}}$ to configurations in which the soft legs of $K_{b_{m'}}$ connect to the external legs only through two-point functions.
Other possible configurations, including soft higher-order $n$-point functions connected to the bias kernel via the external legs, must also be included, but they correct other diagrammatic structures in \refeq{rencond}. 

In the soft limit \refeq{rencond1L}, the one-loop result \refeq{1Lcorr_raw} factorizes into two independent contributions, discussed below.
The first depends only on the initial conditions through the primordial correlator $G^{(s+2)}$, which we evaluate using the soft limits derived in \refsec{soft_limits}.
The second is the soft limit of the bias kernel $K_{b_{m'}}^{(n-s+2)}$, which is fixed by the symmetries of gravitational evolution.

The $s$-soft limit of the primordial $(s+2)$-point function in \refeq{multi-soft} is
\be\label{eq:Ghardsoft}
\boxed{G^{(s+2)}_{\d\cdots\d\, X_1 X_2}
(\vk_1,\dots,\vk_s,\vp,-\vp-\vk_{1\dots s})  \Big|^{{}^{\rm soft}_{\rm 2pt}}
\;\xrightarrow[\,k_i\ll p\,]{}\;
\lambda_{\rm NL}^{(s+2)} P_{X_1 X_2}(p)
 \prod_{i=1}^s P_{\Psi\d}(k_i)\,,
}\ee
where we have also used \refeq{Ppsidelta} and $\lambda_{\rm NL}^{(s+2)}$ is given by \refeq{lambdas}. 
We have retained the term in which each soft leg factorizes into a two-point function, as in \refeq{multi-soft}, since the remaining terms contribute to diagrams other than those displayed in \refeq{1Lcorr_raw}.
Moreover, the same soft limits of primordial $n$-point functions appear in more general inflationary scenarios, and here we use single-source local PNG as an example, with the limits derived in \refsec{soft_limits}.
Therefore, the $q$ $\Psi$ legs among $X_1$ and $X_2$ are integrated over in the loop and appear in $P_{X_1X_2}(p)$.
After taking the $k_i\to0$ limit of \refeq{1Lcorr_raw} and comparing with the tree-level structure in \refeq{treelevel1L}, counting the $\Psi$ legs gives the selection rule
\be \label{eq:fixed_condition_m}
\boxed{m=m'+s-q\,,}
\ee
which is enforced by \refeq{rencond1L}.
Thus, the $G^{(s+2)}$ interaction increases the number of external $\Psi$ legs by $s$, while the integrated $\Psi$ legs reduce it by $q$.

We now turn to the soft contraction of the bias kernel, which at leading order in derivatives is given by
\be
\label{eq:biaskernelexpansionsinglehard}
\boxed{
\begin{aligned}
&
K_{b_{m'}}^{(n-s+2)}
\left(\vp,-\vp-\vk_{1\dots s},\vk_{s+1},\dots,\vk_n\right)
\Big|^{p\to\infty}_{\text{av}_{\hat{\bf p}}}
\\[-2pt]
&\qquad =
\left[
\binom{m'}{q}
\binom{n+2-s-m'}{2-q}
\right]^{-1}
\,s^{\rm 1L}_{b_{m'} a_{m'-q+s};\,q}\,
K_{a_{m'-q+s}}^{(n)}
\left(\vk_1,\dots,\vk_n\right)\,.
\end{aligned}
}
\ee
Here, $s=m-m'+q$, $\text{av}_{\hat{\bf p}}$ denotes the angular average over the direction of $\vp$ and we have not identified the $\Psi$ legs into the kernel arguments. Also \refeq{biaskernelexpansionsinglehard} is valid for every $(q,s)$ pair, and no sum over them is implicit.
The coefficients $s^{\rm 1L}_{b_{m'} a_{m'-q+s};\,q}$ are given in \refapp{tables}.
Tables for these coefficients are provided in \reftab{singlehardcoeff_a0b_0} for the Gaussian case, in \reftab{singlehardcoeff_a0b_q0} for $m'=0$ and non-Gaussian contractions, while the case $m'=1$ is given in \reftab{singlehardcoeff_a1b0_q1}, \reftab{singlehardcoeff_a1b1}, and \reftab{singlehardcoeff_a1b23}. For $m'=2$, see \reftab{singlehardcoeff_a2b0_q2}, \reftab{singlehardcoeff_a2b1}, and \reftab{singlehardcoeff_a2b23}, and for $m'=3$, \reftab{singlehardcoeff_a3b}.
We truncate the computation at fifth order in the bias expansion, counting $\Psi$ as first order, with three external linear fields and PNG up to order 3, which describes a non-vanishing 5-point function (see \refapp{tables} for more details). This keeps bias terms up to $b_{a_3}$, $\lambda^{(3)}_{\rm NL}b_{a_2}$, $\lambda^{(4)}_{\rm NL}b_{a_1}$, and $\lambda^{(5)}_{\rm NL}b_{a_0}$.

Two properties of $s^{\rm 1L}_{b_{m'} a_{m'-q+s};\,q}$ are worth emphasizing. First, these coefficients are independent of the perturbative order $n$, because \refeq{biaskernelexpansionsinglehard} is a field-level relation between the two operators. Second, specifying the operator indices alone does not uniquely determine the coefficients since one must also specify the number of integrated legs $q$, or equivalently $s$, as related by \refeq{fixed_condition_m}.
Finally, note that $s=q=m'=0$ recovers the Gaussian case, with the prefactor in \refeq{biaskernelexpansionsinglehard} becoming $\binom{n+2}{2}^{-1}=2/[(n+2)(n+1)]$.
In this limit, $G^{(2)}_{\d\d}(\vp,-\vp)\to\Plin(p)$, and \refeq{biaskernelexpansionsinglehard} reduces to Eq.~(3.28) of \cite{Bakx:2025cvu}, with $s^{\rm 1L}_{b_0a_0;\,0}=s^{\rm 1L}_{ba}$ the coefficient matrix introduced in \cite{Rubira:2023vzw}.

\begin{figure}[t]
    \centering
    \includegraphics[width=0.7\textwidth]
    {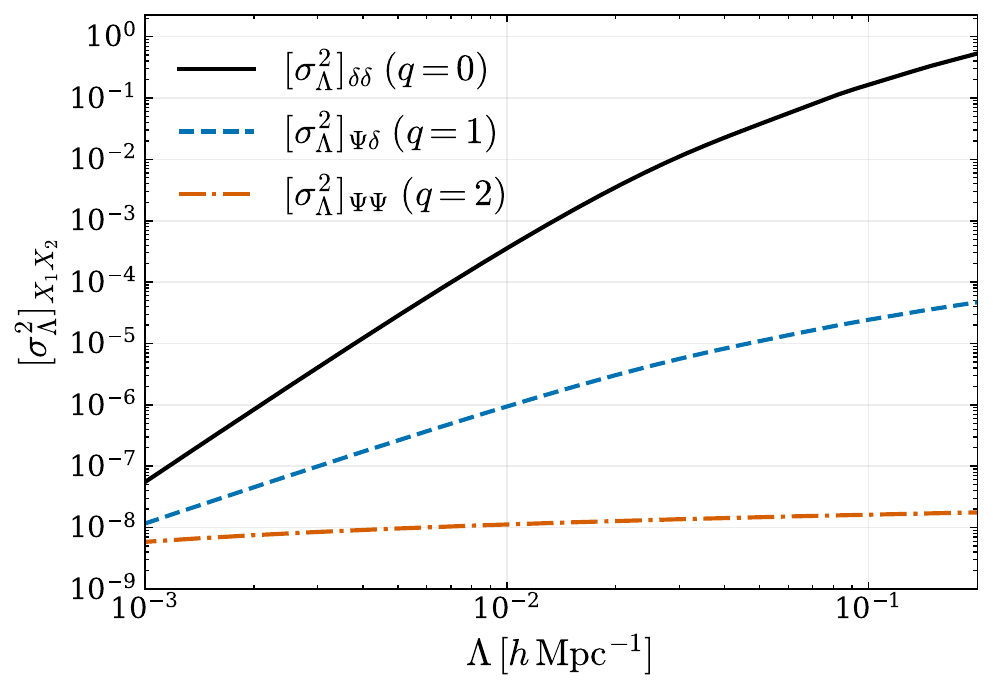}
    \caption{Generalized variances $[\sigma_\Lambda^2]_{X_1X_2}$ \refeq{sigmatildedef} for $\Lambda$CDM cosmology, as a function of the cutoff $\Lambda$ at $z=0$. The curves correspond to $q=0$ (black solid), $q=1$ (blue dashed), and $q=2$ (orange dash-dotted), where $q$ defined in \refeq{q1l} counts the $\Psi$ fields in the pair $(X_1,X_2)$.}
    \label{fig:generalized_variance}
\end{figure}

Using \refeq{Ghardsoft} together with \refeq{biaskernelexpansionsinglehard}, the one-loop correlator reduces, in the soft external limit, to
\ba\label{eq:1Lcorr_reduced}
&\langle{\cal O}_{b_{m'}}\delta_L({\bm k}_1)\cdots\delta_L({\bm k}_n)\rangle_{\text{1L}}'
\Big|_{\vk_i\to 0}^{{}^{\rm soft}_{\rm 2pt}}  =
\sum_{s=0}^{s_{\rm max}}
\sum_{\substack{X_1,X_2\\
q_{\rm min}(s)\le q\le q_{\rm max}}}
\frac{m!(n-m)!}{s!}\,
s^{\rm 1L}_{b_{m'} a_{m};\,q}\, \lambda_{\rm NL}^{(s+2)}
\left[\sigma_\Lambda^{2}\right]_{X_1X_2}
\vs
&\hspace{1.2cm}\times
K_{a_m}^{(n)}(\vk_1,\dots,\vk_m;\vk_{m+1},\dots,\vk_n)
\prod_{i=1}^{m} P_{\Psi\d}(k_i)
\prod_{j=m+1}^{n} P_{\d\d}(k_j)
+{\rm Sym.}\,,
\ea
where we have defined the generalized (dimensionless) variance
\be\label{eq:sigmatildedef}
\left[\sigma_\Lambda^{2}\right]_{X_1 X_2}
\equiv
\int_{p<\Lambda}\! P_{X_1 X_2}(p)\,.
\ee
The Gaussian case is recovered for $(X_1,X_2)=(\d,\d)$, with $\sigma_\Lambda^2 = \left[\sigma_\Lambda^{2}\right]_{\d\d}$, which controls the mixing of operators $\O_{b_{m'}}$ into operators $\O_{a_m}$ with $m=m'$, including the renormalization of Gaussian operators by other Gaussian operators \cite{Rubira:2023vzw,Bakx:2025cvu,Bakx:2026rmd}.  We plot the generalized variances in Fig.~\ref{fig:generalized_variance}, where we note that when increasing $q$, the variance is flattened compared to $\sigma^2$ valid for $q=0$, which has more power on small scales.

Comparing \refeq{1Lcorr_reduced} with the tree-level expression \refeq{treelevel1L}, and using the one-loop renormalization condition \refeq{rencond1L}, we find 
\be\label{eq:Z1L}
\boxed{\;
Z_{b_{m'} a_m}^\text{1L}
=
-
\sum_{s=0}^{s_{\rm max}}
\sum_{\substack{X_1,X_2\\
q_{\rm min}(s)\le q\le q_{\rm max}}}
\,
\frac{1}{s!}\,\lambda_{\rm NL}^{(s+2)}
\left[\sigma_\Lambda^{2}\right]_{X_1 X_2}\,
s^{\rm 1L}_{b_{m'} a_{m};\,q}
\,\delta^{\rm K}_{m,m'+s-q}
\;,}
\ee
where the Kronecker delta imposes the selection rule $m=m'+s-q$ from \refeq{fixed_condition_m}.
We have explicitly verified that the single-hard limits of the bias kernels in \refeq{biaskernelexpansionsinglehard} close on the operator basis listed in \refsec{basis}, so that no additional operators are required. 
We depict the renormalization of a loop diagram and the term absorbing its cutoff dependence as
\ba \label{eq:diagslimits} 
\raisebox{-0.0cm}{\includegraphicsbox[scale=1.0]{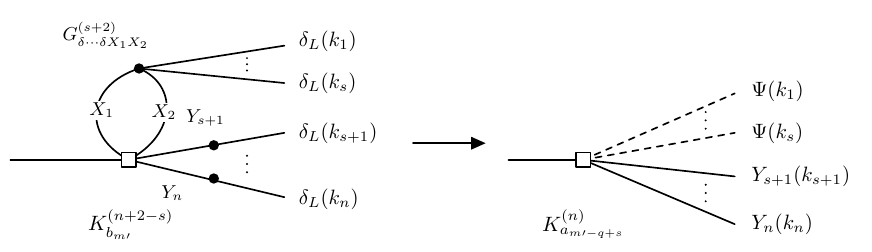}} \,,
\ea
recapping that the $Y_i = {\delta, \Psi}$ legs are these that do not enter the hard loop.

\subsubsection{Examples}
We now consider some illustrative examples of \refeq{Z1L}.
For the renormalized Gaussian operator, $\O_{b_0}=\O_b$, \refeq{fixed_condition_m} reduces to $m=s$ (noting that $q=0$).
Thus, truncating at the four-point function interactions, such that $s=0,1,2$, gives
\bea
[\O_{b_0}]
&=&
\O_{b_0}
+
Z_{b_0 a_0}^{\rm 1L}\O_{a_0}
+
Z_{b_0 a_1}^{\rm 1L}\O_{a_1}
+
Z_{b_0 a_2}^{\rm 1L}\O_{a_2}
+\cdots
\\
&=&
\O_{b_0}
-
\left[\sigma_\Lambda^2\right]_{\d\d}
s^{\rm 1L}_{b_0 a_0;\,0}\O_{a_0}
-
4f_{\rm NL}
\left[\sigma_\Lambda^2\right]_{\d\d}
s^{\rm 1L}_{b_0 a_1;\,0}\O_{a_1}
-
6g_{\rm NL}
\left[\sigma_\Lambda^2\right]_{\d\d}
s^{\rm 1L}_{b_0 a_2;\,0}\O_{a_2}
+\cdots\,. \nonumber
\eea

For operators $\O_{b_1}$ containing one $\Psi$, \refeq{fixed_condition_m} gives $m=1+s-q$, with $q=0$ for $\d\d$ loop legs and $q=1$ for the ordered $\Psi\d$ loop pair.
Truncating again at the primordial four-point function gives
\bea
[\O_{b_1}]
&=&
\O_{b_1}
+
Z_{b_1 a_0}^{\rm 1L}\O_{a_0}
+
Z_{b_1 a_1}^{\rm 1L}\O_{a_1}
+
Z_{b_1 a_2}^{\rm 1L}\O_{a_2}
+
Z_{b_1 a_3}^{\rm 1L}\O_{a_3}
+\cdots
\\
&=&
\O_{b_1}
-
\left[\sigma_\Lambda^2\right]_{\Psi\d}
s^{\rm 1L}_{b_1 a_0;\,1}\O_{a_0}
-
\left(
\left[\sigma_\Lambda^2\right]_{\d\d}
s^{\rm 1L}_{b_1 a_1;\,0}
+
4f_{\rm NL}
\left[\sigma_\Lambda^2\right]_{\Psi\d}
s^{\rm 1L}_{b_1 a_1;\,1}
\right)\O_{a_1}
\nonumber\\
&&\hspace{-0.5cm}
-
\left(
4f_{\rm NL}
\left[\sigma_\Lambda^2\right]_{\d\d}
s^{\rm 1L}_{b_1 a_2;\,0}
+
6g_{\rm NL}
\left[\sigma_\Lambda^2\right]_{\Psi\d}
s^{\rm 1L}_{b_1 a_2;\,1}
\right)\O_{a_2}
-
6g_{\rm NL}
\left[\sigma_\Lambda^2\right]_{\d\d}
s^{\rm 1L}_{b_1 a_3;\,0}\O_{a_3}
+\cdots\,. \nonumber
\eea
The absence of a factor of two in the mixed $\Psi\d$ terms follows from counting the mixed hard pair only once. 
Here, we define
\be
\left[\sigma_\Lambda^2\right]_{\Psi\d}
\equiv
\int_{p<\Lambda} P_{\Psi\d}(p) =\int_{p<\Lambda} \frac{1}{M(p)}\Plin(p)\,.
\ee
The coefficients entering this example are listed in \reftab{singlehardcoeff_a1b0_q1}, \reftab{singlehardcoeff_a1b1} and \reftab{singlehardcoeff_a1b23}.

To conclude, the one-loop renormalization is sensitive to the $(n-2)$-soft structure of operator kernels through \refeq{biaskernelexpansionsinglehard} and to the corresponding multi-soft structure of primordial $n$-point functions through \refeq{Ghardsoft}, so primordial shapes enhanced in these configurations give the dominant contributions.
The mixing matrix $Z_{b_{m'} a_m}^\text{1L}$ in \refeq{Z1L} relates different orders $\Psi^m$ and $\Psi^{m'}$, with $m=m'+s-q$ according to \refeq{fixed_condition_m}, where soft primordial correlators increase the number of external $\Psi$ legs by $s$, while $\Psi$ fields integrated over in the loop reduce it by $q$. 
Primordial higher-point functions therefore generate new non-Gaussian terms, parametrically controlled by generalized loop variances in \refeq{sigmatildedef}, where hard $\Psi$ legs instead replace $P_{\d\d}(p)$ by $P_{\Psi\d}(p)=[1/M(p)]P_{\d\d}(p)$ or $P_{\Psi\Psi}(p)=[1/M^2(p)]P_{\d\d}(p)$. The same transfer-function dependence on the external soft legs gives rise to the familiar scale-dependent bias \cite{2008PhRvD..77l3514D,2008ApJ...677L..77M}.
The relative importance of the different blocks of $Z^{\rm 1L}_{BA}$ is thus controlled by the PNG amplitude $\lambda^{(s+2)}_{\rm NL}$ and the factor $1/M(p)$ for each $\Psi$ leg running in the loop.

\subsection{Two-loop}
\label{sec:renorm_2L_det}

We now extend the previous one-loop discussion to two loops. 
Expanding the renormalization condition \refeq{rencond} to two-loop order gives
\ba\label{eq:rencond2L}
0
=&\,
\langle {\cal O}_{c_{m''}}\delta_L({\bm k}_1)\cdots\delta_L({\bm k}_n)
\rangle_\text{2L}
\big|_{{\bm k}_i\to 0}
+
Z_{c_{m''} b_{m'}}^{\rm 1L}
\langle {\cal O}_{b_{m'}}\delta_L({\bm k}_1)\cdots\delta_L({\bm k}_n)
\rangle_\text{1L}
\big|_{{\bm k}_i\to 0}
\vs
&+
Z_{c_{m''} a_m}^{\rm 2L}
\langle {\cal O}_{a_m}\delta_L({\bm k}_1)\cdots\delta_L({\bm k}_n)
\rangle_\text{tree}
\big|_{{\bm k}_i\to 0}\,,
\ea
where sums over the repeated extended indices $b_{m'}$ and $a_m$ are understood. The first term is the bare two-loop contribution from the operator ${\cal O}_{c_{m''}}$, the second term corresponds to the one-loop contribution \refeq{1Lcorr_reduced}, and the last term is the tree-level contribution of \refeq{treelevel1L}.

At two loops there are two distinct contraction topologies.
The first topology contains two hard pairs, $(X_1,X_2)$ and $(X_3,X_4)$, where $X_i\in\{\d,\Psi\}$, and already appears for Gaussian initial conditions \cite{Bakx:2025cvu,Bakx:2026rmd}. The second contains a single connected block with three hard legs $(X_1,X_2,X_3)$ and is present when the hard primordial bispectrum is nonzero. 
Diagrammatically, the two structures are  
\ba \label{eq:diagslimits2loop} 
\raisebox{-0.0cm}{\includegraphicsbox[scale=1.0]{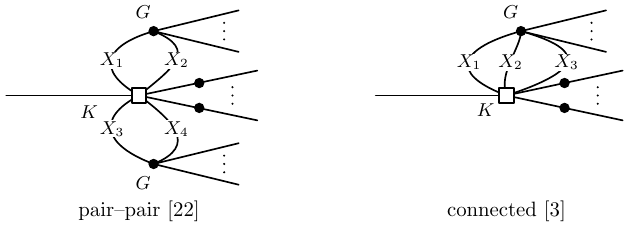}}\,.
\ea
We refer to these as the {\it pair-pair topology} and the {\it connected topology}. 
For either topology, we generalize \refeq{q1l} by writing
\be \label{eq:q1l_general}
q \equiv \#\{\Psi\;{\rm among}\;X_i\}\,.
\ee
For the pair-pair topology, we further denote by $q_1$ and $q_2$ the numbers of $\Psi$ legs in $(X_1,X_2)$ and $(X_3,X_4)$, respectively, such that $q=q_1+q_2$. Each pair is summed once over the three inequivalent possibilities $(\Psi,\Psi)$, $(\Psi,\d)$, and $(\d,\d)$. For the connected topology, the hard triple is summed once over $(\Psi,\Psi,\Psi)$, $(\Psi,\Psi,\d)$, $(\Psi,\d,\d)$, and $(\d,\d,\d)$.

To make the momentum dependence of the two topologies explicit, we first recall the schematic form of the one-loop contribution from \refeq{1Lcorr_raw}, suppressing sums, combinatorial factors, and external power spectra:
\ba
\int_{p<\Lambda}
K_{b_{m'}}^{(n+2-s)}
\!\left(
\vp,-\vp-\vk_{1\dots s},
\vk_{s+1},\dots,\vk_n
\right)
G^{(s+2)}_{\d\cdots\d\,X_1X_2}
\!\left(
\vk_1,\dots,\vk_s,
\vp,-\vp-\vk_{1\dots s}
\right)
\;,
\ea
with $\vk_{i\dots j}\equiv\vk_i+\cdots+\vk_j$, and $s$ counts the external legs attached to the primordial correlator.
At two loops, we denote the mixing matrices associated with the pair-pair and connected topologies by $Z^{\rm 2L,\,[22]}_{c_{m''}a_m}$ and $Z^{\rm 2L,\,[3]}_{c_{m''}a_m}$, respectively.
Schematically, these contributions scale as the integrals below, where the soft limit of the external momenta, matching onto the tree-level structure of ${\cal O}_{a_m}$, and sums over the allowed $q$ and $s$ configurations are understood:
\ba
&Z^{\rm 2L,\,[3]}_{c_{m''}a_m}\sim
\int_{p_1<\Lambda,p_2<\Lambda}
K_{c_{m''}}^{(n+3-s)}
\!\left(
\vp_1,\vp_2,-\vp_1-\vp_2-\vk_{1\dots s},
\vk_{s+1},\dots,\vk_n
\right)
\\
&\qquad \qquad \qquad
\times G^{(s+3)}_{\d\cdots\d\,X_1X_2X_3}
\!\left(
\vk_1,\dots,\vk_s,
\vp_1,\vp_2,-\vp_1-\vp_2-\vk_{1\dots s}
\right)
\;,
\vs[1ex]
&Z^{\rm 2L,\,[22]}_{c_{m''}a_m}\sim
\int_{p_1<\Lambda,p_2<\Lambda}
K_{c_{m''}}^{(n+4-s)}
\!\left(
\vp_1,-\vp_1-\vk_{1\dots s_1},
\vp_2,-\vp_2-\vk_{s_1+1\dots s},
\vk_{s+1},\dots,\vk_n
\right)
\vs
&
\qquad \qquad \qquad \times G^{(s_1+2)}_{\d\cdots\d\,X_1X_2}
\!\left(
\vk_1,\dots,\vk_{s_1},
\vp_1,-\vp_1-\vk_{1\dots s_1}
\right) \\
& \qquad \qquad \qquad \times G^{(s_2+2)}_{\d\cdots\d\,X_3X_4}
\!\left(
\vk_{s_1+1},\dots,\vk_s,
\vp_2,-\vp_2-\vk_{s_1+1\dots s}
\right). \nonumber
\ea
For the pair-pair topology, $s=s_1+s_2$, with $s_1$ and $s_2$ counting the external legs attached to the first and second primordial correlators, respectively. We analyze each topology separately.

\subsubsection{The pair-pair topology}

For the pair-pair topology, we assign $s_1$ external soft legs to the first hard pair through $G^{(s_1+2)}_{\d\cdots\d\,X_1X_2}$ and $s_2$ external soft legs to the second through $G^{(s_2+2)}_{\d\cdots\d\,X_3X_4}$. Each primordial correlator therefore probes the same multi-soft configuration as at one loop, with the soft limit given by \refeq{Ghardsoft}. Counting the external $\Psi$ legs and comparing with \refeq{treelevel1L} gives
\be\label{eq:mdef2l_pairpair}
m=m''+s_1+s_2-q\,.
\ee
At leading order in derivatives, the corresponding kernel limit is
\be\label{eq:doublehard_pairpair}
\boxed{
\begin{aligned}
&K_{c_{m''}}^{(n+4-s_1-s_2)}
\Bigl(
\vp_1,-\vp_1-\vk_{1\dots s_1},
\vp_2,
-\vp_2-\vk_{s_1+1\dots s_1+s_2},
\vk_{s_1+s_2+1},\dots,\vk_n
\Bigr)
\Big|_{\text{av}_{\hat{\bf p}_1,\hat{\bf p}_2}}^{\substack{p_1,p_2\gg k_i\\r=p_1/p_2\ {\rm fixed}}}
\\[2pt]
&\quad=
\mathcal{A}\;
s^{\rm 2L,[22]}_{c_{m''}a_{m''-q+ s_1+s_2};\,q_1q_2}(r)
K_{a_{m''-q+s_1+s_2}}^{(n)}
\left(\vk_1,\dots,\vk_n
\right)\,,
\end{aligned}
}
\ee
valid for {\it every} $(q_1,q_2,s_1,s_2)$, with no sum implicit, and with $\mathcal{A}$ a combinatorial prefactor given by
\ba
\mathcal{A}=\left[
\binom{m''}{q_1}
\binom{m''-q_1}{q_2}
\binom{n+4-s_1-s_2-m''}{2-q_1}
\binom{n+2-s_1-s_2-m''+q_1}{2-q_2}
\right]^{-1}\;.
\ea
Here $r=p_1/p_2$, and the angular average is over both hard directions. Also the coefficients $s^{\rm 2L,[22]}_{c_{m''}a_{m''-q+s_1+s_2};\,q_1q_2}(r)$ specify which $q_1$ and $q_2$ legs carry a $\Psi$ leg.
Moreover, as in \refeq{1Lcorr_raw}, the ordering of the arguments of $K^{(n+4-s_1-s_2)}_{c_{m''}}$ has been suppressed, and it is understood that the arguments carrying $\vp_1$ and $\vp_2$ are of type $(X_1,X_2)$ and $(X_3,X_4)$, respectively.
Therefore, relabeling $n-s_1-s_2+4\to n$ we find that the two-loop pair-pair case leads to an $(n-4)$-soft limit of the bias kernels. The kernel limit removes $q=q_1+q_2$ $\Psi$ legs, while the two primordial soft limits add $s_1+s_2$ external $\Psi$ legs, consistently with \refeq{mdef2l_pairpair}. 

\begin{figure}[t]
    \centering
    \includegraphics[width=0.7\textwidth]
    {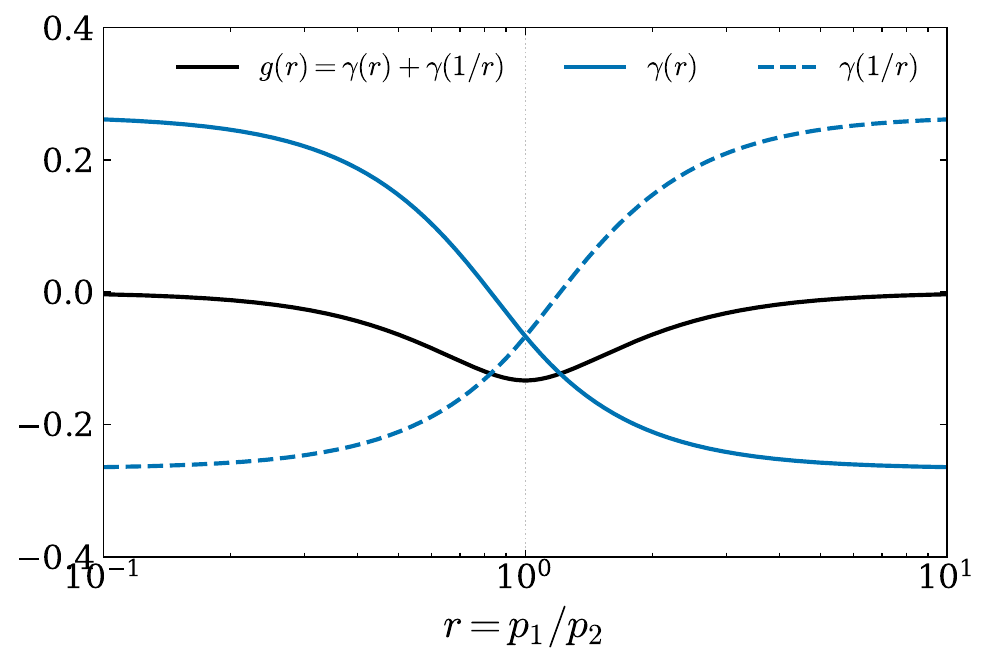}
    \caption{Functions controlling the dependence of the two-loop pair-pair kernel coefficients on the hard-momentum ratio $r=p_1/p_2$, defined in \refeq{gammafunc} and \refeq{doublehardfuncg}. Exchanging the two hard pairs, $r\to1/r$, interchanges the blue curves and leaves $g(r)$ invariant.}
    \label{fig:ggamas}
\end{figure}

Finally, we note some properties of $s^{\rm 2L,[22]}_{c_{m''}a_{m''-q+s_1+s_2};\,q_1q_2}$.
The exchange of the two hard pairs implies
\be
s^{\rm 2L,[22]}_{c_{m''}a_{m''-q+s_1+s_2};\,q_1q_2}(r)
=
s^{\rm 2L,[22]}_{c_{m''}a_{m''-q+s_1+s_2};\,q_2q_1}(1/r)\,.
\ee
Computing the coefficients $s^{\rm 2L,[22]}_{c_{m''}a_{m''-q+s_1+s_2};\,q_1q_2}$ up to $q \leq 2$, we find that the $s^{\rm 2L,[22]}$ functions can be further decomposed into the constant matrices $\hat{s}^{\rm 2L,[22]}$ and $\tilde{s}^{\rm 2L,[22]}$ as 
\ba \label{eq:consist22}
s^{\rm 2L,[22]}_{c_{m''}a_{m''+s_1+s_2};\,00}(r)
&=
\hat{s}^{\rm 2L,[22]}_{c_{m''}a_{m''+s_1+s_2}{;\,00}}
\,\,\,\,\,+
\tilde{s}^{\rm 2L,[22]}_{c_{m''}a_{m''+s_1+s_2}{;\,00}}
\;\,\,\,\,g(r)\;,\vs
s^{\rm 2L,[22]}_{c_{m''}a_{m''-1+s_1+s_2};\,10}(r)
&=
\hat{s}^{\rm 2L,[22]}_{c_{m''}a_{m''-1+s_1+s_2}{;\,10}}
+
\tilde{s}^{\rm 2L,[22]}_{c_{m''}a_{m''-1+s_1+s_2}{;\,10}}
\,\gamma(r)\;,\vs
s^{\rm 2L,[22]}_{c_{m''}a_{m''-1+s_1+s_2};\,01}(r)
&=
\hat{s}^{\rm 2L,[22]}_{c_{m''}a_{m''-1+s_1+s_2}{;\,10}}
+
\tilde{s}^{\rm 2L,[22]}_{c_{m''}a_{m''-1+s_1+s_2}{;\,10}}
\,\gamma(1/r)\;,\vs
s^{\rm 2L,[22]}_{c_{m''}a_{m''-2+s_1+s_2};\,11}(r)
&=
\hat{s}^{\rm 2L,[22]}_{c_{m''}a_{m''-2+s_1+s_2}{;\,11}}
+
\tilde{s}^{\rm 2L,[22]}_{c_{m''}a_{m''-2+s_1+s_2}{;\,11}}
\,g(r)\;,\vs
s^{\rm 2L,[22]}_{c_{m''}a_{m''-2+s_1+s_2};\,20}(r)
&=
\hat{s}^{\rm 2L,[22]}_{c_{m''}a_{m''-2+s_1+s_2}{;\,20}}
+
\tilde{s}^{\rm 2L,[22]}_{c_{m''}a_{m''-2+s_1+s_2}{;\,20}}
\,\gamma(r)\;,\vs
s^{\rm 2L,[22]}_{c_{m''}a_{m''-2+s_1+s_2};\,02}(r)
&=
\hat{s}^{\rm 2L,[22]}_{c_{m''}a_{m''-2+s_1+s_2}{;\,20}}
+
\tilde{s}^{\rm 2L,[22]}_{c_{m''}a_{m''-2+s_1+s_2}{;\,20}}
\,\gamma(1/r)\;,
\ea
such that the full $r$ dependence is absorbed in {\it one single function} 
\bea \label{eq:gammafunc}
\boxed{\gamma(r) \equiv \frac{15r^6-55r^4+9r^2+15}{240r^2}-\frac{(r^2-1)^4}{64r^3}\log\left[\left(\frac{1+r}{1-r}\right)^2\right]}\;,
\eea
that sets the full parametric running of the pair-pair two-loop RG in the non-Gaussian case. 
Note that the second and third lines share the same constant coefficients $\hat{s}^{\rm 2L,[22]}$ and $\tilde{s}^{\rm 2L,[22]}$ with $q=1$, while the last two lines share those with $q=2$.
Tables for $\hat s^{\rm 2L,[22]}$ and $\tilde s^{\rm 2L,[22]}$ are provided in \reftab{twoloopcoeff_Gauss} for the Gaussian case, in \reftab{twoloopcoeff_a_000} for $m''=0$ with PNG and in \reftab{twoloopcoeff_aPsi} for $m''=1$.
For $m''=2$, \reftab{twoloopcoeff_aPsi2} lists the $q=1$ coefficients together with $s^{\rm 2L,[22]}_{;\,11}$ and $s^{\rm 2L,[22]}_{;\,20}=s^{\rm 2L,[22]}_{;\,02}$, which we find to be independent of $r$.
At two loops, we truncate at one external linear field and at PNG of order 2.
This keeps bias terms up to $b_{a_2}$, $\lambda^{(3)}_{\rm NL}b_{a_1}$, and $\lambda^{(4)}_{\rm NL}b_{a_0}$.
We also find $\tilde{s}^{\rm 2L,[22]}_{c_{m''}a_{m''-2+s_1+s_2}{;\,11}} =  \tilde{s}^{\rm 2L,[22]}_{c_{m''}a_{m''-2+s_1+s_2}{;\,20}} = 0$.
The function $g(r)$, which appears in the Gaussian case \cite{Bakx:2025cvu}, can be expressed in terms of the symmetrized $\gamma$
\ba\label{eq:doublehardfuncg}
  g(r) &= \gamma(r)+\gamma(1/r) \vs 
  &= \frac{15r^8 - 40r^6 + 18r^4- 40r^2 + 15}{240r^4}
  -\frac{(r^2-1)^4(r^2+1)}{64r^5}\log\left(\frac{1+r}{1-r}\right)^2\,,
\ea
with the property $g(p/q)=g(q/p)=\gamma(p/q)+\gamma(q/p)$, which we plot in \reffig{ggamas}. In other words, the function $g(r)$ arises as a limiting case when the resulting calculation has symmetry under $p\leftrightarrow q$. 

Provided the two hierarchical hard limits commute, as in the Gaussian calculation of \cite{Bakx:2025cvu,Bakx:2026rmd}, the two-loop coefficient reduces to two successive one-loop contractions 
\be\label{eq:commutationoflimitsrelation}
\begin{aligned}
s^{\rm 2L,[22]}_{c_{m''}a_{m''-q+s_1+s_2};\,q_1q_2}(\infty)
&=
\frac{\binom{m''-q_1}{q_2}}{\binom{m''-q_1+s_1}{q_2}}s^{{\rm 1L}}_{c_{m''}b_{m''+s_1-q_1};\,q_1}
s^{{\rm 1L}}_{b_{m''+s_1-q_1}a_{m''-q+s_1+s_2};\,q_2}\,,
\\
s^{\rm 2L,[22]}_{c_{m''}a_{m''-q+s_1+s_2};\,q_1q_2}(0)
&=\frac{\binom{m''-q_2}{q_1}}{\binom{m''-q_2+s_2}{q_1}}
s^{{\rm 1L}}_{c_{m''}b_{m''+s_2-q_2};\,q_2}
s^{{\rm 1L}}_{b_{m''+s_2-q_2}a_{m''-q+s_1+s_2};\,q_1}\,,
\end{aligned}
\ee
where the intermediate operator label is summed over and, once again, $q = q_1+q_2$.
These relations generalize the consistency relations in Eq.~(4.19) of \cite{Bakx:2025cvu}, derived for the Gaussian case at leading order in gradients, and Eq.~(3.26) of \cite{Bakx:2026rmd}, which also includes higher-gradient operators.
The first line, for instance, expresses the factorization of the two-loop iteration ${\cal O}_{c_{m''}} \xrightarrow{\rm 2L} {\cal O}_{a_{m''-q+s_1+s_2}}$ into two successive one-loop iterations, ${\cal O}_{c_{m''}}  \xrightarrow{\rm 1L}  {\cal O}_{b_{m''+s_1-q_1}}\xrightarrow{\rm 1L} {\cal O}_{a_{m''-q+s_1+s_2}}$, in the hierarchical two-loop regions in which one of the integrated momenta is larger than the other.

\subsubsection{The connected topology}

For the connected topology, the third hard momentum is fixed by momentum conservation, $\vp_3=-\vp_1-\vp_2-\vk_{1\dots s}$, so that only two hard loop momenta are independent. The relevant primordial correlator is the $s$-soft limit of an $(s+3)$-point function, as in \refeq{Gthreehard}:
\be\label{eq:Gdoublehard_general}
\boxed{
\begin{aligned}
&G^{(s+3)}_{\d\cdots\d\,X_1X_2X_3}
(\vk_1,\dots,\vk_s,\vp_1,\vp_2,-\vp_1-\vp_2-\vk_{1\dots s})
\Big|^{{}^{\rm soft}_{\rm 2pt}}
\\[-2pt]
&\qquad\xrightarrow[\,k_i\ll p_1,p_2,p_3\,]{}
\mathcal{B}_{X_1X_2X_3,s+3}(\vp_1,\vp_2,\vp_3)
\prod_{i=1}^s P_{\Psi\d}(k_i)\,,
\end{aligned}
}
\ee
where we used \refeq{Ppsidelta}, with $\mathcal{B}_{X_1X_2X_3,s+3}$ defined in \refeq{BXXX}. This topology therefore probes primordial shapes enhanced in the $(n-3)$-soft limit. 
Comparing once more with \refeq{treelevel1L}, the connected topology selects
\be\label{eq:mdef2l_connected}
m=m''+s-q\,.
\ee
Extending to higher loops through connected topologies thus probes higher cumulants of the primordial fields, obtained by integrating the corresponding hard response functions, such as $\mathcal{B}$, over the loop momenta.

The corresponding leading-derivative kernel limit is
\be\label{eq:doublehard_threehard}
\boxed{
\begin{aligned}
&K_{c_{m''}}^{(n+3-s)}
\left(
\vp_1,\vp_2,-\vp_1-\vp_2-\vk_{1\dots s},
\vk_{s+1},\dots,\vk_n
\right)
\Big|^{\substack{p_1,p_2,p_{12}\gg k_i\\r=p_1/p_2\ {\rm fixed}}}
\\[2pt]
&\quad=
\left[
\binom{m''}{q}
\binom{n+3-s-m''}{3-q}
\right]^{-1}
s^{\rm 2L,[3]}_{c_{m''}a_{m''-q+s};\,X_1X_2X_3}(r,\mu)K_{a_{m''-q+s}}^{(n)}
\left( \vk_{1},\dots,\vk_n \right)\,,
\end{aligned}
}
\ee
with $r=p_1/p_2$ and $\mu=\hat{\vp}_1\cdot\hat{\vp}_2$. 
The labels $X_1X_2X_3$ on the connected kernel coefficients specify the fields carrying $\vp_1,\vp_2,\vp_3$. Also \refeq{doublehard_threehard} is valid for every $(X_1,X_2,X_3,s)$, and no sum over them is implicit. 
Unlike the pair-pair coefficient, $s^{\rm 2L,[3]}$ generally depends on the full hard-triangle shape, including the angle between $\vp_1$ and $\vp_2$, and its angular average cannot in general be taken independently.

\begin{figure}[t]
    \centering
    \includegraphics[width=1.0\textwidth]
    {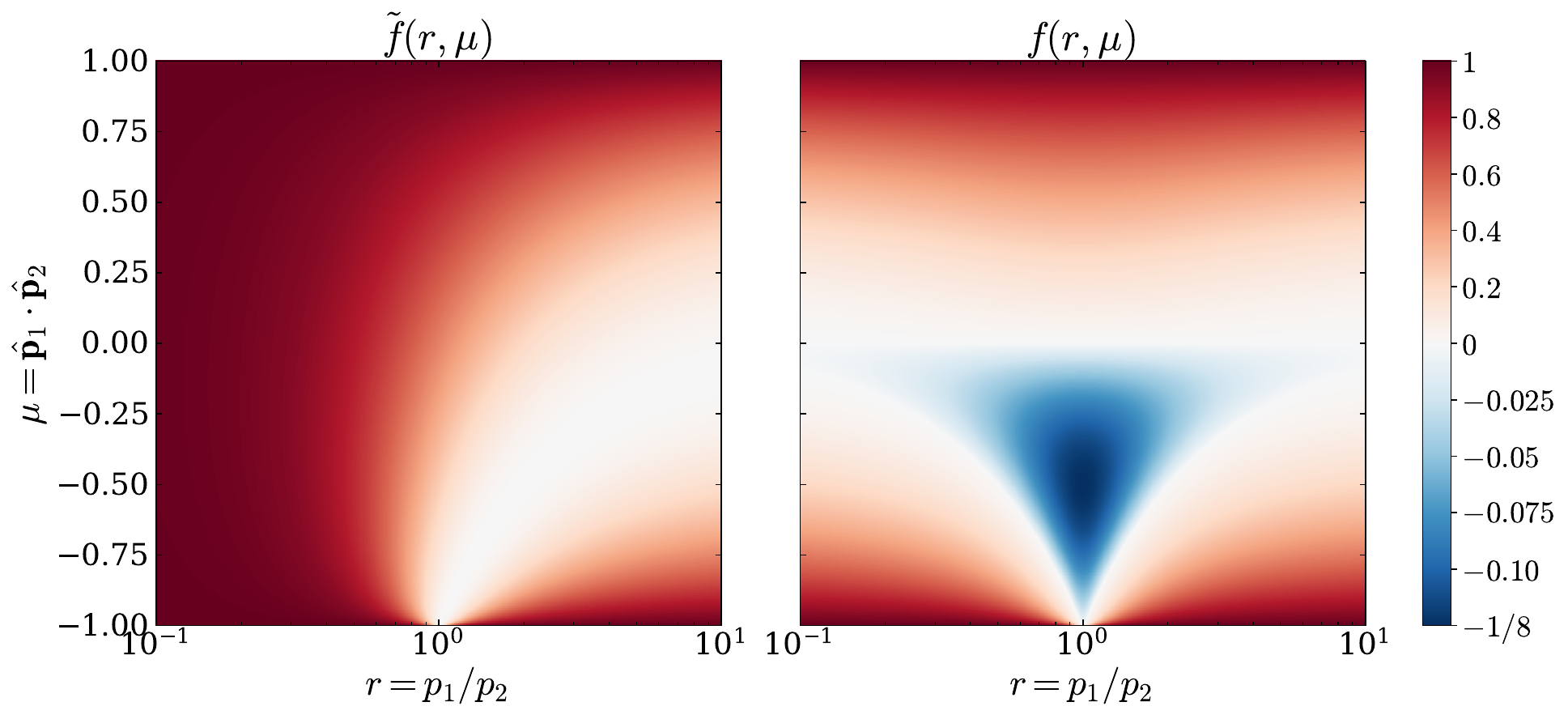}
    \caption{Momentum dependence of $\tilde f(r,\mu)$ (left) and $f(r,\mu)$ (right), defined in \refeq{ftilde} and \refeq{doublehardfuncf}, respectively, with $r=p_1/p_2$ and $\mu=\hat{\vp}_1\cdot\hat{\vp}_2$. }
    \label{fig:ffunc}
\end{figure}

Evaluating the coefficients $s^{\rm 2L,[3]}_{c_{m''}a_{m''-q+s};X_1X_2X_3}$ for $q\leq1$, we find that their momentum dependence can be written using two constant matrices, $\hat{s}^{\rm 2L,[3]}$ and $\tilde{s}^{\rm 2L,[3]}$, as follows:
\begin{align} \label{eq:consist3top}
s^{\rm 2L,[3]}_{c_{m''}a_{m''-q+s};\d\d\d}(r,\mu)
&=
\hat{s}^{\rm 2L,[3]}_{c_{m''}a_{m''-q+s};{0}}
+\tilde{s}^{\rm 2L,[3]}_{c_{m''}a_{m''-q+s};{0}}\,f(r,\mu),
\vs
s^{\rm 2L,[3]}_{c_{m''}a_{m''-q+s};\Psi\d\d}(r,\mu)
&=
\hat{s}^{\rm 2L,[3]}_{c_{m''}a_{m''-q+s};{1}}
+\tilde{s}^{\rm 2L,[3]}_{c_{m''}a_{m''-q+s};{1}}\,\tilde{f}(r,\mu),
\vs
s^{\rm 2L,[3]}_{c_{m''}a_{m''-q+s};\d\Psi\d}(r,\mu)
&=
\hat{s}^{\rm 2L,[3]}_{c_{m''}a_{m''-q+s};{1}}
+\tilde{s}^{\rm 2L,[3]}_{c_{m''}a_{m''-q+s};{1}}\,
\tilde{f}(r^{-1},\mu),
\vs
s^{\rm 2L,[3]}_{c_{m''}a_{m''-q+s};\d\d\Psi}(r,\mu)
&=
\hat{s}^{\rm 2L,[3]}_{c_{m''}a_{m''-q+s};{1}}
+\tilde{s}^{\rm 2L,[3]}_{c_{m''}a_{m''-q+s};{1}}\,\mu^2\,,
\end{align}
where we define
\begin{equation}\label{eq:ftilde}
    \boxed{\tilde{f}(r,\mu)
    \equiv \frac{(1+r\mu)^2}{1+r^2+2r\mu}\,.}
\end{equation}
Note that the last three lines of \refeq{consist3top} share the same constant coefficients $\hat{s}^{\rm 2L,[3]}$ and $\tilde{s}^{\rm 2L,[3]}$.
These $\hat{s}^{\rm 2L,[3]}$ and $\tilde{s}^{\rm 2L,[3]}$ matrices are given in \reftab{twoloopcoeff_a_000} for $m''=0$ and in \reftab{twoloopcoeff_aPsi} for $m''=1$.
The momenta dependence is entirely encoded in $\tilde{f}(r,\mu)$ and $\tilde{f}(r^{-1},\mu)$, which are interchanged under $\vp_1\leftrightarrow\vp_2$\footnote{Note that the functions $\tilde{f}(r,\mu),\;\tilde{f}(r^{-1},\mu), \;\mu^2$ are just the squared cosines of the angles between the 3 available momenta $\mathbf{p}_i\cdot\mathbf{p}_j/p_ip_j$.}. Together with $\mu^2$, these functions determine the full momentum dependence of the connected two-loop RG contributions, which arise only for non-Gaussian initial conditions.
The function $f(r,\mu)$ also appears in the Gaussian stochastic sector~\cite{Bakx:2025cvu} and can be expressed in terms of these functions as
\begin{equation}
\label{eq:doublehardfuncf}
\boxed{
\begin{aligned}
f(r,\mu)
&= \frac{1}{2}\left[
   \tilde{f}(r,\mu)+\tilde{f}(r^{-1},\mu)+\mu^2-1
   \right]
\\
&= \frac{\mu(1+r\mu)(r+\mu)}{1+r^2+2r\mu}
= \frac{
    [(\vp_1+\vp_2)\cdot\vp_1]
    [(\vp_1+\vp_2)\cdot\vp_2]
    [\vp_1\cdot\vp_2]
   }{p_1^2 p_2^2 |\vp_1+\vp_2|^2}\,.
\end{aligned}
}
\end{equation}
We display $f(r,\mu)$ and $\tilde f(r,\mu)$ in \reffig{ffunc}.

The three hierarchical limits of the connected topology are $r\to\infty$, $r\to0$, and $(r,\mu)\to(1,-1)$, where respectively $p_2$, $p_1$, or $p_3$ becomes much smaller than the other two loop momenta while remaining much larger than the external momenta.
In each hierarchy, two sequences arise because the harder loop can produce either one intermediate $\Psi$ leg or that leg together with the $s$ external $\Psi$ legs. We first consider the sequence in which the harder loop produces only the intermediate $\Psi$ leg, while the second loop generates the $s$ external $\Psi$ legs. In the following hierarchical relations, we use $q_X\equiv\delta^{\rm K}_{X\Psi}$ to count whether a hard field is of $\Psi$ type.
The corresponding kernel relations for this contraction channel are
\ba \label{eq:seq_3top_1}
&
\left\langle
s^{\rm 2L,[3]}_{c_{m''}a_{m''-q+s};\,{X_1X_2X_3}}
\right\rangle_{\mu}(r\to\infty)
=
\frac{\binom{m''-q+q_{X_2}}{{q_{X_2}}}}
{\binom{q}{{q_{X_2}}}
 \binom{3-q}{1-{q_{X_2}}}
 \binom{m''-q+{q_{X_2}}+1}{{q_{X_2}}+1}}
\\
&\hspace{6cm}\times\,
s^{\rm 1L}_{c_{m''}b_{m''-{q_{X_1}}+1};\,{q_{X_1}}}
s^{\rm 1L}_{b_{m''-{q_{X_1}}+1}a_{m''-q+s};\,q-{q_{X_1}}+1}\,,
\vs
&
\left\langle
s^{\rm 2L,[3]}_{c_{m''}a_{m''-q+s};\,{X_1X_2X_3}}
\right\rangle_{\mu}(r\to0)
=
\frac{\binom{m''-q+{q_{X_1}}}{{q_{X_1}}}}
{\binom{q}{{q_{X_1}}}
 \binom{3-q}{1-{q_{X_1}}}
 \binom{m''-q+{q_{X_1}}+1}{{q_{X_1}}+1}}
\nonumber\\
&\hspace{6cm}\times\,
s^{\rm 1L}_{c_{m''}b_{m''-{q_{X_2}}+1};\,{q_{X_2}}}
s^{\rm 1L}_{b_{m''-{q_{X_2}}+1}a_{m''-q+s};\,q-{q_{X_2}}+1}\,,
\vs
&
\left\langle
s^{\rm 2L,[3]}_{c_{m''}a_{m''-q+s};\,{X_1X_2X_3}}
\right\rangle_{\mu_3}(p_3/p_1\to0)
=
\frac{\binom{m''-q+{q_{X_3}}}{{q_{X_3}}}}
{\binom{q}{q_{X_3}}
 \binom{3-q}{1-{q_{X_3}}}
 \binom{m''-q+{q_{X_3}}+1}{{q_{X_3}}+1}}
\nonumber\\
&\hspace{6cm}\times\,
s^{\rm 1L}_{c_{m''}b_{m''-{q_{X_3}}+1};\,{q_{X_3}}}
s^{\rm 1L}_{b_{m''-{q_{X_3}}+1}a_{m''-q+s};\,q-{q_{X_3}}+1} 
\,, \nonumber
\ea
where $q={q_{X_1}}+{q_{X_2}}+{q_{X_3}}$, $\mu_3=\vp_1\cdot\vp_3/(p_1p_3)$, the intermediate operator label $b$ is summed and the brackets denote angular averaging.
In the second sequence, the harder loop produces $s+1$ $\Psi$ legs, with $s$ external legs and one intermediate leg. The latter participates in the remaining loop, which produces no additional external $\Psi$ legs. The sequential limits for this contraction channel are
\ba \label{eq:seq_3top_2}
&
\left\langle
s^{\rm 2L,[3]}_{c_{m''}a_{m''-q+s};\,X_1X_2X_3}
\right\rangle_{\mu}(r\to\infty)
=
\frac{\binom{m''-q+q_{X_2}}{q_{X_2}}}
{\binom{q}{q_{X_2}}
 \binom{3-q}{1-q_{X_2}}
 \binom{m''-q+q_{X_2}+s+1}{q_{X_2}+1}}
\\
&\hspace{4cm}
\times\,
s^{\rm 1L}_{c_{m''}b_{m''-q+q_{X_2}+s+1};\,q-q_{X_2}}
s^{\rm 1L}_{b_{m''-q+q_{X_2}+s+1}a_{m''-q+s};\,q_{X_2}+1}
\,,
\vs
&
\left\langle
s^{\rm 2L,[3]}_{c_{m''}a_{m''-q+s};\,X_1X_2X_3}
\right\rangle_{\mu}(r\to0)
=
\frac{\binom{m''-q+q_{X_1}}{q_{X_1}}}
{\binom{q}{q_{X_1}}
 \binom{3-q}{1-q_{X_1}}
 \binom{m''-q+q_{X_1}+s+1}{q_{X_1}+1}}
\nonumber\\
&\hspace{4cm}
\times\,
s^{\rm 1L}_{c_{m''}b_{m''-q+q_{X_1}+s+1};\,q-q_{X_1}}
s^{\rm 1L}_{b_{m''-q+q_{X_1}+s+1}a_{m''-q+s};\,q_{X_1}+1}
\,,
\vs
&
\left\langle
s^{\rm 2L,[3]}_{c_{m''}a_{m''-q+s};\,X_1X_2X_3}
\right\rangle_{\mu_3}(p_3/p_1\to0)
=
\frac{\binom{m''-q+q_{X_3}}{q_{X_3}}}
{\binom{q}{q_{X_3}}
 \binom{3-q}{1-q_{X_3}}
 \binom{m''-q+q_{X_3}+s+1}{q_{X_3}+1}}
\nonumber\\
&\hspace{4cm}
\times\,
s^{\rm 1L}_{c_{m''}b_{m''-q+q_{X_3}+s+1};\,q-q_{X_3}}
s^{\rm 1L}_{b_{m''-q+q_{X_3}+s+1}a_{m''-q+s};\,q_{X_3}+1}
\,, \nonumber
\ea
noticing the extra $s$ factors in the right-hand side compared to \refeq{seq_3top_1}.

\subsubsection{Full two-loop renormalization}

Combining the primordial soft limits with the kernel limits derived above, we can express the bare two-loop correlator in terms of tree-level operator structures as
\ba\label{eq:2Lcorr_reduced}
&\langle{\cal O}_{c_{m''}}\delta_L({\bm k}_1)\cdots\delta_L({\bm k}_n)
\rangle_{\rm 2L}'
\Big|_{\vk_i\to0}^{{}^{\rm soft}_{\rm 2pt}} =
\sum_{a_m}m!(n-m)!
\left[
Z^{\rm 2L,\,[22]}_{c_{m''} a_m}
+
Z^{\rm 2L,\,[3]}_{c_{m''} a_m}
\right]
\vs
&\hspace{1cm}\times
K_{a_m}^{(n)}(\vk_1,\dots,\vk_m;\vk_{m+1},\dots,\vk_n)
\prod_{i=1}^{m}P_{\Psi\d}(k_i)
\prod_{j=m+1}^{n}P_{\d\d}(k_j)
+{\rm Sym.}\,.
\ea
The derivation follows the same steps as at one loop, so we give only the resulting coefficients. 
The pair-pair contribution is
\ba\label{eq:I2L_22}
Z^{\rm 2L,\,[22]}_{c_{m''} a_m}
&=
\frac{1}{2}
\sum_{\substack{s_1,s_2\ge0}}
\sum_{\substack{X_1,\dots,X_4}}
\frac{\lambda_{\rm NL}^{(s_1+2)}
\lambda_{\rm NL}^{(s_2+2)}}{s_1!s_2!}
\delta^{\rm K}_{m,m''+s_1+s_2-q}
\vs
&\quad\times
\int_{p_1,p_2<\Lambda}
P_{X_1X_2}(p_1)
P_{X_3X_4}(p_2)
s^{\rm 2L,[22]}_{c_{m''}a_m;\,q_1q_2}(r)\,,
\ea
where $q=q_1+q_2$, the Kronecker symbol enforces the counting relation \refeq{mdef2l_pairpair}, and each hard pair is summed over the three inequivalent assignments specified after \refeq{q1l_general}.
The connected contribution is
\ba\label{eq:I2L_3}
Z^{\rm 2L,\,[3]}_{c_{m''} a_m}
&=
\sum_{s\ge0}
\sum_{\substack{X_1,X_2,X_3}}
\frac{1}{s!}
\delta^{\rm K}_{m,m''+s-q}
\int_{p_1,p_2<\Lambda}
\mathcal{B}_{X_1X_2X_3,s+3}(\vp_1,\vp_2)
s^{\rm 2L,[3]}_{c_{m''}a_m;\,{X_1X_2X_3}}(r,\mu)\,,
\ea
where the hard triple is summed over the four inequivalent assignments specified after \refeq{q1l_general} and the Kronecker symbol now enforces \refeq{mdef2l_connected}. 

To obtain the full two-loop mixing matrix, we insert \refeq{2Lcorr_reduced} into \refeq{rencond2L}, and match onto \refeq{treelevel1L} to obtain
\be\label{eq:Z2L_general}
\boxed{
Z_{c_{m''} a_m}^{\rm 2L}
=
Z_{c_{m''} b_{m'}}^{\rm 1L}
Z_{b_{m'} a_m}^{\rm 1L}
-
Z^{\rm 2L,\,[22]}_{c_{m''} a_m}
-
Z^{\rm 2L,\,[3]}_{c_{m''} a_m}\,.
}
\ee
The first term arises from the one-loop insertion and ensures consistent treatment of the overlap between single-hard and double-hard limits, as we discuss in \refsec{running}. 
Note that \refeq{Z2L_general} reduces to Eq.~(4.16) of \cite{Bakx:2025cvu} in the Gaussian limit.

\subsubsection{Examples}

For $\O_{c_j}$ with $j=1,2$, mixing into $\d$ requires $(q,s)=(j,0)$, while mixing into $\Psi$ allows $(q,s)=(j-1,0)$ or $(j,1)$, where $s=s_1+s_2$ for the pair-pair topology. Thus, mixing into $\Psi$ can either retain one original $\Psi$ leg or generate one through a primordial soft limit
\ba
Z^{\rm 2L,[22]}_{c_1\d}
&=
\int_{p_1,p_2<\Lambda}
P_{\d\d}(p_1)P_{\d\d}(p_2)
\frac{s^{\rm 2L,[22]}_{c_1\d;\,10}(r)}{M(p_1)}\,,
\vs
Z^{\rm 2L,[22]}_{c_1\Psi}
&=
\int_{p_1,p_2<\Lambda}
P_{\d\d}(p_1)P_{\d\d}(p_2)
\left[
\frac12 s^{\rm 2L,[22]}_{c_1\Psi;\,00}(r)
+
\frac{8f_{\rm NL}}{M(p_1)}
s^{\rm 2L,[22]}_{c_1\Psi;\,10}(r)
\right]\,,
\vs
Z^{\rm 2L,[22]}_{c_2\d}
&=
\int_{p_1,p_2<\Lambda}
P_{\d\d}(p_1)P_{\d\d}(p_2)
\left[
\frac{s^{\rm 2L,[22]}_{c_2\d;\,20}(r)}{M^2(p_1)}
+
\frac{s^{\rm 2L,[22]}_{c_2\d;\,11}(r)}
{2M(p_1)M(p_2)}
\right]\,,
\vs
Z^{\rm 2L,[22]}_{c_2\Psi}
&=
\int_{p_1,p_2<\Lambda}
P_{\d\d}(p_1)P_{\d\d}(p_2)
\Bigg[
\frac{s^{\rm 2L,[22]}_{c_2\Psi;\,10}(r)}{M(p_1)}
+
\frac{8f_{\rm NL}}{M^2(p_1)}
s^{\rm 2L,[22]}_{c_2\Psi;\,20}(r)
\vs
&\hspace{4cm}
+
\frac{4f_{\rm NL}}{M(p_1)M(p_2)}
s^{\rm 2L,[22]}_{c_2\Psi;\,11}(r)
\Bigg]\,.
\ea
Note that the terms proportional to $f_{\rm NL}$ additionally include the two possible assignments of the external soft leg.
For the connected topology, the corresponding contributions are
\ba \label{eq:Z3top}
Z^{\rm 2L,[3]}_{c_0\d}
&=
\int_{p_1,p_2<\Lambda}
\mathcal B_{\d\d\d,3}(\vp_1,\vp_2)
s^{\rm 2L,[3]}_{c_0\d;\,{\d\d\d}}(r,\mu)\,,
\vs
Z^{\rm 2L,[3]}_{c_0\Psi}
&=
\int_{p_1,p_2<\Lambda}
\mathcal B_{\d\d\d,4}(\vp_1,\vp_2)
s^{\rm 2L,[3]}_{c_0\Psi;\,{\d\d\d}}(r,\mu)\,,
\vs
Z^{\rm 2L,[3]}_{c_1\d}
&=
\int_{p_1,p_2<\Lambda}
\mathcal B_{\Psi\d\d,3}(\vp_1,\vp_2)
s^{\rm 2L,[3]}_{c_1\d;\,{\Psi\d\d}}(r,\mu)\,,
\vs
Z^{\rm 2L,[3]}_{c_1\Psi}
&=
\int_{p_1,p_2<\Lambda}
\Big[
\mathcal B_{\d\d\d,3}(\vp_1,\vp_2)
s^{\rm 2L,[3]}_{c_1\Psi;\,{\d\d\d}}(r,\mu) +
\mathcal B_{\Psi\d\d,4}(\vp_1,\vp_2)
s^{\rm 2L,[3]}_{c_1\Psi;\,\Psi\d\d}(r,\mu)
\Big]\,.
\ea
For local PNG, we have from \refeq{BXXX}:
\ba
\mathcal{B}_{\d\d\d,3}(\vp_1,\vp_2)
&=
2\fnl M(p_1)M(p_2)M(p_{12})
\vs
&\quad\times
\left[
P_\varphi(p_1)P_\varphi(p_2)
+P_\varphi(p_1)P_\varphi(p_{12})
+P_\varphi(p_2)P_\varphi(p_{12})
\right]\;,
\vs
\mathcal{B}_{\d\d\d,4}(\vp_1,\vp_2)
&=
\left(8\fnl^2+6g_{\rm NL}\right)
M(p_1)M(p_2)M(p_{12})
\vs
&\quad\times
\left[
P_\varphi(p_1)P_\varphi(p_2)
+P_\varphi(p_1)P_\varphi(p_{12})
+P_\varphi(p_2)P_\varphi(p_{12})
\right]\;,
\ea
and the mixed response is given by
\ba
\mathcal{B}_{\Psi\d\d,3}(\vp_1,\vp_2)
&=
\frac{\mathcal{B}_{\d\d\d,3}(\vp_1,\vp_2)}{M(p_1)} =
2\fnl M(p_2)M(p_{12})
\vs
&\quad\times
\left[
P_\varphi(p_1)P_\varphi(p_2)
+P_\varphi(p_1)P_\varphi(p_{12})
+P_\varphi(p_2)P_\varphi(p_{12})
\right]\,.
\ea 
We note that all three possible assignments of $\Psi$ in the bispectrum are equivalent after integrating \refeq{Z3top}.

\subsection{Higher loops}
\label{sec:renorm_higherL_det}

We sketch in this section the renormalization structure when extending to higher loops. At $L$ loops, the deterministic renormalization condition takes the form
\be\label{eq:rencond_higherL}
0=
\sum_{\ell=0}^{L}
Z_{c_{m''} b_{m'}}^{\ell {\rm L}}
\left\langle
{\cal O}_{b_{m'}}\delta_L({\bm k}_1)\cdots\delta_L({\bm k}_n)
\right\rangle_{(L-\ell){\rm L}}
\bigg|_{\vk_i\to 0}\,.
\ee
Here $Z^{0{\rm L}}_{c_{m''} b_{m'}}=\delta^{\rm K}_{c_{m''} b_{m'}}$, the index $c_{m''}$ is fixed, and the repeated extended index $b_{m'}$ is summed.
At each loop order, different topologies contribute.
We denote a topology containing $B$ connected hard blocks by $[h_1h_2\cdots h_B]$, where $h_\alpha$ is the number of hard legs in block $\alpha$. For example,
\be
L=1:\quad [2]\,,
\qquad
L=2:\quad [22]\,,\,[3]\,,
\qquad
L=3:\quad [222]\,,\,[32]\,,\,[4]\,.
\ee
At one loop, the topology $[2]$ probes the two-hard-leg kernel limit \refeq{biaskernelexpansionsinglehard}, together with the primordial response \refeq{Ghardsoft}.
At two loops, the pair-pair topology $[22]$ combines two responses of this type and probes the four-hard-leg kernel limit \refeq{doublehard_pairpair}.
The connected topology $[3]$ instead probes the three-hard-leg kernel limit \refeq{doublehard_threehard} and the primordial response \refeq{Gdoublehard_general}.

It is useful to make this structure explicit for the two extreme topologies that appear at order $L$, which we discuss next.
The fully connected topology $[L+1]$ contains a single hard block and probes a new primordial response at each loop order.
At the opposite extreme, the fully disconnected topology $[2^L]\equiv[22\cdots2]$ contains $L$ two-leg hard blocks, and its primordial contribution factorizes into products of the two-hard-leg responses already present at one loop.

The leading term involving $s$ soft legs of an $(s+h)$-point function having a connected block of $h$ hard legs has the schematic form of \refeq{GXXXX}, which we repeat here for convenience:
\be\label{eq:higherL_hard_response}
\boxed{
\begin{aligned}
&G^{(s+h)}_{\d\cdots\d\,X_1\cdots X_h}
\left(
\vk_1,\ldots,\vk_s,
\vp_1,\ldots,\vp_{h-1},
-\vp_{1\cdots h-1}-\vk_{1\cdots s}
\right)\Big|^{{}^{\rm soft}_{\rm 2pt}}
\\[-2pt]
&\qquad\xrightarrow[\,k_i\ll p_1,\ldots,p_h\,]{}
{\cal G}^{(h)}_{X_1\cdots X_h,s+h}
(\vp_1,\ldots,\vp_{h-1})
\prod_{i=1}^{s}P_{\Psi\d}(k_i)\,,
\end{aligned}
}
\ee
with ${\cal G}^{(h)}_{X_1\cdots X_h,s+h}$ being a response function.

For the fully connected topology $[h]$, the response enters the $(h-1)$-loop mixing together with the corresponding multi-hard bias-kernel coefficient:
\ba\label{eq:IhigherL_h}
Z^{(h-1){\rm L},[h]}_{c_{m''}a_m}
&\supset
\int_{\substack{p_1,\ldots,p_{h-1}<\Lambda}}
{\cal G}^{(h)}_{X_1\cdots X_h,s+h}
(\vp_1,\ldots,\vp_{h-1})
s^{(h-1){\rm L},[h]}_{c_{m''}a_m;\,{X_1\cdots X_h}}
(\vp_1,\ldots,\vp_{h-1})\,,
\ea
where the sums over $s$ and inequivalent hard-field assignments, as well as all combinatorial prefactors, are left implicit.
Here $q$ counts the hard $\Psi$ legs and $m=m''+s-q$.
For $h=2$, ${\cal G}^{(2)}_{X_1X_2,s+2}=\lambda_{\rm NL}^{(s+2)}P_{X_1X_2}$, while ${\cal G}^{(3)}\equiv\mathcal B$ recovers the notation of \refeq{Gdoublehard_general}.
The coefficient $s^{(h-1){\rm L},[h]}_{c_{m''}a_m;\,{X_1\cdots X_h}}$, which depends on the hard-momentum ratios and relative angles, is obtained by matching the simultaneous $h$-hard-leg limit of $K_{c_{m''}}^{(n+h-s)}$ onto $K_{a_m}^{(n)}$.
Therefore, the fully connected topology $[h]$ probes the soft-field responses of connected cumulants of the primordial fields.
For $h>2$, these contributions are typically suppressed by the near-Gaussianity of the primordial fluctuations.

We now turn to the opposite extreme, the fully disconnected topology $[2^L]$. Its primordial contribution is a product of $L$ copies of the two-hard-leg soft limit \refeq{Ghardsoft}, with $s_\alpha$ soft legs entering the $\alpha$th correlator and $\sum_{\alpha=1}^{L}s_\alpha=s$. Hence, this topology contains no new connected primordial hard correlator, since its primordial contribution is entirely determined by products of two-point functions. The corresponding contribution to the $L$-loop mixing is
\ba\label{eq:IhigherL_2L}
Z^{L{\rm L},[2^L]}_{c_{m''}a_m}
&\supset
\int_{p_1,\ldots,p_L<\Lambda}
\left[
\prod_{\alpha=1}^{L}
\lambda_{\rm NL}^{(s_\alpha+2)}
P_{X_{2\alpha-1}X_{2\alpha}}(p_\alpha)
\right]
s^{L{\rm L},[2^L]}_{c_{m''}a_m;\,q_1\cdots q_L}
(r_2,\ldots,r_L)\,,
\ea
where the sums over $s_\alpha$ and $X_i$, the assignments of the soft legs among the hard pairs, and all combinatorial prefactors are left implicit.
Here $q_\alpha$ is the number of $\Psi$ legs in the $\alpha$th hard pair, $q=\sum_\alpha q_\alpha$, $m=m''+s-q$, and $r_\alpha\equiv p_1/p_\alpha$ for $\alpha=2,\ldots,L$, so that $r_2=r$ at two loops.
Note that since the power spectra are independent of the hard directions, the kernel can be angularly averaged independently and no angular dependence is left. Moreover, each hard pair contributes with a generalized variance \refeq{sigmatildedef}, such that in the Gaussian limit, the contribution scales as $(\sigma_\Lambda^2)^L$, suppressing higher loops when $\sigma_\Lambda^2\ll1$.
This is the only non-zero topology for Gaussian initial conditions. At two loops and leading order in gradients, the multi-hard coefficient dependence on $r=p_1/p_2$ is captured by the single function $g(r)$ \cite{Bakx:2025cvu} in the Gaussian case. At next-to-leading order in gradients three independent functions are required \cite{Bakx:2026rmd}.

\newpage

\section{Renormalization of stochastic terms}
\label{sec:stoch}

In addition to the deterministic bias operators in \refeq{bias-expansion}, the bias expansion contains stochastic contributions, which encode the impact of short-scale physics on large-scale observables \cite{Rubira:2024tea, Rubira:2025rqo}.
In the operator language, renormalizing these contributions requires counterterms for products of composite operators \cite{Rubira:2024tea, Bakx:2025cvu, Bakx:2026rmd}. In this section, we examine how PNG modifies the renormalization of stochastic terms, focusing throughout on leading-gradient stochastic operators. We first consider the one-loop renormalization of products of two operators in \refsec{stoch_two_products}, then extend the analysis to products of three or more operators in \refsec{stoch_higher_products}. Finally, we discuss the collapsed trispectrum configuration in \refsec{stoch_higher_loops}, which appears at two loops.


\subsection{One-loop: product of two operators}  \label{sec:stoch_two_products}

Following \cite{Bakx:2025cvu, Bakx:2026rmd}, the product of two extended bias operators can be renormalized as
\ba
\label{eq:stoch_OPE_two}
[\O_{a_m}(\vx)\O_{b_\ell}(\vy)]
=
[\O_{a_m}(\vx)][\O_{b_\ell}(\vy)]
+
Z_{a_m b_\ell c_{m'}}
[\O_{c_{m'}}(\v{X})]
\diracpi(\vx-\vy)\,,
\ea
where $\v{X}=(\vx+\vy)/2$, and the unit operator $\mathbf{1}$ is included in the operator basis when summing over $c_{m'}$. Also note that the matrix $Z$ is dimensionful.
Unlike the deterministic bias basis, for which the mean of the tracer overdensity is fixed to zero, the stochastic basis must contain the unit operator since the mixing of the operator product into $\mathbf{1}$ absorbs the momentum-independent contact contribution and renormalizes the leading stochastic noise in the tracer power spectrum, including shot-noise-like terms.

The contact coefficients are fixed by imposing the large-scale renormalization condition, generalizing \refeq{rencond},
\ba
\label{eq:stoch_rencond_two}
&\left\langle
[\O_{a_m}(\vk)\O_{b_\ell}(\vk')]
\dlin(\vk_1)\cdots\dlin(\vk_n)
\right\rangle
\bigg|_{\vk,\vk',\vk_i\to0}
\vs
&\hspace{4.5cm}=
\left\langle
[\O_{a_m}(\vk)][\O_{b_\ell}(\vk')]
\dlin(\vk_1)\cdots\dlin(\vk_n)
\right\rangle_{\rm tree}
\bigg|_{\vk,\vk',\vk_i\to0}\,.
\ea
After the renormalization of the individual operators has been performed as outlined in \refsec{renorm}, the remaining one-loop UV-sensitive contribution is the stochastic part. 
Substituting \refeq{stoch_OPE_two} into \refeq{stoch_rencond_two} and expanding $Z_{a_m b_\ell c_{m'}}$ to one-loop order gives
\ba
\label{eq:stoch_rencond_two_1L}
0&=
\left\langle
[\O_{a_m}(\vk)][\O_{b_\ell}(\vk')]
\dlin(\vk_1)\cdots\dlin(\vk_n)
\right\rangle_{\rm 1L}
\bigg|_{\vk,\vk',\vk_i\to0} \vs
&\hspace{3cm}+
Z_{a_m b_\ell c_{m'}}^{\rm 1L}
\left\langle
[\O_{c_{m'}}(\vk+\vk')]
\dlin(\vk_1)\cdots\dlin(\vk_n)
\right\rangle_{\rm tree}\bigg|_{\vk,\vk',\vk_i\to0}\,.
\ea
As in the deterministic discussion, we restrict below to configurations in which the operator legs not entering the primordial correlators contract with the external $\dlin$ fields through two-point functions.

Let the hard legs entering the two primordial correlators carry field labels $(X_1,X_2)$ and $(X_3,X_4)$, with $X_i\in\{\d,\Psi\}$, and $X_1$ and $X_3$ labeling the hard legs belonging to $\O_{a_m}$, and $X_2$ and $X_4$ those belonging to $\O_{b_\ell}$.  Analogously to \refeq{q1l}, we count the hard $\Psi$ legs integrated from each bias kernel as
\bea
q_1 \equiv
\#\{\Psi\;{\rm among}\;X_1,X_3\}\,, \quad
q_2
\equiv
\#\{\Psi\;{\rm among}\;X_2,X_4\}\,, \quad
q
\equiv
q_1+q_2\,,
\label{eq:qstoch1L}
\eea
with $0\leq q_1\leq\min(2,m)$ and $0\leq q_2\leq\min(2,\ell)$. The legs of $\O_{a_m}$ that contract directly with external fields therefore contain $m-q_1$ powers of $\Psi$, while those of $\O_{b_\ell}$ contain $\ell-q_2$ powers.

At one loop, the two operators are connected by two hard contractions.
We assign $s_1$ soft external legs to the first primordial correlator and $s_2$ soft external legs to the second. 
Of the remaining $n-s_1-s_2$ external legs, $t$ contract directly with $\O_{a_m}$ and the remaining $n-s_1-s_2-t$ with $\O_{b_\ell}$.
We display a representative ordering in which the first $s_1$ momenta enter the first primordial correlator, the following $s_2$ momenta enter the second, and the following $t$ momenta contract directly with $\O_{a_m}$. 
For convenience, we define the corresponding momentum sums
\bea
\vk_{S_1}
&\equiv&
\vk_{1\cdots s_1}\,,
\qquad
\vk_{S_2}
\equiv
\vk_{s_1+1\cdots s_1+s_2}\,,
\\
\vk_A
&\equiv&
\vk_{s_1+s_2+1\cdots s_1+s_2+t}\,,
\qquad
\vk_B
\equiv
\vk_{s_1+s_2+t+1\cdots n}\,. \nonumber
\eea
These momentum sums, together with the two operator momenta, satisfy
\be
\vk+\vk'
+\vk_{S_1}+\vk_{S_2}+\vk_A+\vk_B
=0\,.
\ee
Thus, only one of $\vk$ and $\vk'$ is independent, so we write the hard arguments in terms of $\vk$, with the dependence on $\vk'$ fixed by the momentum-conservation condition above. With these conventions, the one-loop stochastic contribution is given by
\ba
\label{eq:stoch_corr_two_1L_raw}
&
\left\langle
\O_{a_m}(\vk)\O_{b_\ell}(\vk')
\dlin(\vk_1)\cdots\dlin(\vk_n)
\right\rangle_{\rm 1L}'
\Big|^{{}^{\rm soft}_{\rm 2pt}}
\vs
&=
\sum_{s_1,s_2}
\sum_{t=0}^{n-s_1-s_2}
\sum_{\substack{X_1,\dots,X_4\in\{\d,\Psi\}\\
q_1=\#\Psi\ {\rm in}\ X_1,X_3\\
q_2=\#\Psi\ {\rm in}\ X_2,X_4}}
{\cal C}^{(n)}_{s_1s_2t;\,q_1q_2}
\left[
\prod_{i=s_1+s_2+1}^{s_1+s_2+t}
P_{Y_i\d}(k_i)
\right]_{m-q_1}
\left[
\prod_{j=s_1+s_2+t+1}^{n}
P_{Y_j\d}(k_j)
\right]_{\ell-q_2}
\vs
&\hspace{0.8cm}\times
\int_{p<\Lambda}
K_{a_m}^{(t+2)}
\left(
-\vk_{s_1+s_2+1},\dots,-\vk_{s_1+s_2+t},
\vp+\vk+\vk_A,-\vp
\right)
\vs
&\hspace{0.8cm}\times
K_{b_\ell}^{(n-s_1-s_2-t+2)}
\left(
-\vk_{s_1+s_2+t+1},\dots,-\vk_n,
-\vp-\vk-\vk_A-\vk_{S_2},
\vp-\vk_{S_1}
\right)
\vs
&\hspace{0.8cm}\times
G^{(s_1+2)}_{\d\cdots\d\,X_1X_2}
\left(
\vk_1,\dots,\vk_{s_1},
-\vp,\vp-\vk_{S_1}
\right)
\vs
&\hspace{0.8cm}\times
G^{(s_2+2)}_{\d\cdots\d\,X_3X_4}
\left(
\vk_{s_1+1},\dots,\vk_{s_1+s_2},
\vp+\vk+\vk_A,
-\vp-\vk-\vk_A-\vk_{S_2}
\right)
+{\rm Sym.}\,,
\ea
with the loop integral represented by the diagram
\ba
\raisebox{-0.0cm}{\includegraphicsbox[scale=0.7]{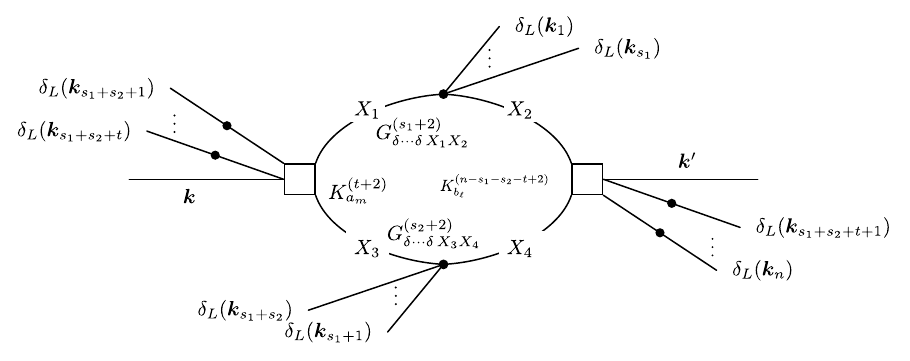}}\,,
\ea
and where we restrict once more to configurations in which the soft legs of the bias kernels connect to the external legs only through two-point functions.  The ${\cal C}^{(n)}_{s_1s_2t;\,q_1q_2}$ coefficient gives the multiplicity of the displayed contraction, and ${\rm Sym.}$ sums over inequivalent assignments of the external momenta to the two kernels and primordial correlators. Similar to the deterministic case, the two correlators generate $s_1+s_2$ external $\Psi$ insertions, whereas the $q$ hard $\Psi$ legs remain integrated over in the loop, such that
\be
\label{eq:stoch_m_rule_two}
m'=m+\ell+s_1+s_2-q\,,
\ee
in analogy to the selection rule \refeq{fixed_condition_m}. 

To evaluate \refeq{stoch_corr_two_1L_raw} in the soft limit required by \refeq{stoch_rencond_two}, we use the corresponding limits of the bias kernels and primordial correlators.
We first discuss the bias kernel limit, described by the hard limit of the kernel product $K_{a_m}^{(t+2)}K_{b_\ell}^{(n-s_1-s_2-t+2)}$.
We define
\be
u\equiv n-s_1-s_2\,,
\qquad
r\equiv m+\ell-q\,,
\ee
where $r$ is the number of powers of $\Psi$ remaining after the hard contractions but before the soft PNG legs are added.
The stochastic one-loop coefficient $s^{{\rm 1L}}_{a_m b_\ell c_r;\,q_1q_2}$ is then defined by
\ba
\label{eq:stoch_kernel_two}
&
\frac{1}{r!(u-r)!}
\left[
\sum_{t=0}^{u}
{\cal C}^{(u)}_{s_1s_2t;\,q_1q_2}
\Bigl(
K_{a_m}^{(t+2)}K_{b_\ell}^{(u-t+2)}
+{\rm Sym.}
\Bigr)
\right]_{\operatorname{av}_{\hat{\bf p}}}^{p\gg k,k',k_i}
\nonumber\\
&\hspace{5cm}=
s^{{\rm 1L}}_{a_m b_\ell c_r;\,q_1q_2}
K_{c_r}^{(u)}
(\vk_{s_1+s_2+1},\dots,\vk_n)\,,
\ea
where we keep only the leading terms in the gradient expansion and we omit momentum arguments. The sum over $t$ accounts for the possible numbers of soft external legs attached directly to $\O_{a_m}$, and ${\rm Sym.}$ sums over the remaining inequivalent assignments of these $u$ legs to the two kernels.

We next consider the primordial-correlator part. Applying \refeq{Ghardsoft} to both primordial correlators in \refeq{stoch_corr_two_1L_raw} gives 
\ba \label{eq:soft_primordial_stoch}
&\left.
G^{(s_1+2)}_{\d\cdots\d\,X_1X_2}
G^{(s_2+2)}_{\d\cdots\d\,X_3X_4}
\right|^{{}^{\rm soft}_{\rm 2pt}}
\xrightarrow{\,k_i\ll p\,}
\lambda_{\rm NL}^{(s_1+2)}
\lambda_{\rm NL}^{(s_2+2)}
P_{X_1X_2}(p)P_{X_3X_4}(p)
\prod_{i=1}^{s_1+s_2}P_{\Psi\d}(k_i)\,,
\ea
where we omit the momentum arguments on the left side. 
The hard factors remain inside the single loop integral, motivating the definition of the dimensionful quantity
\be
\label{eq:stoch_Sigma2_def}
\boxed{\left[\Sigma_\Lambda^2\right]_{X_1X_2|X_3X_4}
\equiv
\int_{p<\Lambda}
P_{X_1X_2}(p)P_{X_3X_4}(p)}\,,
\ee
that for $X_i=\d$ reduces to the Gaussian stochastic moment
\be \label{eq:Sigma}
\left[\Sigma_\Lambda^2\right]_{\d\d|\d\d}
=
\int_{p<\Lambda}P_{\d\d}^2(p)\,.
\ee
This reproduces the generalized moment controlling the Gaussian stochastic terms introduced in \cite{Rubira:2024tea}.

Using \refeq{stoch_kernel_two} and \refeq{soft_primordial_stoch}, the one-loop correlator \refeq{stoch_corr_two_1L_raw} reduces in the soft limit to 
\ba
\label{eq:stoch_corr_two_1L_reduced}
&
\left\langle
\O_{a_m}(\vk)\O_{b_\ell}(\vk')
\dlin(\vk_1)\cdots\dlin(\vk_n)
\right\rangle_{\rm 1L}'
\Big|_{\vk,\vk',\vk_i\to0}^{{}^{\rm soft}_{\rm 2pt}}
\vs
&=
m'!(n-m')!
\sum_{s_1,s_2}
\sum_{\bm X}^{\rm allowed}
\frac{1}{s_1!s_2!}
\lambda_{\rm NL}^{(s_1+2)}
\lambda_{\rm NL}^{(s_2+2)}
\left[\Sigma_\Lambda^2\right]_{X_1X_2|X_3X_4}
s^{{\rm 1L}}_{a_m b_\ell c_{m+\ell-q};\,q_1q_2}
\vs
&\hspace{1cm}\times
K_{c_{m'}}^{(n)}(\vk_1,\dots,\vk_n)
\prod_{i=1}^{m'}P_{\Psi\d}(k_i)
\prod_{j=m'+1}^{n}P_{\d\d}(k_j)
\delta^{\rm K}_{m',m+\ell+s_1+s_2-q}
+{\rm Sym.}\,.
\ea
Comparing with the tree-level term in \refeq{stoch_rencond_two_1L} gives 
\be
\label{eq:stoch_Z_two_1L}
\boxed{
\begin{aligned}
Z_{a_m b_\ell c_{m'}}^{\rm 1L}
&=-
\sum_{s_1,s_2}
\sum_{\bm X}^{\rm allowed}
\frac{\lambda_{\rm NL}^{(s_1+2)}
      \lambda_{\rm NL}^{(s_2+2)}}{s_1!s_2!}
\\
&\quad\times
\left[\Sigma_\Lambda^2\right]_{X_1X_2|X_3X_4}
s^{{\rm 1L}}_{a_m b_\ell c_{m+\ell-q};\,q_1q_2}
\delta^{\rm K}_{m',m+\ell+s_1+s_2-q}\,,
\end{aligned}
}
\ee
where $\sum_{\bm X}^{\rm allowed}$ restricts the sum to $q_1\leq m$, $q_2\leq\ell$ and $0\leq m'=m+\ell+s_1+s_2-q\leq n$ and the sums over $s_1$ and $s_2$ also obey $s_1+s_2\leq n$.

After considering the renormalization of contact operators, we make explicit how these contact terms enter observable correlators.  For the tracer power spectrum, only mixing into the unit operator contributes at leading order in gradients. Therefore, the local contact contribution at any loop order has the form \cite{Bakx:2025cvu,Bakx:2026rmd}
\ba
P_{gg}^{\rm stoch}(k)
&= [c_{\mathbf 1}]+O(k^2)\,.
\ea
At one-loop order, the renormalized constant is related to the bare one by
\ba
[c_{\mathbf 1}]&=c_{\mathbf 1}
-b_{[a_m]}b_{[b_\ell]}
Z_{a_m b_\ell\mathbf 1}^{\rm 1L}
+\cdots\,.
\ea
Therefore, PNG can shift the value of $[c_{\mathbf 1}]$ through generalized moments $[\Sigma_\Lambda^2]_{\d\Psi|\d\d}$ defined in \refeq{stoch_Sigma2_def}.
For the bispectrum, we evaluate the condition \refeq{stoch_rencond_two} at $n=1$. The contact term can therefore mix into a non-trivial linear operator, yielding the leading stochastic contribution
\be \label{eq:Bgggstoch}
B_{ggg}^{\rm stoch}(\vk_1,\vk_2,\vk_3)
=
[c_A]b_B P_{AB}(k_3)
+2\ {\rm perms.}
+[d_{\mathbf 1}]\,,
\ee
where following \cite{Bakx:2026rmd}, we define
\be \label{eq:stoch_cA_def}
[c_A]
\equiv c_{[A]}
-b_{[a_m]}b_{[b_\ell]}Z_{a_m b_\ell A}\,,
\qquad \textrm{with} \qquad
c_{[A]}
\equiv c_B\bigl([1+Z]^{-1}\bigr)_{BA}\,.
\ee
Thus, while the Gaussian case contains terms such as $[c_\d]b_\d P_{\d\d}$, PNG allows additional terms, for example $[c_\Psi]b_\d P_{\Psi\d}$. The constant $[d_{\mathbf 1}]$ arises from the product of three operators and is discussed in the following subsection.

\subsection{One-loop: products of three or more operators}
\label{sec:stoch_higher_products}

Products of three or more operators are required to renormalize stochastic contributions to tracer correlators beyond the power spectrum. For three operators, the corresponding operator product expansion reads \cite{Bakx:2025cvu,Bakx:2026rmd}
\ba
\label{eq:stoch_OPE_three}
[\O_{a_m}(\vx)\O_{b_\ell}(\vy)\O_{c_t}(\vz)]
&=
[\O_{a_m}(\vx)][\O_{b_\ell}(\vy)][\O_{c_t}(\vz)]
\nonumber\\
&\quad+
Z_{a_m b_\ell d_u}
[\O_{d_u}(\v{X}_{xy})\O_{c_t}(\vz)]
\diracpi(\vx-\vy)
+2\ {\rm perms.}
\nonumber\\
&\quad+
Z_{a_m b_\ell c_t d_u}
[\O_{d_u}(\v{X}_{xyz})]
\diracpi(\vx-\vy)
\diracpi(\vx-\vz)\,,
\ea
with $\v{X}_{xy}=(\vx+\vy)/2$ and $\v{X}_{xyz}=(\vx+\vy+\vz)/3$.
The terms containing a single Dirac delta subtract pairwise contact contributions and are therefore completely fixed by the two-operator renormalization discussed in the previous subsection.
The last term introduces the new coefficient $Z_{a_m b_\ell c_t d_u}$, which removes the remaining contribution localized at $\vx=\vy=\vz$.
As before, the sum over $d_u$ includes the unit operator that when inserted into tracer correlators contributes to the constant stochastic bispectrum contribution $[d_{\mathbf 1}]$ in \refeq{Bgggstoch}, whereas terms with nontrivial $\O_{d_u}$ first contribute to the trispectrum.
This expression generalizes directly to products of more than three operators.

The renormalization proceeds as in \refsec{stoch_two_products} and we focus here on the structure of the bias-kernel and primordial-correlator limits entering the calculation. The diagram linking all three operators through the same hard loop is given by
\ba
 \raisebox{-0.0cm}{\includegraphicsbox[scale=0.7]{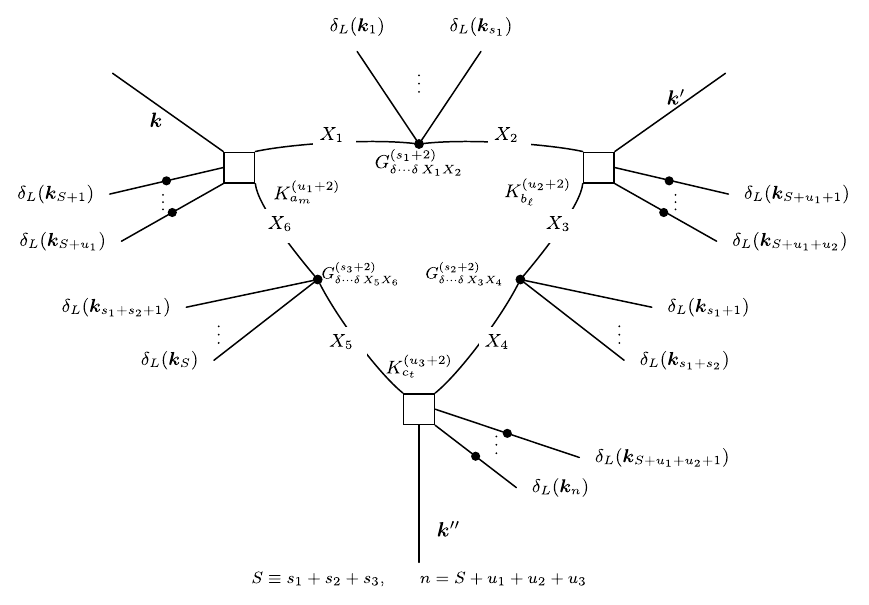}}.
\ea

We begin with the soft limit of the bias kernels. For a product of $J$ operators $\O_{A_\alpha}$, let $q_\alpha$ count the hard $\Psi$ legs integrated from the $\alpha$-th kernel. The kernel reduction takes the schematic form
\ba
\label{eq:stoch_kernel_J_schematic}
\left[
\prod_{\alpha=1}^{J}
K_{A_\alpha}^{(u_\alpha+2)}
\left(
\dots,\vp_\alpha,-\vp_{\alpha+1}
\right)
\right]_{\operatorname{av}_{\hat{\bf p}}}^{p\gg k_i}
&\qquad=
s^{{\rm 1L}}_{A_1\cdots A_J d_r;\,q_1\cdots q_J}
K_{d_r}^{(u)}(\ldots)\,,
\ea
with constants $s^{{\rm 1L}}_{A_1\cdots A_J d_r;\,q_1\cdots q_J}$ and where $u=\sum_{\alpha=1}^J u_\alpha$ counts the soft kernel legs and $r$ counts the $\Psi$ fields remaining after the hard contractions.
The momenta $\vp_\alpha$ label successive segments of the same loop, with $\vp_{J+1}\equiv\vp_1$, and differ only by soft external momenta. Choosing $\vp_1=\vp$, we average over the common loop direction $\hat{\vp}$ at fixed $p$. As in \refeq{stoch_kernel_two}, symmetrization over the soft external legs and the corresponding multiplicities are understood.
For three operators, $u=u_1+u_2+u_3$, and this becomes
\ba
\label{eq:stoch_kernel_three_schematic}
&
\left[
K_{a_m}^{(u_1+2)}
\left(\ldots,\vp_1,-\vp_2\right)
K_{b_\ell}^{(u_2+2)}
\left(\ldots,\vp_2,-\vp_3\right)
K_{c_t}^{(u_3+2)}
\left(\ldots,\vp_3,-\vp_1\right)
\right]_{\operatorname{av}_{\hat{\bf p}}}^{p\gg k_i}
\nonumber\\
&\qquad=
s^{{\rm 1L}}_{a_m b_\ell c_t d_r;\,q_1q_2q_3}
K_{d_r}^{(u)}(\ldots)\,,
\ea
where the resulting operator contains
\be
\label{eq:stoch_m_rule_three_grav}
r=m+\ell+t-q\,,
\ee
powers of $\Psi$, with $q=q_1+q_2+q_3$ counting the hard $\Psi$ legs integrated from the three kernels in the order shown.

For the initial-condition contributions, let $s_\alpha$ denote the number of soft legs attached to the $\alpha$-th primordial correlator and define $S=\sum_{\alpha=1}^J s_\alpha$.
The product of the primordial correlators then has the schematic soft limit
\ba
\label{eq:stoch_G_J_schematic}
&
\left.
\prod_{\alpha=1}^{J}
G^{(s_\alpha+2)}_{
\d\cdots\d\,X_{2\alpha-1}X_{2\alpha}}
\left(\ldots,\vp_\alpha,-\vp_\alpha+\ldots\right)
\right|^{{}^{\rm soft}_{\rm 2pt}}
\nonumber\\
&\hspace{1cm}\longrightarrow
\left[
\prod_{\alpha=1}^{J}
\lambda_{\rm NL}^{(s_\alpha+2)}
\right]
\prod_{\alpha=1}^{J}
P_{X_{2\alpha-1}X_{2\alpha}}(p)
\prod_{i=1}^{S}P_{\Psi\d}(k_i)\,,
\ea
where we suppress the soft-momentum assignments and the corresponding shifts of the hard momenta.
For three operators, the hard power spectra combine into the generalized moment
\be
\label{eq:stoch_Sigma3_schematic}
\boxed{
\begin{aligned}
\left[\Sigma_\Lambda^3\right]_{X_1X_2|X_3X_4|X_5X_6}
&\equiv
\int_{\vp}^{\Lambda}
P_{X_1X_2}(p)
P_{X_3X_4}(p)
P_{X_5X_6}(p)\,.
\end{aligned}
}
\ee
Combining the kernel and primordial soft limits dresses $\O_{d_r}$ with $S$ additional powers of $\Psi$.
Its total number of $\Psi$ fields is therefore
\be
\label{eq:stoch_m_rule_three}
m'=m+\ell+t+S-q\,.
\ee
More generally, the soft limits of products of $J$  primordial $n$-point functions will be sensitive to 
\be
\label{eq:stoch_SigmaJ_schematic}
\left[\Sigma_\Lambda^J\right]_{\bm X}
\equiv
\int_{\vp}^{\Lambda}
\prod_{\alpha=1}^{J}
P_{X_{2\alpha-1}X_{2\alpha}}(p)\,,
\ee
with $\bm X=(X_1,X_2|\cdots|X_{2J-1},X_{2J})$.

Following the notation of \cite{Bakx:2026rmd}, $[c_A]$, $[d_A]$, and $[e_A]$ denote the renormalized stochastic amplitudes associated with the primitive contacts of products of two, three, and four operators, respectively.
For the bispectrum, the new three-operator contact contributes only through the unit operator, as in \refeq{Bgggstoch}.
For the trispectrum, the corresponding stochastic contribution is
\bea
\label{eq:stoch_trispectrum_higher_products}
T_{gggg}^{\rm stoch}
&=&
[c_A]b_Bb_C
B_{ABC}(\vk_{12},\vk_3,\vk_4)
+5\ {\rm perms.}
\nonumber\\
&&+
[d_A]b_B P_{AB}(k_4)
+3\ {\rm perms.}
\nonumber\\
&&+
[c_A][c_B]P_{AB}(k_{12})
+2\ {\rm perms.}
+[e_{\mathbf 1}]
+ O(k_i^2)\,,
\eea
where the coefficients associated with the three- and four-operator contacts are defined by \cite{Bakx:2026rmd}
\ba
[d_E]
&= d_{[E]}-b_A b_B b_C Z_{ABCE}
-3[c_F]b_C Z_{FCE}\,,
\qquad
d_{[E]}=d_C([1+Z]^{-1})_{CE}\,,\\
[e_{\mathbf 1}]
&= e_{\mathbf 1}
-b_A b_B b_C b_D Z_{ABCD\mathbf 1}
-6[c_E]b_Cb_D Z_{ECD\mathbf 1} -4[d_E]b_D Z_{ED\mathbf 1}
-3[c_E][c_F]Z_{EF\mathbf 1}\,. \nonumber
\ea

\subsection{Two-loop: the collapsed configuration}
\label{sec:stoch_higher_loops}
At higher loop orders, the stochastic sector contains an increasing number of distinct topologies. Representative two-loop topologies for operator pairs are
\be
\label{eq:diagslimits2loop_a}
\raisebox{-0.0cm}{\includegraphicsbox[scale=1.0]{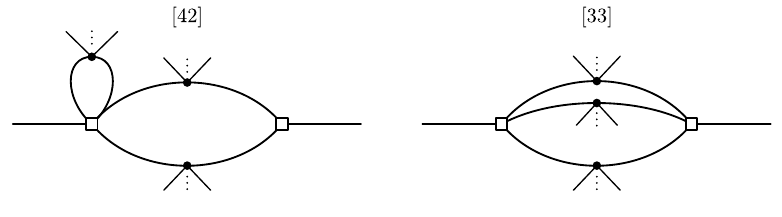}}\,,
\ee
whose systematic treatment is beyond the scope of this work.
We briefly comment on one configuration that appears when including PNGs, that is the configuration involving the collapsed, or counter-collinear, limit of the primordial trispectrum:
\be
\label{eq:diagslimits2loop_b}
\raisebox{-0.0cm}{\includegraphicsbox[scale=1.0]{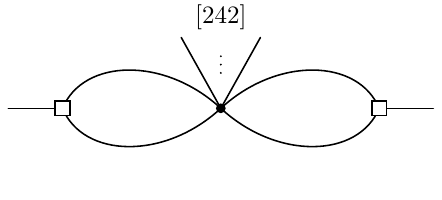}}\,.
\ee
This primordial contribution requires a connected trispectrum and is therefore absent for Gaussian initial conditions. Similar configurations involving higher primordial correlators arise at higher loop orders and for products of more operators.

For a product of two operators, the collapsed configuration arises when each bias kernel contains a nearly back-to-back hard pair, with a soft momentum $\vK$ exchanged between the pairs. The corresponding contribution has the schematic form
\bea
\label{eq:stoch_collapse_raw}
&&
\int_{\vp_1,\vp_2}^{\Lambda}
K_{a_m}(\vp_1,-\vp_1+\vK,\ldots)
K_{b_\ell}(\vp_2,-\vp_2-\vK,\ldots)
\nonumber\\
&&\hspace{3cm}\times
T_{X_1X_2X_3X_4}
(\vp_1,-\vp_1+\vK,\vp_2,-\vp_2-\vK)\,,
\eea
where $\vK$ is the total momentum of the first hard pair which equals minus that of the second, fixed by a soft combination of the external momenta. Its magnitude $k_{\rm coll}\equiv|\vK|$ is much smaller than the independently integrated hard momenta $p_1$ and $p_2$.

For single-source local PNG, the leading soft-exchange contribution factorizes into two squeezed responses connected by the long-mode power spectrum. In terms of primordial correlators, this limit corresponds to correlating two small-scale power spectra modulated by the long wavelength field $\Psi(\vk_{\rm coll})$ \cite{Assassi:2015fma}
\ba
\label{eq:stoch_collapse_limit}
&
T_{X_1X_2X_3X_4}
(\vp_1,-\vp_1+\vK,\vp_2,-\vp_2-\vK)
\nonumber\\
&\qquad\xrightarrow{k_{\rm coll}\ll p_1,p_2}
\langle P_{X_1X_2}(p_1)\Big|_{\Psi(\vk_{\rm coll})} P_{X_3X_4}(p_2)\Big|_{\Psi(\vk_{\rm coll})}\rangle \,.
\ea
For example, for hard density legs, \refeq{Ghardsoft} gives
\be
T_{\d\d\d\d}\big|_{\rm coll}
\sim
\bigl[\lambda_{\rm NL}^{(3)}\bigr]^2
P_{\d\d}(p_1)P_{\d\d}(p_2)
P_{\Psi\Psi}(k_{\rm coll})\,,
\ee
so the collapsed amplitude is fixed by the product of the two squeezed responses. 
For a local collapsed shape whose hard dependence factorizes into two power spectra, the corresponding unweighted hard moment is
\ba
\label{eq:stoch_collapse_moment}
\left[\Sigma_\Lambda^{\rm coll}\right]_{X_1X_2|X_3X_4}
&\equiv
\int_{\vp_1,\vp_2}^{\Lambda}
P_{X_1X_2}(p_1)P_{X_3X_4}(p_2) =
\left[\sigma_\Lambda^2\right]_{X_1X_2}
\left[\sigma_\Lambda^2\right]_{X_3X_4}\,.
\ea
For single-source local PNG, \refeq{stoch_collapse_limit} and the factorized hard moment in \refeq{stoch_collapse_moment} show that this contribution is reproduced by the one-loop deterministic renormalization of the two individual operators, as pointed out by \cite{Assassi:2015fma} and no additional operators are therefore required for the leading collapsed contribution. In more general inflationary scenarios, the collapsed covariance need not be fixed by the squeezed responses and can require additional stochastic operators describing long-wavelength modulations of the short-scale power \cite{Assassi:2015fma}.

\newpage

\section{Running of bias parameters}
\label{sec:running}

We now derive the renormalization-group (RG) equations for the deterministic bias parameters. In the renormalized-operator approach of \refsec{renorm}, the cutoff-dependent contributions are absorbed by the mixing matrix $Z_{BA}$, leaving renormalized operators and biases that are independent of the cutoff. Here we keep the cutoff dependence of the smoothed bare operators ${\cal O}_A(\Lambda)$ explicit, which leads to the running of the bare bias parameters $b_A(\Lambda)$ and allows us to connect loop predictions evaluated at different cutoffs through the RG trajectories we derive below, as also confirmed by lattice simulations \cite{Huang:2026zvi}. 

We derive the flow directly from the cutoff independence of tracer correlation functions, following \cite{Bakx:2025cvu,Bakx:2026rmd}. 
For a formulation using the partition function, see \cite{Rubira:2023vzw, Rubira:2024tea, Nikolis:2024kbx}.\footnote{
In Paper I \cite{Nikolis:2024kbx}, we used the Wilson--Polchinski approach to renormalize the bias coefficients by adding an interaction vertex to the partition function. Although exact, this approach is difficult to extend to higher loop orders and higher orders in PNG. At higher loops, the shell method used in the partition-function approach requires integrating out multiple shell regions with distinct momentum support. The soft-limit method \cite{Bakx:2025cvu,Bakx:2026rmd} we adopt in this work naturally generalizes to PNG and relates bias renormalization to the soft limits of bias kernels and thus to the projection of operators onto the lower-order basis. A further difficulty of the partition-function formulation is that interaction vertices beyond the bispectrum do not map one-to-one onto connected primordial $n$-point functions. We bypass this difficulty here by working directly with the primordial correlators.

Finally, applying our $n$-point function approach to the renormalization of stochastic terms, we find that the one-loop contributions arise from the squeezed $n$-point function limits rather than from the non-local kernel with two hard arguments.}  
Throughout this section we work at leading order in the gradient expansion and take all external momenta to satisfy $k_i\ll\Lambda$.
Since the smoothing scale $\Lambda$ is arbitrary, the tracer field must be independent of it,
\be\label{eq:RG_operator_condition}
0=
\frac{{\rm d}}{{\rm d}\Lambda}\delta_{g,{\rm det}}
=
\frac{{\rm d}b_A}{{\rm d}\Lambda}{\cal O}_A
+b_A\frac{{\rm d}{\cal O}_A}{{\rm d}\Lambda}\,,
\ee
so correlating it with $n>0$ long-wavelength linear fields gives
\be\label{eq:RG_Npoint_condition}
0=
\frac{{\rm d}b_A}{{\rm d}\Lambda}
\left\langle
{\cal O}_A\delta_L(\vk_1)\cdots\delta_L(\vk_n)
\right\rangle
+b_A\frac{{\rm d}}{{\rm d}\Lambda}
\left\langle
{\cal O}_A\delta_L(\vk_1)\cdots\delta_L(\vk_n)
\right\rangle\,,
\ee
such that the external fields do not contribute to the derivative because $k_i<\Lambda$. Moreover, the case $n=0$ only fixes the coefficient of the unit operator through $\langle\delta_g\rangle=0$ and does not affect the running of the non-trivial deterministic operators.

Expanding the bias running in loops, we write
\be\label{eq:RG_beta_expansion}
\frac{{\rm d}b_A}{{\rm d}\Lambda}
=
\sum_{L=1}^{\infty}
\left.\frac{{\rm d}b_A}{{\rm d}\Lambda}\right|_{L{\rm L}}\,,
\ee
such that at total loop order $L$, \refeq{RG_Npoint_condition} becomes
\ba\label{eq:RG_condition_higherL}
0={}&
\sum_{\ell=1}^{L}
\left.\frac{{\rm d}b_A}{{\rm d}\Lambda}\right|_{\ell{\rm L}}
\left\langle
{\cal O}_A\delta_L(\vk_1)\cdots\delta_L(\vk_n)
\right\rangle_{(L-\ell){\rm L}}+b_A\frac{{\rm d}}{{\rm d}\Lambda}
\left\langle
{\cal O}_A\delta_L(\vk_1)\cdots\delta_L(\vk_n)
\right\rangle_{L{\rm L}}\,.
\ea
The relation between bare bias and the renormalized biases defined by the \refeq{rencond} scheme is given taking $\Lambda\to0$, such that $b_A(\Lambda\to0)=b_{[A]}$ \cite{Rubira:2023vzw,Huang:2026zvi}.  In the following, we first derive the one-loop running in the presence of PNG and then extend the result to two loops.

\subsection{One-loop}
\label{sec:running_1L_det}

At one loop, \refeq{RG_condition_higherL} reduces to
\be\label{eq:RG_condition_1L}
0=
\left.\frac{{\rm d}b_A}{{\rm d}\Lambda}\right|_{\rm 1L}
\left\langle
{\cal O}_A\delta_L(\vk_1)\cdots\delta_L(\vk_n)
\right\rangle_{\rm tree}
+b_A\frac{{\rm d}}{{\rm d}\Lambda}
\left\langle
{\cal O}_A\delta_L(\vk_1)\cdots\delta_L(\vk_n)
\right\rangle_{\rm 1L}\,,
\ee
such that the one-loop renormalization condition \refeq{rencond1L} implies
\ba\label{eq:RG_1L_matching}
&
\left\langle
{\cal O}_{b_{m'}}\delta_L(\vk_1)\cdots\delta_L(\vk_n)
\right\rangle_{\rm 1L}
\bigg|_{\vk_i\to0} =-
Z^{\rm 1L}_{b_{m'}a_m}
\left\langle
{\cal O}_{a_m}\delta_L(\vk_1)\cdots\delta_L(\vk_n)
\right\rangle_{\rm tree}
\bigg|_{\vk_i\to0}\,.
\ea
Substituting this relation into \refeq{RG_condition_1L} and matching the independent tree-level kernels gives
\be\label{eq:RG_beta1L_matrix}
\left.\frac{{\rm d}b_{a_m}}{{\rm d}\Lambda}\right|_{\rm 1L}
=
b_{b_{m'}}(\Lambda)
\frac{{\rm d}Z^{\rm 1L}_{b_{m'}a_m}}{{\rm d}\Lambda}\,.
\ee
Inserting the one-loop mixing matrix \refeq{Z1L}, the flow becomes
\be\label{eq:RG_beta1L_PNG}
\boxed{
\begin{aligned}
\left.\frac{{\rm d}b_{a_m}}{{\rm d}\Lambda}\right|_{\rm 1L}
={}&-
\sum_{b_{m'}}b_{b_{m'}}(\Lambda)
\sum_{s}
\sum_{\substack{X_1,X_2\\
q = \#\Psi}}
\frac{1}{s!}
s^{\rm 1L}_{b_{m'}a_{m'-q+s};\,q}
\delta^{\rm K}_{m,m'+s-q} \lambda_{\rm NL}^{(s+2)}
\frac{{\rm d}}{{\rm d}\Lambda}
\left[\sigma_\Lambda^2\right]_{X_1X_2}
\,,
\end{aligned}
}
\ee
where $q$ is defined in \refeq{q1l} and counts the $\Psi$ fields among the two hard legs, and the ranges of $s$ and $q$ are those defined below \refeq{q1l}. 
The Kronecker delta in \refeq{RG_beta1L_PNG} fixes the selection rule \refeq{fixed_condition_m}.
The derivative of the generalized variance \refeq{sigmatildedef} generalizes the Gaussian case
\be\label{eq:RG_variance_shell}
\frac{{\rm d}\sigma_\Lambda^2}{{\rm d}\Lambda}
\equiv
\frac{\Lambda^2}{2\pi^2}\Plin(\Lambda)\,,
\ee
to 
\be\label{eq:RG_generalized_variance_sigma}
\frac{{\rm d}}{{\rm d}\Lambda}
\left[\sigma_\Lambda^2\right]_{X_1X_2}
=
\frac{1}{M^q(\Lambda)}
\frac{{\rm d}\sigma_\Lambda^2}{{\rm d}\Lambda}
=
\frac{\Lambda^2}{2\pi^2}
P_{X_1X_2}(\Lambda)
=
\frac{1}{M^q(\Lambda)}
\frac{\Lambda^2}{2\pi^2}\Plin(\Lambda)\,.
\ee

The one-loop running of the Gaussian bias parameters is
\begin{equation} \label{eq:Gaussrunning}
\frac{db_{a_0}}{d\L}
= -\frac{\L^2\Plin(\L)}{2\pi^2}
\Biggl(
b_{b_0}s^{\rm 1L}_{b_0a_0;\,0}
+\frac{b_{b_1}}{M(\L)}s^{\rm 1L}_{b_1a_0;\,1}
+\frac{b_{b_2}}{M^2(\L)}s^{\rm 1L}_{b_2a_0;\,2}
\Biggr)\,,
\end{equation}
where the first term is the usual Gaussian mixing \cite{Rubira:2023vzw, Bakx:2025cvu}, obtained by integrating out two hard $\delta$ legs of a Gaussian operator. The second and third terms mix operators with one and two $\Psi$ fields into the Gaussian sector by integrating out hard $\Psi\delta$ and $\Psi\Psi$ pairs, respectively, whose power spectra introduce the factors $M^{-1}(\L)$ and $M^{-2}(\L)$ relative to $\Plin(\L)$.\footnote{Note once more that in our convention for the bias expansion \refeq{bias-expansion}, the PNG amplitudes $\lambda_{\rm NL}$ are absorbed into their bias coefficients, whereas they are usually kept explicit, including in \cite{Nikolis:2024kbx}.} All three contributions have $s=0$, coming from two-point function contributions and all $\Psi$ fields in the source operator are integrated out. 
Note from \reffig{generalized_variance} that the Gaussian term $b_{b_0}s^{\rm 1L}_{b_0a_0;\,0}$, coming from $\left[\sigma_\Lambda^2\right]_{\delta \delta}$ is the dominant one on small scales by many orders of magnitude.

We truncate the PNG dependence by keeping contributions $b_{b_{m'}}\lambda_{\rm NL}^{(s+2)}$ with $m'+s\leq3$, where $m'$ counts the $\Psi$ fields already present in the source operator and $s$ counts the additional soft $\Psi$ fields generated by the primordial $(s+2)$-point function.
Independently, we keep bias operators through fifth order in the linear fields.
The corresponding running of the non-Gaussian bias parameters is
\ba \label{eq:bPNG_run}
\frac{db_{a_1}}{d\L}
&= -\frac{\L^2\Plin(\L)}{2\pi^2}
\Biggl(
b_{b_1}s^{\rm 1L}_{b_1a_1;\,0}
+\frac{b_{b_2}}{M(\L)}s^{\rm 1L}_{b_2a_1;\,1}
+\frac{b_{b_3}}{M^2(\L)}s^{\rm 1L}_{b_3a_1;\,2}
\\
&\hspace{4cm}
+\lambda_{\rm NL}^{(3)}b_{b_0}s^{\rm 1L}_{b_0a_1;\,0}
+\frac{\lambda_{\rm NL}^{(3)}b_{b_1}}{M(\L)}
 s^{\rm 1L}_{b_1a_1;\,1}
+\frac{\lambda_{\rm NL}^{(3)}b_{b_2}}{M^2(\L)}
 s^{\rm 1L}_{b_2a_1;\,2}
\Biggr),
\vs[1ex]
\frac{db_{a_2}}{d\L}
&= -\frac{\L^2\Plin(\L)}{2\pi^2}
\Biggl(
b_{b_2}s^{\rm 1L}_{b_2a_2;\,0}
+\frac{b_{b_3}}{M(\L)}s^{\rm 1L}_{b_3a_2;\,1}
+\lambda_{\rm NL}^{(3)}b_{b_1}s^{\rm 1L}_{b_1a_2;\,0}
\vs
&\hspace{4cm}
+\frac{\lambda_{\rm NL}^{(3)}b_{b_2}}{M(\L)}
 s^{\rm 1L}_{b_2a_2;\,1}
+\frac{\lambda_{\rm NL}^{(4)}}{2}b_{b_0}s^{\rm 1L}_{b_0a_2;\,0}
+\frac{\lambda_{\rm NL}^{(4)}b_{b_1}}{2M(\L)}
 s^{\rm 1L}_{b_1a_2;\,1}
\Biggr),
\vs[1ex]
\frac{db_{a_3}}{d\L}
&= -\frac{\L^2\Plin(\L)}{2\pi^2}
\Biggl(
b_{b_3}s^{\rm 1L}_{b_3a_3;\,0}
+\lambda_{\rm NL}^{(3)}b_{b_2}s^{\rm 1L}_{b_2a_3;\,0}
+\frac{\lambda_{\rm NL}^{(4)}}{2}b_{b_1}s^{\rm 1L}_{b_1a_3;\,0}
+\frac{\lambda_{\rm NL}^{(5)}}{6}b_{b_0}s^{\rm 1L}_{b_0a_3;\,0}
\Biggr), \nonumber
\ea
such that $b_{a_1}$ is directly sensitive to $4\fnl=\lambda^{(3)}_{\rm NL}$, $b_{a_2}$ to both $\fnl$ and $12g_{\rm NL}=\lambda^{(4)}_{\rm NL}$ and $b_{a_3}$ also to $\lambda^{(5)}_{\rm NL}$.

\begin{figure}[t]
    \centering
    \includegraphics[width=0.95\linewidth]{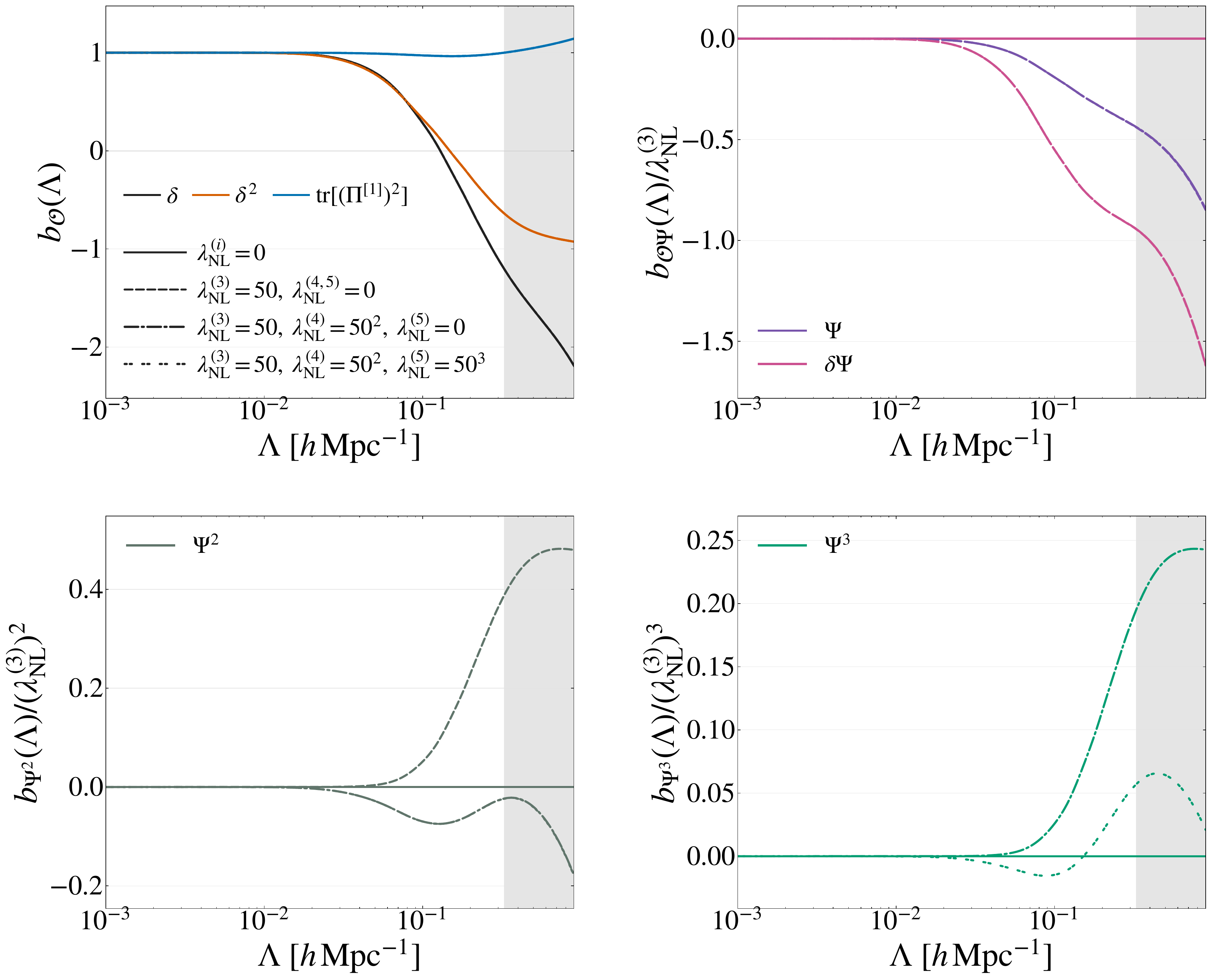}
    \caption{Running of Gaussian terms (top-left), and terms with one (top-right), two (bottom-left) and three (bottom-right) powers of $\Psi$ for different values of $4\fnl=\lambda_{\rm NL}^{(3)}$, $12g_{\rm NL}=\lambda_{\rm NL}^{(4)}$ and $\lambda_{\rm NL}^{(5)}$. We set initial conditions at $\Lambda_\star = 0$, with $b^\star_{A}=b_{A}(\Lambda_\star)$ and fixing $b^\star_{a_m}=0$ for $m\geq1$, $b^\star_{a}=1$ for ${\rm ord}({\cal O}_{a}) \leq 2$ and $b^\star_{a}=0$ for ${\rm ord}({\cal O}_{a}) \geq 3$. We normalize the bias parameters as $b_{a_n}/\fnl^n$, such that they assume $O(1)$ values. The shaded region corresponds to $\sigma^2_\L\gtrsim 1$.}
    \label{fig:1Lrunning_IC1}
\end{figure}

\reffig{1Lrunning_IC1} shows the running of the bias parameters for different values of $\fnl$, $g_{\rm NL}$, and $\lambda_{\rm NL}^{(5)}$, with $g_{\rm NL}=O(\fnl^2)$ and $\lambda_{\rm NL}^{(5)}=O(\fnl^3)$. We impose initial conditions at $\Lambda_\star=0$, defining $b^\star_A\equiv b_A(\Lambda_\star)$, and set $b^\star_{a_m}=0$ for $m\geq1$, $b^\star_a=1$ for ${\rm ord}({\cal O}_a)\leq2$, and $b^\star_a=0$ for ${\rm ord}({\cal O}_a)\geq3$, so that only Gaussian bias parameters are initially nonzero. The solid lines show the Gaussian case, $\lambda_{\rm NL}^{(i)}=0$ for $i\geq3$, in which no non-Gaussian bias parameters are generated. For nonzero primordial couplings, the $b_{b_0}$ terms on the right-hand side of \refeq{bPNG_run} generate non-Gaussian bias parameters even though their initial values vanish. For $\fnl\neq0$, we plot $b_{a_n}/\fnl^n$ to factor out the expected PNG scaling. The resulting coefficients are $O(1)$, consistent with the usual parametrization $b_\Psi\to\fnl b_\Psi$, in which the PNG amplitude is factored out rather than absorbed into the bias coefficient as in \refeq{bias-expansion}. For these initial conditions, the Gaussian bias parameters (top-left panel) are almost insensitive to PNG because the generalized variances $\left[\sigma_\Lambda^2\right]_{\Psi\delta}$ and $\left[\sigma_\Lambda^2\right]_{\Psi\Psi}$ entering \refeq{Gaussrunning} are orders of magnitude smaller than $\left[\sigma_\Lambda^2\right]_{\delta\delta}$, as shown in \reffig{generalized_variance}.\footnote{These initial conditions differ from those in \cite{Nikolis:2024kbx}, where the bias parameters are initialized at $\Lambda=0.2\,h\,{\rm Mpc}^{-1}$ and evolved towards $\Lambda\to0$. In particular, those initial conditions excite eigenmodes that grow towards the infrared, leading to substantially different running as $\Lambda\to0$.
}

In contrast, \reffig{1Lrunning_IC1} shows that the running of the PNG bias parameters is sensitive to the primordial amplitudes $\lambda_{\rm NL}^{(i)}$, which enter \refeq{bPNG_run} through terms with no inverse powers of $M(\Lambda)$. The running of $b_{a_1}$ (top-right panel) depends on $\lambda_{\rm NL}^{(3)}$, that of $b_{a_2}$ (bottom-left panel) additionally depends on $\lambda_{\rm NL}^{(4)}$, and that of $b_{a_3}$ (bottom-right panel) also depends on $\lambda_{\rm NL}^{(5)}$.

The solution of the one-loop RG equation generates iterated higher-loop contributions. To illustrate this, consider as an example the running of a Gaussian bias parameter $b_{a_0}$ with $b^\star_{a_m}=0$ for $m\geq1$. Expanding around a reference scale $\Lambda_\star$, we write
\begin{equation}\label{eq:RG_solution_expansion_a0}
b_{a_0}(\Lambda)
=
b^\star_{a_0}
+b^{(1)}_{a_0}(\Lambda)
+b^{(2)}_{a_0}(\Lambda)
+\cdots,
\qquad
b^\star_{a_m}\equiv b_{a_m}(\Lambda_\star),
\end{equation}
where the superscript counts successive insertions of the one-loop mixing matrix, each $b^{(n)}$ associate with (the hierarchical regions of) different loop contributions, with  $b^\star_{a_0}$ corresponding to the tree-level term.\footnote{Note that the expansion can be organized both by the number of one-loop insertions, each integrating a pair of hard linear fields, as we do here, and by inverse powers of $M$, which track the number of $\Psi$ legs integrated over in each contraction.}
For the first iteration, under the initial conditions $b^\star_{a_m}=0$ for $m\geq1$, only the Gaussian operators contribute, and hence
\begin{equation}\label{eq:RG_solution_1L_a0}
b_{a_0}^{(1)}(\Lambda)
= 
-\int_{\Lambda_\star}^{\Lambda}{\rm d}\Lambda_1\,
\frac{{\rm d}\sigma_{\Lambda_1}^2}{{\rm d}\Lambda_1}
\sum_{b_0}
b^\star_{b_0}\,
s^{\rm 1L}_{b_0a_0;0}\,,
\end{equation}
which is the usual one-loop correction sourced by the initial Gaussian bias parameters.
The second iteration is obtained by inserting the first-order running of the intermediate biases back into the flow of $b_{a_0}$
\begin{align}\label{eq:RG_solution_1L1L_a0}
b_{a_0}^{(2)}(\Lambda)
={}&
\int_{\Lambda_\star}^{\Lambda}{\rm d}\Lambda_1\,
\frac{{\rm d}\sigma_{\Lambda_1}^2}{{\rm d}\Lambda_1}
\int_{\Lambda_\star}^{\Lambda_1}{\rm d}\Lambda_2\,
\frac{{\rm d}\sigma_{\Lambda_2}^2}{{\rm d}\Lambda_2}\Bigg\{\sum_{b_0,c_0}
b^\star_{c_0}\,
s^{\rm 1L}_{b_0a_0;0}
s^{\rm 1L}_{c_0b_0;0}\\
&\hspace{-1.5cm}
+
4f_{\rm NL}
\frac{1}
{M(\Lambda_1)}
\sum_{b_1,c_0}
b^\star_{c_0}\,
s^{\rm 1L}_{b_1a_0;1}
s^{\rm 1L}_{c_0b_1;0}
+
\frac{12g_{\rm NL}}{2}
\frac{1}
{M^2(\Lambda_1)}
\sum_{b_2,c_0}
b^\star_{c_0}\,
s^{\rm 1L}_{b_2a_0;2}
s^{\rm 1L}_{c_0b_2;0} + \dots
\Bigg\}, \nonumber
\end{align}
where the dots represent higher-in-$\Psi$ terms and each of the three terms in \refeq{RG_solution_1L1L_a0} corresponds to the $(n){\rm pt}$ function channels
\begin{align}\label{eq:RG_solution_cascades_a0}
{\cal O}_{c_0}
&\xrightarrow[\Lambda_2]{(2){\rm pt}}
{\cal O}_{b_0}
\xrightarrow[\Lambda_1]{(2){\rm pt}}
{\cal O}_{a_0},
\notag\\
{\cal O}_{c_0}
&\xrightarrow[\Lambda_2]{(3){\rm pt}}
{\cal O}_{b_1}
\xrightarrow[\Lambda_1]{(2){\rm pt}}
{\cal O}_{a_0},
\notag\\
{\cal O}_{c_0}
&\xrightarrow[\Lambda_2]{(4){\rm pt}}
{\cal O}_{b_2}
\xrightarrow[\Lambda_1]{(2){\rm pt}}
{\cal O}_{a_0}.
\end{align}
Thus, even if the PNG bias parameters vanish at $\Lambda_\star$, the first step of the flow can generate operators containing higher powers in $\Psi$ fields through the primordial $n$-point functions, while another RG step can integrate out their hard $\Psi$ legs and mix them back into the Gaussian sector. Integrating out $\Psi$ legs suppresses the channel by negative powers of $M$. 

\begin{figure}[t]
    \centering
    \includegraphics[width=0.95\linewidth]{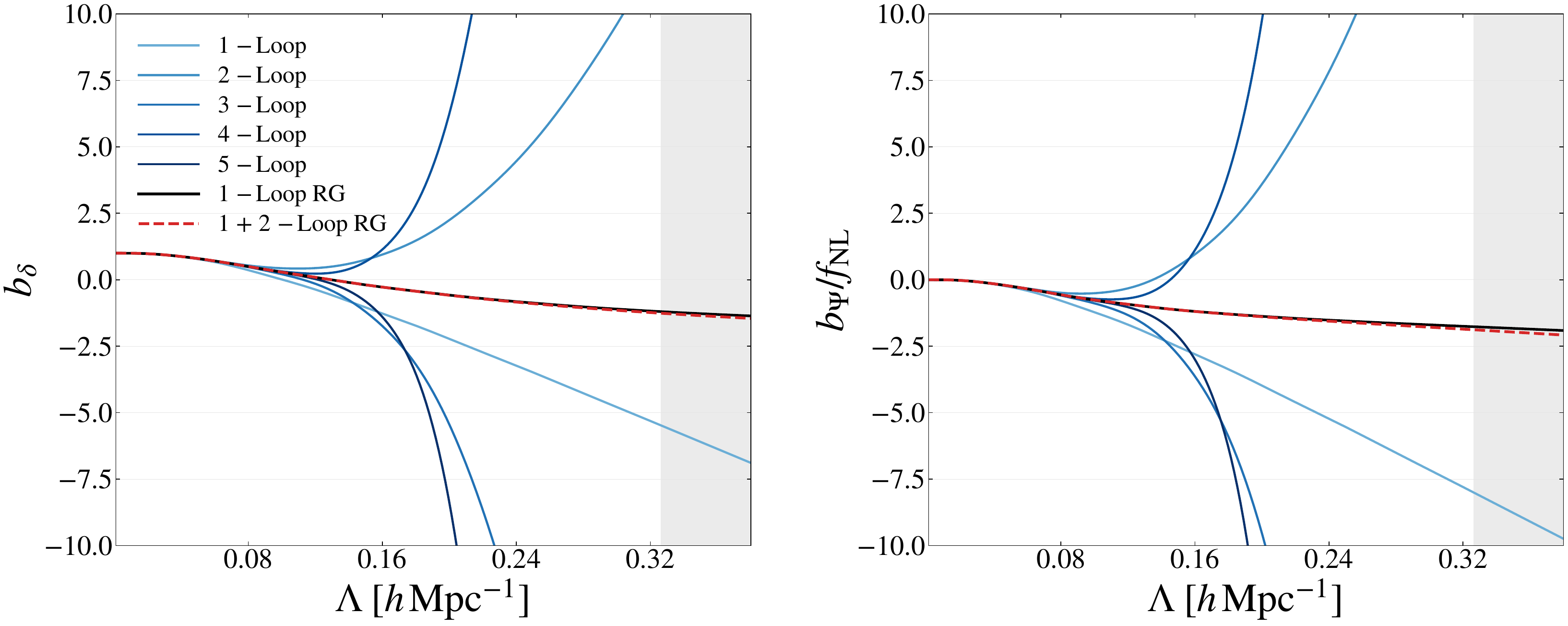}
    \caption{Iterative solution for $b_\d$ in the left and $b_\Psi$ in the right panel, as given by \refeq{RG_solution_expansion_a0}, with each $b^{(n)}_{a_0}$ term associated to an $n$-loop term. The full one-loop RG solution is shown in black while the full two-loop RG result is shown in red. We consider $\fnl = 50$ and keep $\lambda_{\rm NL}^{(i)} = 0$ for $i\geq 4$, taking the same initial conditions as in \reffig{1Lrunning_IC1}. The gray area indicates the region where $\sigma^2_\Lambda > 1$.}
    \label{fig:iterativesol}
\end{figure}

We show each term of the iterative solution \refeq{RG_solution_expansion_a0} for $b_\delta$ in the left panel of \reffig{iterativesol}, while the right panel shows similar terms for the expansion of the $b_\Psi$ solution. The full RG solution resumming all these terms is given in black.  Note that the one-loop perturbative result captures the leading running at low $\Lambda$, while the full one-loop RG solution adds significant contributions at higher $\Lambda$, demonstrating the importance of these iterated contributions in both the Gaussian and non-Gaussian terms.

\subsection{Two-loop}
\label{sec:running_2L_det}

\subsubsection{Two-loop running}

At two loops, cutoff independence requires
\ba\label{eq:RG_condition_2L}
0={}&
\left.\frac{{\rm d}b_A}{{\rm d}\Lambda}\right|_{\rm 2L}
\left\langle
{\cal O}_A\delta_L(\vk_1)\cdots\delta_L(\vk_n)
\right\rangle_{\rm tree}
+
\left.\frac{{\rm d}b_A}{{\rm d}\Lambda}\right|_{\rm 1L}
\left\langle
{\cal O}_A\delta_L(\vk_1)\cdots\delta_L(\vk_n)
\right\rangle_{\rm 1L}
\vs
&+
b_A\frac{{\rm d}}{{\rm d}\Lambda}
\left\langle
{\cal O}_A\delta_L(\vk_1)\cdots\delta_L(\vk_n)
\right\rangle_{\rm 2L}\,.
\ea
Using the one-loop matching relation \refeq{RG_1L_matching}, the two-loop renormalization condition \refeq{rencond2L} gives
\ba\label{eq:RG_2L_matching}
&\left\langle
{\cal O}_{c_{m''}}\delta_L(\vk_1)\cdots\delta_L(\vk_n)
\right\rangle_{\rm 2L}
\bigg|_{\vk_i\to0}
\vs
&\qquad=
\left(
Z^{\rm 1L}_{c_{m''}b_{m'}}
Z^{\rm 1L}_{b_{m'}a_m}
-
Z^{\rm 2L}_{c_{m''}a_m}
\right)
\left\langle
{\cal O}_{a_m}\delta_L(\vk_1)\cdots\delta_L(\vk_n)
\right\rangle_{\rm tree}
\bigg|_{\vk_i\to0}\,.
\ea
Substituting this expression and the one-loop flow \refeq{RG_beta1L_matrix} into \refeq{RG_condition_2L}, and matching the independent tree-level structures, yields
\be\label{eq:RG_beta2L_matrix}
\left.\frac{{\rm d}b_{a_m}}{{\rm d}\Lambda}\right|_{\rm 2L}
=
b_{c_{m''}}(\Lambda)
\left[
\frac{{\rm d}Z^{\rm 2L}_{c_{m''}a_m}}{{\rm d}\Lambda}
-
Z^{\rm 1L}_{c_{m''}b_{m'}}
\frac{{\rm d}Z^{\rm 1L}_{b_{m'}a_m}}{{\rm d}\Lambda}
\right].
\ee
Inserting the two-loop mixing matrix \refeq{Z2L_general}, the derivative of the product of one-loop counterterms partially cancels the second term in \refeq{RG_beta2L_matrix}, leaving 
\ba\label{eq:RG_beta2L_Zsub}
\left.\frac{{\rm d}b_{a_m}}{{\rm d}\Lambda}\right|_{\rm 2L}
&=
b_{c_{m''}}(\Lambda)
\left[
\frac{{\rm d}Z^{\rm 1L}_{c_{m''}b_{m'}}}{{\rm d}\Lambda}
Z^{\rm 1L}_{b_{m'}a_m}
-
\frac{{\rm d}}{{\rm d}\Lambda}
\left(
Z^{\rm 2L,[22]}_{c_{m''}a_m}
+
Z^{\rm 2L,[3]}_{c_{m''}a_m}
\right)
\right]
\vs
&\equiv
\left.\frac{{\rm d}b_{a_m}}{{\rm d}\Lambda}\right|_{\rm 1L,1L}
+
\left.\frac{{\rm d}b_{a_m}}{{\rm d}\Lambda}\right|_{\rm 2L,[22]}
+
\left.\frac{{\rm d}b_{a_m}}{{\rm d}\Lambda}\right|_{\rm 2L,[3]}\,.
\ea
Here the first contribution arises from two successive one-loop contractions ({\it one-loop squared}), with the derivative acting on the first mixing matrix. The remaining contributions are the cutoff derivatives of the bare pair-pair and connected coefficients defined in \refeq{I2L_22} and \refeq{I2L_3}, respectively.

The red curves in \reffig{iterativesol} show the full two-loop RG solution for $b_\delta$ and $b_\Psi$ obtained from \refeq{RG_beta2L_Zsub}. Note that while the one-loop RG solution captures substantial corrections through iterated one-loop contributions, the additional change from one-loop to two-loop RG evolution is comparatively small and becomes significant only at large $\Lambda$, in the non-perturbative regime $\sigma_\Lambda^2>1$. This suggests that the dominant higher-loop contributions, arising from hierarchical momentum regions, are already captured by the one-loop-iterated RG solution, as in the Gaussian case \cite{Bakx:2025cvu,Bakx:2026rmd}. We now discuss how these hierarchical regions are subtracted from the two-loop solution \refeq{RG_beta2L_Zsub}.

\subsubsection{Hierarchical limits}

We now identify the part of \refeq{RG_beta2L_Zsub} that is not already generated by successive one-loop running. 
In the language of the method of regions \cite{Beneke:1997zp}, the hierarchical configurations $k_i\ll p\ll\Lambda$ reproduce two successive one-loop contractions, such that the one-loop squared term subtracts these contributions from the two-loop flow.
For Gaussian initial conditions (taking $m''=m=0,\;s_1=s_2=0,\;q_1=q_2=q=0$), this cancellation follows from $s^{\rm 2L,[22]}_{c_0a_0;\,00}(0) =s^{\rm 1L}_{c_0b_0;\,0}s^{\rm 1L}_{b_0a_0;\,0}$. The [3] topology is absent in that case, and the sum of the one-loop squared and the [22] term \refeq{RG_beta2L_Zsub} gives \cite{Bakx:2025cvu,Bakx:2026rmd}
\ba\label{eq:gausssubtraction}
\left.\frac{{\rm d}b_{a_0}}{{\rm d}\Lambda}\right|_{\rm 2L}
&=
-b_{c_0}(\Lambda)
\frac{{\rm d}\sigma_\Lambda^2}{{\rm d}\Lambda}
\int_{p<\Lambda}\Plin(p)
\left[
s^{\rm 2L,[22]}_{c_0a_0;\,00}(p/\Lambda)
-s^{\rm 2L,[22]}_{c_0a_0;\,00}(0)
\right]\,,
\ea
so the bracket vanishes as $p/\Lambda\to0$ after using \refeq{commutationoflimitsrelation}. 

We generalize this expression to the non-Gaussian case, in which we find that the hierarchical cancellation depends on the integrated $\Psi$ legs. To have a clear contribution from each topology, we write
\be
Z^{\rm 1L}_{b_{m'}a_m}
=
\sum_q Z^{\rm 1L}_{b_{m'}a_m;\,q}\,,
\ee
where $Z^{\rm 1L}_{b_{m'}a_m;\,q}$ denotes the contribution with $q$ integrated $\Psi$ legs, so that the number of generated $\Psi$ legs is fixed by $s=m-m'+q$.
For the two successive contractions, we have
\be
s_1=m'-m''+q_1,
\qquad
s_2=m-m'+q_2\,.
\ee
The first contraction leaves $m''-q_1$ original $\Psi$ legs and generates $s_1$ new ones, giving $m'=m''-q_1+s_1$ intermediate $\Psi$ legs in total.
For $q_2\leq2$, the second contraction can:
\begin{enumerate}
    \item integrate no $\Psi$ legs, or integrate one or two original $\Psi$ legs;
    \item integrate one newly generated $\Psi$ leg, either alone ($q_2=1$) or together with one original $\Psi$ leg ($q_2=2$);
    \item integrate two newly generated $\Psi$ legs ($q_2=2$).
\end{enumerate}
As we show below, the first two cases contribute to the subtraction of hierarchical regions of the $[22]$ and $[3]$ topologies, respectively.
The third case constitutes the remainder ({\it `rest'}), which is not matched by hierarchical limits of the bare topologies.
For convenience, we introduce the combinatorial weights
\be \label{eq:weights}
w^{[22]}_{m''m';\,q_1q_2}
\equiv
\frac{\binom{m''-q_1}{q_2}}
     {\binom{m'}{q_2}},
\qquad
w^{[3]}_{m''m';\,q_1q_2}
\equiv
\frac{s_1\binom{m''-q_1}{q_2-1}}
     {\binom{m'}{q_2}}\,,
\qquad
w^{[{\rm rest}]}_{m''m';\,q_1q_2}
\equiv
\frac{\binom{s_1}{2}\binom{m''-q_1}{q_2-2}}
     {\binom{m'}{q_2}}\,,
\ee
that select contractions in which the second step integrates zero, one, or two newly generated $\Psi$ legs, respectively.
Note that they satisfy $w^{[22]}+w^{[3]}+w^{[{\rm rest}]}=1$.
We separate the successive one-loop term according to the origin of the $\Psi$ legs integrated by the second contraction:
\be
\left.\frac{{\rm d}b_{a_m}}{{\rm d}\Lambda}\right|_{\rm 1L,1L}
=
\left.\frac{{\rm d}b_{a_m}}{{\rm d}\Lambda}
\right|_{\rm 1L,1L;\,[22]}
+
\left.\frac{{\rm d}b_{a_m}}{{\rm d}\Lambda}
\right|_{\rm 1L,1L;\,[3]} +
\left.\frac{{\rm d}b_{a_m}}{{\rm d}\Lambda}
\right|_{\rm 1L,1L;\,[{\rm rest}]}\,,
\ee
where the last term contains sequences such as $\O\to\O\Psi^2\to\dlin$ and is not matched by hierarchical limits of the bare $[22]$ and $[3]$ topologies.
The three contributions are
\ba
\left.\frac{{\rm d}b_{a_m}}{{\rm d}\Lambda}
\right|_{\rm 1L,1L;\,[22]}
&\equiv
b_{c_{m''}}(\Lambda)
\sum_{b_{m'}} \sum_{q_1}\sum_{q_2=0}^{2}
w^{[22]}_{m''m';\,q_1q_2}
\frac{{\rm d}Z^{\rm 1L}_{c_{m''}b_{m'};\,q_1}}
     {{\rm d}\Lambda}
Z^{\rm 1L}_{b_{m'}a_m;\,q_2}\,,
\vs
\left.\frac{{\rm d}b_{a_m}}{{\rm d}\Lambda}
\right|_{\rm 1L,1L;\,[3]}
&\equiv
b_{c_{m''}}(\Lambda)
\sum_{b_{m'}} \sum_{q_1}\sum_{q_2=1}^{2}
w^{[3]}_{m''m';\,q_1q_2}
\frac{{\rm d}Z^{\rm 1L}_{c_{m''}b_{m'};\,q_1}}
     {{\rm d}\Lambda}
Z^{\rm 1L}_{b_{m'}a_m;\,q_2}\,,
\vs
\left.\frac{{\rm d}b_{a_m}}{{\rm d}\Lambda}
\right|_{\rm 1L,1L;\,[{\rm rest}]}
&\equiv
b_{c_{m''}}(\Lambda)
\sum_{b_{m'}} \sum_{q_1}
w^{[{\rm rest}]}_{m''m';\,q_1\,2}
\frac{{\rm d}Z^{\rm 1L}_{c_{m''}b_{m'};\,q_1}}
     {{\rm d}\Lambda}
Z^{\rm 1L}_{b_{m'}a_m;\,2}\,.
\ea
Below we show how the first two contributions subtract the hierarchical regions of the $[22]$ and $[3]$ diagrams, respectively.

We first consider the subtraction of hierarchical regions of the $[22]$ topology.
Let $q_1$ and $q_2$ denote the numbers of $\Psi$ legs in the hard pairs $(X_1,X_2)$ and $(X_3,X_4)$, respectively, with $q=q_1+q_2$.
Choosing the shell pair as $p_1=\Lambda$ and writing $p\equiv p_2<\Lambda$, the leading hierarchical contribution corresponds to $r=\Lambda/p\to\infty$ and reads
\ba
\frac{db_{a_m}}{d\L}\Big|^{p/\L\to 0}_{{\rm 2L},[22]} &=-
b_{c_{m''}}(\Lambda)
\frac{{\rm d}\sigma_\Lambda^2}{{\rm d}\Lambda}
\sum_{s_1,s_2\geq0}^{s_1+s_2\leq 1}
\sum_{\substack{X_1,\ldots,X_4\\q_2\leq2}}
\frac{
\lambda_{\rm NL}^{(s_1+2)}
\lambda_{\rm NL}^{(s_2+2)}
}{s_1!s_2!}
\vs
&\quad\times
\frac{\delta^{\rm K}_{m,m''+s_1+s_2-q}}{M^{q_1}(\Lambda)}
\int_{p<\Lambda}
\frac{\Plin(p)}{M^{q_2}(p)}
\left[
s^{\rm 2L,[22]}_{c_{m''}a_m;\,q_1q_2}
\left(\infty\right)\right]\;. \vs
&=-
b_{c_{m''}}(\Lambda)
\frac{{\rm d}\sigma_\Lambda^2}{{\rm d}\Lambda}
\sum_{s_1,s_2\geq0}^{s_1+s_2\leq 1}
\sum_{\substack{X_1,\ldots,X_4\\q_2\leq 2}}
\frac{
\lambda_{\rm NL}^{(s_1+2)}
\lambda_{\rm NL}^{(s_2+2)}
}{s_1!s_2!}w^{[22]}_{m''m';\,q_1q_2}
\\
&\quad \times \frac{\delta^{\rm K}_{m,m''+s_1+s_2-q}}{M^{q_1}(\Lambda)}
\int_{p<\Lambda}
\frac{\Plin(p)}{M^{q_2}(p)}
\left[s^{{\rm 1L}}_{c_{m''}b_{m''+s_1-q_1};\,q_1}
s^{{\rm 1L}}_{b_{m''+s_1-q_1}a_{m''-q+s_1+s_2};\,q_2}\right]. \nonumber
\ea
The second equality follows from the sequential limits in \refeq{commutationoflimitsrelation}, whose combinatorial prefactor is the weight $w^{[22]}_{m''m';\,q_1q_2}$ defined in \refeq{weights}, with $m'=m''-q_1+s_1$. The weighted successive one-loop term therefore cancels the leading hierarchical contribution, giving
\be\label{eq:RG_beta2L_explicit}
\boxed{
\begin{aligned}
\left.\frac{{\rm d}b_{a_m}}{{\rm d}\Lambda}
\right|_{\rm 1L,1L;\,[22]}
+
\left.\frac{{\rm d}b_{a_m}}{{\rm d}\Lambda}
\right|_{\rm 2L,[22]}
={}&
-
b_{c_{m''}}(\Lambda)
\frac{{\rm d}\sigma_\Lambda^2}{{\rm d}\Lambda}
\sum_{s_1,s_2\geq0}^{s_1+s_2\leq1}
\sum_{\substack{X_1,\ldots,X_4\\q_2\leq 2}}
\frac{
\lambda_{\rm NL}^{(s_1+2)}
\lambda_{\rm NL}^{(s_2+2)}
}{s_1!s_2!}
\\
&\hspace{-5cm}\times
\frac{\delta^{\rm K}_{m,m''+s_1+s_2-q}}{M^{q_1}(\Lambda)}
\int_{p<\Lambda}
\frac{\Plin(p)}{M^{q_2}(p)}
\left[
s^{\rm 2L,[22]}_{c_{m''}a_m;\,q_1q_2}
\left(\frac{\Lambda}{p}\right)
-
s^{\rm 2L,[22]}_{c_{m''}a_m;\,q_1q_2}(\infty)
\right],
\end{aligned}
}
\ee
such that the expression in the square brackets vanishes as $p/\Lambda\to0$, removing the contribution from hierarchically separated scales.
Note that the pair-exchange symmetry also allows the bracket to be written with the smaller momentum ratio as $s^{\rm 2L,[22]}_{c_{m''}a_m;\,q_2q_1}(p/\Lambda) - s^{\rm 2L,[22]}_{c_{m''}a_m;\,q_2q_1}(0)$. This expression generalizes \refeq{gausssubtraction} in the presence of PNGs.

We next consider the connected topology, matching correlators with one external linear density field.
Its bare contribution is 
\ba\label{eq:RG_beta2L_explicit3}
\left.\frac{{\rm d}b_{a_m}}{{\rm d}\Lambda}
\right|_{\rm 2L,[3]}
={}&
-b_{c_{m''}}(\Lambda)
\sum_{s=0}^{1}
\sum_{\substack{X_1,X_2,X_3\\q=\#\Psi}}
\frac{\delta^{\rm K}_{m,m''+s-q}}{s!}
\vs
&\quad\times
\frac{{\rm d}}{{\rm d}\Lambda}
\int_{p_1,p_2<\Lambda}
{\cal B}_{X_1X_2X_3,s+3}(\vp_1,\vp_2)
s^{\rm 2L,[3]}_{c_{m''}a_m;\,X_1X_2X_3}
(\vp_1,\vp_2),
\ea
which can be decomposed as
\ba
\left.\frac{{\rm d}b_{a_m}}{{\rm d}\Lambda}
\right|_{\rm 2L,[3]}
={}&
-b_{c_{m''}}(\Lambda)
\sum_{s=0}^{1}
\sum_{\substack{X_1,X_2,X_3\\q=\#\Psi}}
\frac{\delta^{\rm K}_{m,m''+s-q}}{s!}
\frac{\Lambda^2}{2\pi^2}
\vs
&\times
\int_0^\Lambda\frac{dp\,p^2}{2\pi^2}
\int_{-1}^{1}\frac{d\mu}{2}
\Bigg[
{\cal B}_{X_1X_2X_3,s+3}(\Lambda,p,\mu)
s^{\rm 2L,[3]}_{c_{m''}a_m;\,X_1X_2X_3}
(\Lambda/p,\mu)
\vs
&\quad+
{\cal B}_{X_1X_2X_3,s+3}(p,\Lambda,\mu)
s^{\rm 2L,[3]}_{c_{m''}a_m;\,X_1X_2X_3}
(p/\Lambda,\mu)
\Bigg].
\ea
These integrals contain three hierarchical regions: $p_1\ll p_2,p_3$, $p_2\ll p_1,p_3$, and $p_3\ll p_1,p_2$, with the smaller hard momentum still much larger than the external momenta. 
For example, for local PNG the region $p_1=p\ll p_2\simeq p_3\simeq\Lambda$ gives, apart from the common prefactors,
\be
\int_0^\Lambda\frac{dp\,p^2}{2\pi^2}
\frac{2C_{s+3}\Plin(\Lambda)\Plin(p)}
     {M^{\eta_1+1}(p)M^{\eta_2+\eta_3}(\Lambda)}
\left\langle
s^{\rm 2L,[3]}_{c_{m''}a_m;\,X_1X_2X_3}(0,\mu)
\right\rangle_\mu\,,
\ee
where $\eta_i=\delta^{\rm K}_{X_i,\Psi}$, $q=\eta_1+\eta_2+\eta_3$, $C_3=2f_{\rm NL}$, and $C_4=8f_{\rm NL}^2+6g_{\rm NL}$.
The folded region $p_3\ll p_1,p_2$ appears in both shell terms and with $p=p_3$ and $\mu_3=\hat{\vp}_1\cdot\hat{\vp}_3$, their combined leading contribution is
\be
\int_0^\Lambda\frac{dp\,p^2}{2\pi^2}
\frac{2C_{s+3}\Plin(\Lambda)\Plin(p)}
     {M^{\eta_3+1}(p)M^{\eta_1+\eta_2}(\Lambda)}
\left\langle
s^{\rm 2L,[3]}_{c_{m''}a_m;\,X_3X_2X_1}(0,\mu_3)
\right\rangle_{\mu_3}\,.
\ee
For local PNG, the sequential limits \refeq{seq_3top_1} and \refeq{seq_3top_2}, together with the weights \refeq{weights}, identify the three leading hierarchical contributions with $-\left.\frac{{\rm d}b_{a_m}}{{\rm d}\Lambda}\right|_{\rm 1L,1L;\,[3]}$.
Their cancellation therefore gives
\be
\boxed{
\begin{aligned}
&\left.\frac{{\rm d}b_{a_m}}{{\rm d}\Lambda}
\right|_{\rm 2L,[3]}
+
\left.\frac{{\rm d}b_{a_m}}{{\rm d}\Lambda}
\right|_{\rm 1L,1L;\,[3]}
=
-b_{c_{m''}}(\Lambda)
\sum_{s=0}^{1}
\sum_{\substack{X_1,X_2,X_3\\q=\#\Psi}}
\frac{\delta^{\rm K}_{m,m''+s-q}}{s!}
\\
&\quad\times
\frac{\Lambda^2}{2\pi^2}
\int_0^\Lambda\frac{dp\,p^2}{2\pi^2}
\Bigg\{
\int_{-1}^{1}\frac{d\mu}{2}
\Big[
{\cal B}_{X_1X_2X_3,s+3}(\Lambda,p,\mu)
s^{\rm 2L,[3]}_{c_{m''}a_m;\,X_1X_2X_3}(\Lambda/p,\mu)
\\
&\qquad\qquad+
{\cal B}_{X_1X_2X_3,s+3}(p,\Lambda,\mu)
s^{\rm 2L,[3]}_{c_{m''}a_m;\,X_1X_2X_3}(p/\Lambda,\mu)
\Big]
\\
&\qquad-
2C_{s+3}\Plin(\Lambda)\Plin(p)
\Bigg[
\frac{
\left\langle
s^{\rm 2L,[3]}_{c_{m''}a_m;\,X_1X_2X_3}(0,\mu)
\right\rangle_\mu
}{
M^{\eta_1+1}(p)M^{\eta_2+\eta_3}(\Lambda)
}
\\
&\qquad\qquad+
\frac{
\left\langle
s^{\rm 2L,[3]}_{c_{m''}a_m;\,X_1X_2X_3}(\infty,\mu)
\right\rangle_\mu
}{
M^{\eta_2+1}(p)M^{\eta_1+\eta_3}(\Lambda)
}
+
\frac{
\left\langle
s^{\rm 2L,[3]}_{c_{m''}a_m;\,X_3X_2X_1}(0,\mu_3)
\right\rangle_{\mu_3}
}{
M^{\eta_3+1}(p)M^{\eta_1+\eta_2}(\Lambda)
}
\Bigg]
\Bigg\}.
\end{aligned}
}
\ee

\newpage
\section{Discussion and conclusions}
\label{sec:disc}

In this work, we develop the general structure of galaxy bias renormalization in the presence of primordial non-Gaussianity, including both deterministic and stochastic contributions. We focus on scalar tracers, presenting explicit examples at one and two loops.

We show that the ultraviolet sensitivity of biased tracers factorizes into two parts: the soft limits of primordial correlators, discussed in \refsec{soft_limits}, and the soft limits of bias kernels, exemplified by \refeq{biaskernelexpansionsinglehard}, \refeq{doublehard_pairpair} and \refeq{doublehard_threehard}.
The former encode the primordial initial conditions, while the latter encode gravitational evolution and the structure of the bias expansion.
Soft limits thus provide an organizing principle for renormalization in LSS and a direct connection between bias renormalization and analyses of inflationary soft limits and consistency relations \cite{maldacena:2003,Paolo_Creminelli_2004,Assassi_2012,Joyce_2015,Mirbabayi_2016}.

For the one-loop deterministic part, two hard legs are integrated over, and a primordial $n$-point function therefore enters through the $(n-2)$-soft limit given for the local case in \refeq{multi-soft}.
The mixing matrix \refeq{Z1L} extends the standard renormalization sourced by the squeezed bispectrum at leading order in $\fnl$ \cite{Assassi:2015fma,Nikolis:2024kbx} to primordial correlators of arbitrary order.
We explicitly perform the renormalization for PNG of the local type, including a non-vanishing five-point function and bias operators through fifth order.
The non-Gaussian corrections are parametrized by the generalized variances defined in \refeq{sigmatildedef} (see \refapp{tables} for coefficients).
For the mixing $\O_{b_{m'}}\to\O_{a_m}$, an $(s+2)$-point primordial correlator generates $s$ external powers of $\Psi$, while integrating over $q$ hard $\Psi$ legs removes $q$ powers.

The two-loop renormalization of deterministic coefficients in \refsec{renorm_2L_det} is given by the two topologies shown in \refeq{diagslimits2loop}.
The pair-pair topology $[22]$, which is also present for Gaussian initial conditions \cite{Bakx:2025cvu,Bakx:2026rmd}, combines two of the same two-hard primordial responses that appear at one loop.
The pair-pair kernels \refeq{doublehard_pairpair} for the bias operators considered can be described by a single functional dependence via $\gamma(r)$ given in \refeq{gammafunc}, with $r=p_1/p_2$.
This function is just the unsymmetrized version of the function $g(r)$ that appears in the Gaussian calculation ($\gamma(r)+\gamma(1/r)=g(r)$) \cite{Bakx:2025cvu}.
The connected topology $[3]$ instead probes the new three-hard response \refeq{Gdoublehard_general}, corresponding to an $(n-3)$-soft limit of the primordial correlator.
Their contributions, \refeq{I2L_22} and \refeq{I2L_3}, enter the two-loop mixing matrix \refeq{Z2L_general}.
The $[3]$-topology kernel can be described in terms of the function $\tilde{f}(r,\mu)$ given in \refeq{ftilde}. This function is related to the function $f(r,\mu)$ found in the stochastic sector of \cite{Bakx:2025cvu}. We further discuss in \refsec{renorm_higherL_det} the general renormalization structure at arbitrary loop order by organizing each contribution in terms of its connected hard blocks.

The fact that two-loop renormalization reduces to just two general momentum- and angle-dependent functions, one per topology, allows us to impose strong consistency conditions from sequential loop limits [see \refeq{commutationoflimitsrelation}, \refeq{seq_3top_1} and \refeq{seq_3top_2}], and in some cases allows us to bypass the two-loop calculation completely. It would be interesting to understand whether these functions are related to fundamental symmetries, since that could simplify the computation of higher-loop coefficients.

We also consider the renormalization of stochastic terms in the presence of PNG in \refsec{stoch}, extending the formalism for the renormalization of stochastic operators presented in \cite{Rubira:2024tea,Bakx:2025cvu,Bakx:2026rmd}.
We provide the general renormalization structure for products of two and three operators at one loop, with the parametric dependence of a product of $J$ operators given in \refeq{stoch_SigmaJ_schematic}.
We further discuss the collapsed configuration that appears at two loops.

Finally, we extend the $\fnl$ RG formulation of Paper~I \cite{Nikolis:2024kbx}, providing one- and two-loop RG equations for the full non-Gaussian bias basis.
These equations generalize the Gaussian RG results of \cite{Rubira:2023vzw,Bakx:2025cvu,Bakx:2026rmd}, including the dependence of the bias running on $\fnl$, $g_{\rm NL}$, and higher primordial couplings $\lambda_{\rm NL}^{(i)}$. At two loops, we find the leading hierarchical contributions of the $[22]$ and $[3]$ topologies cancel against the corresponding weighted one-loop-squared terms and the matching is determined by the origin of the $\Psi$ legs integrated over in the second contraction.

Our paper provides a general framework for renormalizing bias operators in the presence of PNG arising from non-vanishing primordial $n$-point functions.
Although we provide explicit examples for single-source local PNG, our results in terms of soft limits of primordial $n$-point functions are directly applicable to more general inflationary models.
Restricting our analysis to a non-zero bispectrum, we expect our two-loop results to extend to models involving the exchange of a scalar field coupled to the curvature perturbations, such as quasi-single-field inflation \cite{Chen_2010}, by considering $\Psi=k^\D\varphi$ as in \refeq{Psi-def} and appropriately modifying the generalized variances (cf.~\cite{Assassi:2015fma,Nikolis:2024kbx}).
For the connected topology, this extension also requires the model-specific bispectrum in its full kinematic range.
For higher-order PNG, we expect the RG coefficients and the momentum structure to differ in models of inflation where PNG is sourced by the coupling between multiple fields, in models with higher-spin PNG (as considered at one loop in \cite{Assassi:2015fma,Nikolis:2024kbx}), or in models with other interesting features such as parity violation (see \cite{Kurita:2025hmp} for galaxy shapes and \cite{Azyzy:2025zif} for the matter trispectrum).

\acknowledgments

We thank Mathias Garny and Benjamin Wallisch for useful discussions and Fabian Schmidt for his participation in the early stages of this project and comments in the draft. We are particularly indebted to Denis Werth for his support. CN is supported by the Swedish Research Council (VR) under grants 2019-02337 and 2024-04289 and is further supported by a scholarship award from the ``A.G. Leventis foundation''. CN thanks the Max Planck Institute for Astrophysics and the University Observatory of Munich (USM) for hospitality during the completion of this work.

\textit{AI usage statement}: We used GPT Sol-6.1 and Claude Opus-5.5 for optimizing code initially written by the authors, for assistance with scripts performing numerical computations, improve sentence formulations and to perform grammar checks in the paper. The authors have verified all results and take full responsibility for the content of this paper.

\appendix

\section{Displaced PNG operators}
\label{app:disp}

In this appendix, we give the configuration-space expansion of the displaced PNG field used in \refsec{kernels}. The Lagrangian and Eulerian positions are related by $\vx=\vq+\vdisp(\vq)$, so that
\be
  \Psi(\vq) = \Psi(\vx-\vdisp(\vq))\,,
\ee
expanding the displacement perturbatively as
\be
  \vdisp(\vq)=\vdisp^{(1)}(\vq)+\vdisp^{(2)}(\vq)+\vdisp^{(3)}(\vq)+\cdots\,.
\ee
Because the displacement is itself evaluated at the Lagrangian position, it must likewise be Taylor-expanded about $\vx$:
\be
  \vdisp(\vq)=\vdisp(\vx-\vdisp(\vq))
  =\vdisp(\vx)-s^j(\vq)\partial_j\vdisp(\vx)+\cdots\,.
\ee
These relations determine the perturbative expansion of the displaced field $\Psi(\vq)$.

As discussed in \refsec{kernels}, the Fourier transform of each contribution $\Psi^{(n)}(\vq)$ defines the displacement kernel $\mathcal{S}^{(n)}$ through \refeq{SPsi_def}. The expressions below give the corresponding configuration-space contributions.
Through fourth order,
\bea
\Psi^{(1)}(\vq)
&=& \Psi^{(1)}(\vx)\,,
\vs
\Psi^{(2)}(\vq)
&=& -s^{(1)}_i\partial^i\Psi^{(1)}\,,
\vs
\Psi^{(3)}(\vq)
&=& -s^{(2)}_i\partial^i\Psi^{(1)}
+\frac{1}{2}s^{(1)}_is^{(1)}_j\partial^i\partial^j\Psi^{(1)}
+s^{(1)}_j\partial^j s^{(1)}_i\,\partial^i\Psi^{(1)}\,,
\vs
\Psi^{(4)}(\vq)
&=& -s^{(3)}_i\partial^i\Psi^{(1)}
+s^{(1)}_j\partial^j s^{(2)}_i\,\partial^i\Psi^{(1)}
+s^{(2)}_j\partial^j s^{(1)}_i\,\partial^i\Psi^{(1)}
-\frac{1}{2}s^{(1)}_j s^{(1)}_k
  \partial^j\partial^k s^{(1)}_i\,\partial^i\Psi^{(1)}
\vs
&&
+s^{(2)}_i s^{(1)}_j\partial^i\partial^j\Psi^{(1)}
-s^{(1)}_k\partial^k s^{(1)}_j
  \partial^j s^{(1)}_i\,\partial^i\Psi^{(1)}
\\
&&
-s^{(1)}_k\partial^k s^{(1)}_i
  s^{(1)}_j\partial^i\partial^j\Psi^{(1)}
-\frac{1}{6}s^{(1)}_i s^{(1)}_j s^{(1)}_k
  \partial^i\partial^j\partial^k\Psi^{(1)}\,,
\nonumber
\eea
and the fifth-order contribution is
\bea
\Psi^{(5)}(\vq)
&=&
-s^{(4)}_i\partial^i\Psi^{(1)}
+s^{(1)}_j\partial^j s^{(3)}_i\,\partial^i\Psi^{(1)}
+s^{(3)}_j\partial^j s^{(1)}_i\,\partial^i\Psi^{(1)}
+s^{(2)}_j\partial^j s^{(2)}_i\,\partial^i\Psi^{(1)}
\vs
&&
-s^{(1)}_k\partial^k s^{(1)}_j
 \partial^j s^{(2)}_i\,\partial^i\Psi^{(1)}
-s^{(1)}_k\partial^k s^{(2)}_j
 \partial^j s^{(1)}_i\,\partial^i\Psi^{(1)}
-s^{(2)}_k\partial^k s^{(1)}_j
 \partial^j s^{(1)}_i\,\partial^i\Psi^{(1)}
\vs
&&
+s^{(1)}_l\partial^l s^{(1)}_k
 \partial^k s^{(1)}_j
 \partial^j s^{(1)}_i\,\partial^i\Psi^{(1)}
+\frac{1}{2}s^{(1)}_ks^{(1)}_l
 \partial^k\partial^l s^{(1)}_j
 \partial^j s^{(1)}_i\,\partial^i\Psi^{(1)}
\vs
&&
-\frac{1}{2}s^{(1)}_js^{(1)}_k
 \partial^j\partial^k s^{(2)}_i\,\partial^i\Psi^{(1)}
-s^{(1)}_js^{(2)}_k
 \partial^j\partial^k s^{(1)}_i\,\partial^i\Psi^{(1)}
\vs
&&
+s^{(1)}_js^{(1)}_l
 \partial^l s^{(1)}_k
 \partial^j\partial^k s^{(1)}_i\,\partial^i\Psi^{(1)}
+\frac{1}{6}s^{(1)}_js^{(1)}_ks^{(1)}_l
 \partial^j\partial^k\partial^l s^{(1)}_i\,
 \partial^i\Psi^{(1)}
\vs
&&
+s^{(1)}_is^{(3)}_j
 \partial^i\partial^j\Psi^{(1)}
-s^{(1)}_is^{(1)}_k
 \partial^k s^{(2)}_j\,
 \partial^i\partial^j\Psi^{(1)}
-s^{(1)}_is^{(2)}_k
 \partial^k s^{(1)}_j\,
 \partial^i\partial^j\Psi^{(1)}
\vs
&&
+s^{(1)}_is^{(1)}_l
 \partial^l s^{(1)}_k
 \partial^k s^{(1)}_j\,
 \partial^i\partial^j\Psi^{(1)}
+\frac{1}{2}s^{(1)}_is^{(1)}_ks^{(1)}_l
 \partial^k\partial^l s^{(1)}_j\,
 \partial^i\partial^j\Psi^{(1)}
\vs
&&
+\frac{1}{2}s^{(2)}_is^{(2)}_j
 \partial^i\partial^j\Psi^{(1)}
-s^{(2)}_is^{(1)}_k
 \partial^k s^{(1)}_j\,
 \partial^i\partial^j\Psi^{(1)}
\vs
&&
+\frac{1}{2}
 s^{(1)}_k\partial^k s^{(1)}_i\,
 s^{(1)}_l\partial^l s^{(1)}_j\,
 \partial^i\partial^j\Psi^{(1)}
\vs
&&
-\frac{1}{2}s^{(1)}_is^{(1)}_js^{(2)}_k
 \partial^i\partial^j\partial^k\Psi^{(1)}
+\frac{1}{2}s^{(1)}_is^{(1)}_js^{(1)}_l
 \partial^l s^{(1)}_k\,
 \partial^i\partial^j\partial^k\Psi^{(1)}
\vs
&&
+\frac{1}{24}s^{(1)}_is^{(1)}_js^{(1)}_ks^{(1)}_l
 \partial^i\partial^j\partial^k\partial^l\Psi^{(1)}\,,
\eea
where all fields on the right-hand sides above are evaluated at $\vx$. The superscript on $\Psi^{(n)}(\vq)$ labels the contribution of total perturbative order $n$ in the expansion about $\vx$, with the primordial field $\Psi$ counted as first order.

At first order, the displacement is given by
\be
  \vdisp^{(1)}=-\frac{\nabla}{\nabla^2}\delta^{(1)}\,,
\ee
while at second order \cite{MSZ}
\be
\vdisp^{(2)}
=\frac{3}{14}\frac{\nabla}{\nabla^2}
\left(
{\rm Tr}\left[M^{(1)}M^{(1)}\right]
-{\rm Tr}\left[M^{(1)}\right]^2
\right)\;.
\ee
Higher-order displacements follow from the recursion relations of \cite{MSZ} (see also \cite{Baldauf:2020bsd} for $\vdisp_L^{(3)}$), so the longitudinal and transverse components of the third-order displacement are
\bea
&&
\vdisp_L^{(3)}
=\frac{5}{9}\frac{\nabla}{\nabla^2}
\left(
{\rm Tr}\left[M^{(1)}M^{(2)}\right]
-{\rm Tr}\left[M^{(1)}\right]{\rm Tr}\left[M^{(2)}\right]
\right)
-\frac{1}{18}\frac{\nabla}{\nabla^2}
\epsilon_{ijk}\epsilon_{lmn}
M^{(1)}_{il}M^{(1)}_{jm}M^{(1)}_{kn}\;,
\vs
&&
s_{T,i}^{(3)}
=-\frac{1}{3\nabla^2}\epsilon_{iab}\partial_a
\left(
\epsilon_{bjk}M_{jl}^{(1)}M_{kl}^{(2)}
\right)\;.
\eea
The longitudinal part of the fourth-order displacement is
\ba
\vdisp_L^{(4)}
=&\frac{\nabla}{\nabla^2}
\left[
\frac{7}{11}
\left(
\mathrm{Tr}\left[M^{(1)}M^{(3)}\right]
-\mathrm{Tr}\left[M^{(1)}\right]\mathrm{Tr}\left[M^{(3)}\right]
\right)
\right.
\\
&\left.
+\frac{17}{66}
\left(
\mathrm{Tr}\left[M^{(2)}M^{(2)}\right]
-\mathrm{Tr}\left[M^{(2)}\right]^2
\right)
\right]
-\frac{\nabla}{\nabla^2}
\left[
\frac{13}{66}\epsilon_{ijk}\epsilon_{lmn}
M^{(1)}_{il}M^{(1)}_{jm}M^{(2)}_{kn}
\right]\;,
\nonumber
\ea
and its transverse part is
\bea
&&
s_{T,i}^{(4)}
=-\frac{1}{2\nabla^2}\epsilon_{iab}\partial_a
\left(
\epsilon_{bjk}M_{jl}^{(1)}M_{kl}^{(3)}
\right)\;.
\eea
At each order, the total displacement is $\vdisp^{(n)}=\vdisp_L^{(n)}+\vdisp_T^{(n)}$, where $\nabla\times \vdisp_L^{(n)}=0$ and $\nabla\cdot \vdisp_T^{(n)}=0$. Note that the transverse component first appears at third order.
The deformation tensors are defined by $M_{ij}^{(n)}=\partial_i s_j^{(n)}$, with derivatives taken with respect to the Lagrangian argument of the displacement. Their explicit expressions through third order are
\ba
M_{ij}^{(1)}
=&-\frac{\partial_i\partial_j}{\nabla^2}\delta^{(1)}
\\
M_{ij}^{(2)}
=&\frac{3}{14}\frac{\partial_i\partial_j}{\nabla^2}
\left(
\mathrm{Tr}\left[M^{(1)}M^{(1)}\right]
-\mathrm{Tr}\left[M^{(1)}\right]^2
\right)
\vs
M_{ij}^{(3)}
=&
\frac{5}{9}\frac{\partial_i\partial_j}{\nabla^2}
\left(
\mathrm{Tr}\left[M^{(1)}M^{(2)}\right]
-\mathrm{Tr}\left[M^{(1)}\right]\mathrm{Tr}\left[M^{(2)}\right]
\right)
-\frac{1}{18}\frac{\partial_i\partial_j}{\nabla^2}
\epsilon^{abk}\epsilon^{lmn}
M^{(1)}_{al}M^{(1)}_{bm}M^{(1)}_{kn}
\vs
-&\frac{\partial_i\partial_k}{3\nabla^2}
\left(
M_{ja}^{(1)}M_{ka}^{(2)}
-M_{ja}^{(2)}M_{ka}^{(1)}
\right)\;.
\nonumber
\ea

\section{Tables} \label{app:tables}

This appendix collects the coefficients describing the projection of bias operators onto the lower-order basis under one- and two-loop renormalization, as discussed in Sec.~\ref{sec:renorm}. 

At one loop, the $s$-soft limit of a primordial $(s+2)$-point function generates $s$ additional $\Psi$ fields (with amplitude $\lambda_{\rm NL}^{(s+2)}$) and closure under renormalization therefore requires operators containing powers of $\Psi$, together with their products with the Gaussian bias operators. We denote operators containing $m$ factors of $\Psi$ by $\O_{a_m}$ and organize the one-loop truncation by keeping contributions $b_{a_m}\lambda_{\rm NL}^{(s+2)}$ with $m+s\leq S$, where $S$ is the maximum PNG order considered. In this counting, the bispectrum has PNG order 1, the trispectrum has PNG order 2, and so forth. For a source operator $\O_{a_m}$, this prescription keeps two-hard responses with at most $S-m$ soft legs, corresponding to primordial correlators of order at most $S+2-m$. For $m=S$, only two-point contractions are kept.
For our calculations, we choose $S=3$ at one loop and $S=2$ at two loops.

At one loop, we truncate at $b_{a_3}$, $b_{a_2}\lambda_{\rm NL}^{(3)}$, $b_{a_1}\lambda_{\rm NL}^{(4)}$, and $b_{a_0}\lambda_{\rm NL}^{(5)}$. All lower-order contributions are also of course retained, and therefore the basis includes operators containing up to three $\Psi$ fields. One-loop single-hard projection coefficients are given in \reftab{singlehardcoeff_a0b_0} ($b_0\to a_{0}$),  \reftab{singlehardcoeff_a0b_q0} ($b_0\to a_{1,2,3}$), \reftab{singlehardcoeff_a1b0_q1} ($b_1\to a_0$),
\reftab{singlehardcoeff_a1b1} ($b_1\to a_1$), \reftab{singlehardcoeff_a1b23} ($b_1\to a_{2,3}$),
\reftab{singlehardcoeff_a2b0_q2} ($b_2\to a_0$), \reftab{singlehardcoeff_a2b1} ($b_2\to a_1$),
\reftab{singlehardcoeff_a2b23} ($b_2\to a_{2,3}$), and \reftab{singlehardcoeff_a3b} ($b_3\to a_{1,2,3}$).

At two loops, we count a primordial $n$-point function as PNG order $n-2$.
For the $[22]$ topology, the truncation gives $m+s_1+s_2\leq2$, where $s_1$ and $s_2$ count the soft legs attached to the two hard pairs. For the connected $[3]$ topology, the relevant primordial correlator has $s+3$ legs, so the corresponding condition is $m+s+1\leq2$. This truncation therefore requires operators containing up to two $\Psi$ fields. Two-loop coefficients for the $[22]$ and connected $[3]$ topologies are given in \reftab{twoloopcoeff_Gauss}, \reftab{twoloopcoeff_a_000} and \reftab{twoloopcoeff_aPsi} for source operators containing zero and one $\Psi$ fields; \reftab{twoloopcoeff_aPsi2} gives the $[22]$ coefficients for sources containing two $\Psi$ fields.

Independently of the PNG truncation, we keep bias operators through fifth order in the linear fields. To extend the Gaussian calculations of \cite{Bakx:2025cvu}, we evaluate their projections using up to three external linear density fields (after taking two modes to be hard) at one loop and one external linear density field at two loops (after taking up to four modes to be hard).

\begin{table}[t]
\centering
\renewcommand{\arraystretch}{1.4}
\begin{adjustbox}{max width=\textwidth}
$\begin{array}{c|ccccccc}
 & \multicolumn{7}{c}{s^{\rm 1L}_{b_0a_0;0}} \\
 b_0 \,\backslash\, a_0
 & {\rm tr}\big[\Pi^{[1]}\big] & {\rm tr}\big[\big(\Pi^{[1]}\big)^2\big] & \Big({\rm tr}\big[\Pi^{[1]}\big]\Big)^2 & \Big({\rm tr}\big[\Pi^{[1]}\big]\Big)^3 & {\rm tr}\big[\big(\Pi^{[1]}\big)^2\big]{\rm tr}\big[\Pi^{[1]}\big] & {\rm tr}\big[\big(\Pi^{[1]}\big)^3\big] & {\rm tr}\big[\Pi^{[1]}\Pi^{[2]}\big] \\
\hline
{\rm tr}\big[\Pi^{[1]}\big] & 0 & 0 & 0 & 0 & 0 & 0 & 0 \\
\hline
{\rm tr}\big[\big(\Pi^{[1]}\big)^2\big] & \frac{68}{21} & \frac{254}{2205} & \frac{2624}{735} & \frac{141664}{101871} & \frac{5856}{18865} & \frac{457616}{2546775} & -\frac{2644}{13475} \\
\Big({\rm tr}\big[\Pi^{[1]}\big]\Big)^2 & \frac{68}{21} & \frac{254}{2205} & \frac{2624}{735} & \frac{141664}{101871} & \frac{5856}{18865} & \frac{457616}{2546775} & -\frac{2644}{13475} \\
\hline
\Big({\rm tr}\big[\Pi^{[1]}\big]\Big)^3 & 3 & 0 & \frac{68}{7} & \frac{2624}{245} & \frac{254}{735} & 0 & 0 \\
{\rm tr}\big[\big(\Pi^{[1]}\big)^2\big]{\rm tr}\big[\Pi^{[1]}\big] & \frac{5}{3} & \frac{116}{105} & \frac{176}{35} & \frac{1754}{343} & \frac{46258}{15435} & \frac{1784}{1715} & -\frac{50}{63} \\
{\rm tr}\big[\big(\Pi^{[1]}\big)^3\big] & 1 & \frac{58}{35} & \frac{94}{35} & \frac{3971}{1715} & \frac{4448}{1029} & \frac{2676}{1715} & -\frac{25}{21} \\
{\rm tr}\big[\Pi^{[1]}\Pi^{[2]}\big] & \frac{41}{21} & \frac{76}{63} & \frac{268}{49} & \frac{2489609}{509355} & \frac{61232}{18865} & \frac{3262052}{2546775} & -\frac{19231}{17325} \\
\hline
\Big({\rm tr}\big[\Pi^{[1]}\big]\Big)^4 & 0 & 0 & 6 & \frac{136}{7} & 0 & 0 & 0 \\
{\rm tr}\big[\big(\Pi^{[1]}\big)^3\big]{\rm tr}\big[\Pi^{[1]}\big] & 0 & 1 & 1 & \frac{94}{35} & \frac{152}{35} & \frac{58}{35} & 0 \\
{\rm tr}\big[\big(\Pi^{[1]}\big)^2\big]\Big({\rm tr}\big[\Pi^{[1]}\big]\Big)^2 & 0 & 1 & \frac{7}{3} & \frac{716}{105} & \frac{572}{105} & 0 & 0 \\
\Big({\rm tr}\big[\big(\Pi^{[1]}\big)^2\big]\Big)^2 & 0 & \frac{38}{15} & \frac{4}{15} & \frac{32}{49} & \frac{6184}{735} & \frac{928}{735} & 0 \\
{\rm tr}\big[\Pi^{[1]}\Pi^{[1]}\Pi^{[2]}\big] & 0 & \frac{103}{105} & \frac{17}{15} & \frac{128668}{46305} & \frac{21724}{5145} & \frac{80018}{46305} & \frac{22}{105} \\
{\rm tr}\big[\Pi^{[1]}\big]{\rm tr}\big[\Pi^{[1]}\Pi^{[2]}\big] & 0 & \frac{4}{7} & \frac{18}{7} & \frac{324706}{46305} & \frac{10832}{3087} & \frac{207472}{231525} & \frac{394}{1575} \\
{\rm tr}\big[\Pi^{[1]}\Pi^{[3]}\big] & 0 & \frac{268}{315} & \frac{34}{15} & \frac{940493}{169785} & \frac{152275}{33957} & \frac{1211306}{848925} & -\frac{6614}{17325} \\
{\rm tr}\big[\Pi^{[2]}\Pi^{[2]}\big] & 0 & \frac{466}{735} & \frac{2168}{735} & \frac{348056}{46305} & \frac{10120}{3087} & \frac{9644}{9261} & \frac{1516}{2205} \\
\hline
\Big({\rm tr}\big[\Pi^{[1]}\big]\Big)^5 & 0 & 0 & 0 & 10 & 0 & 0 & 0 \\
{\rm tr}\big[\big(\Pi^{[1]}\big)^3\big]\Big({\rm tr}\big[\Pi^{[1]}\big]\Big)^2 & 0 & 0 & 0 & 1 & 2 & 1 & 0 \\
{\rm tr}\big[\big(\Pi^{[1]}\big)^2\big]\Big({\rm tr}\big[\Pi^{[1]}\big]\Big)^3 & 0 & 0 & 0 & 3 & 3 & 0 & 0 \\
{\rm tr}\big[\big(\Pi^{[1]}\big)^3\big]{\rm tr}\big[\big(\Pi^{[1]}\big)^2\big] & 0 & 0 & 0 & 0 & \frac{7}{5} & \frac{9}{5} & 0 \\
{\rm tr}\big[\Pi^{[1]}\big]\Big({\rm tr}\big[\big(\Pi^{[1]}\big)^2\big]\Big)^2 & 0 & 0 & 0 & \frac{4}{15} & \frac{58}{15} & 0 & 0 \\
\Big({\rm tr}\big[\Pi^{[1]}\big]\Big)^2{\rm tr}\big[\Pi^{[1]}\Pi^{[2]}\big] & 0 & 0 & 0 & \frac{67}{21} & \frac{8}{7} & 0 & 1 \\
{\rm tr}\big[\big(\Pi^{[1]}\big)^2\big]{\rm tr}\big[\Pi^{[1]}\Pi^{[2]}\big] & 0 & 0 & 0 & \frac{38}{147} & \frac{1819}{735} & \frac{232}{735} & \frac{19}{15} \\
{\rm tr}\big[\Pi^{[1]}\big]{\rm tr}\big[\Pi^{[1]}\Pi^{[1]}\Pi^{[2]}\big] & 0 & 0 & 0 & \frac{17}{15} & \frac{143}{105} & \frac{10}{21} & \frac{2}{3} \\
{\rm tr}\big[\Pi^{[1]}\Pi^{[2]}\Pi^{[2]}\big] & 0 & 0 & 0 & \frac{422}{343} & \frac{1037}{1029} & \frac{1493}{5145} & \frac{128}{105} \\
{\rm tr}\big[\Pi^{[1]}\big]{\rm tr}\big[\Pi^{[2]}\Pi^{[2]}\big] & 0 & 0 & 0 & \frac{856}{245} & \frac{626}{735} & 0 & \frac{20}{21} \\
{\rm tr}\big[\Pi^{[1]}\Pi^{[1]}\Pi^{[3]}\big] & 0 & 0 & 0 & \frac{6407}{6615} & \frac{1189}{735} & \frac{5371}{6615} & \frac{1}{5} \\
{\rm tr}\big[\Pi^{[1]}\big]{\rm tr}\big[\Pi^{[1]}\Pi^{[3]}\big] & 0 & 0 & 0 & \frac{2579}{945} & \frac{100}{63} & \frac{272}{675} & \frac{41}{225} \\
{\rm tr}\big[\Pi^{[2]}\Pi^{[3]}\big] & 0 & 0 & 0 & \frac{142094}{46305} & \frac{19994}{15435} & \frac{2411}{9261} & \frac{229}{315} \\
{\rm tr}\big[\Pi^{[1]}\Pi^{[4]}\big] & 0 & 0 & 0 & \frac{523888}{509355} & \frac{98489}{113190} & \frac{1332383}{5093550} & \frac{746}{5775} \\
\hline
\end{array}$
\end{adjustbox}
\caption{Gaussian one-loop coefficients $s^{\rm 1L}_{b_0a_0;0}=s^{\rm 1L}_{ba}$, adapted from \cite{Bakx:2025cvu}. Rows label the renormalized operators $\O_{b_0}$, and columns the operators $\O_{a_0}$ onto which they project.}
\label{tab:singlehardcoeff_a0b_0}
\end{table}

\begin{table}[t]
\centering
\renewcommand{\arraystretch}{1.6}
\resizebox{\textwidth}{!}{%
$\begin{array}{c|cccc|cc|c}
 & \multicolumn{4}{c|}{s^{\rm 1L}_{b_0a_1;0}} & \multicolumn{2}{c|}{s^{\rm 1L}_{b_0a_2;0}} & s^{\rm 1L}_{b_0a_3;0} \\
 b_0 \,\backslash\, a_i
 & \Psi & {\rm tr}\big[\Pi^{[1]}\big]\Psi & \Big({\rm tr}\big[\Pi^{[1]}\big]\Big)^2\Psi & {\rm tr}\big[\big(\Pi^{[1]}\big)^2\big]\Psi
 & \Psi^2 & {\rm tr}\big[\Pi^{[1]}\big]\Psi^2
 & \Psi^3 \\
\hline
 {\rm tr}\big[ \Pi^{[1]} \big] & 0 & 0 & 0 & 0 & 0 & 0 & 0 \\
\hline
 {\rm tr}\big[ \big( \Pi^{[1]} \big)^2 \big] & 1 & \frac{68}{21} & \frac{2624}{735} & \frac{254}{2205} & 1 & \frac{68}{21} & 1 \\
 \Big( {\rm tr}\big[ \Pi^{[1]} \big] \Big)^2 & 1 & \frac{68}{21} & \frac{2624}{735} & \frac{254}{2205} & 1 & \frac{68}{21} & 1 \\
\hline
 \Big( {\rm tr}\big[ \Pi^{[1]} \big] \Big)^3 & 0 & 3 & \frac{68}{7} & 0 & 0 & 3 & - \\
 {\rm tr}\big[ \big( \Pi^{[1]} \big)^2 \big] {\rm tr}\big[ \Pi^{[1]} \big] & 0 & \frac{5}{3} & \frac{176}{35} & \frac{116}{105} & 0 & \frac{5}{3} & - \\
 {\rm tr}\big[ \big( \Pi^{[1]} \big)^3 \big] & 0 & 1 & \frac{94}{35} & \frac{58}{35} & 0 & 1 & - \\
 {\rm tr}\big[\Pi^{[1]} \Pi^{[2]} \big] & 0 & \frac{41}{21} & \frac{268}{49} & \frac{76}{63} & 0 & \frac{41}{21} & - \\
\hline
 {\Big( {\rm tr}\big[ \Pi^{[1]} \big] \Big)^4} & 0 & 0 & 6 & 0 & - & - & - \\
 {{\rm tr}\big[ \big( \Pi^{[1]} \big)^3 \big] {\rm tr}\big[ \Pi^{[1]} \big]} & 0 & 0 & 1 & 1 & - & - & - \\
 {{\rm tr}\big[ \big( \Pi^{[1]} \big)^2 \big] \big({\rm tr}\big[ \Pi^{[1]} \big]\big)^2} & 0 & 0 & \frac{7}{3} & 1 & - & - & - \\
 {\Big( {\rm tr}\big[ \big( \Pi^{[1]} \big)^2 \big] \Big)^2} & 0 & 0 & \frac{4}{15} & \frac{38}{15} & - & - & - \\
 {\rm tr}\big[\Pi^{[1]} \Pi^{[1]} \Pi^{[2]} \big] & 0 & 0 & \frac{17}{15} & \frac{103}{105} & - & - & - \\
 {\rm tr}\big[ \Pi^{[1]} \big] {\rm tr}\big[\Pi^{[1]} \Pi^{[2]} \big] & 0 & 0 & \frac{18}{7} & \frac{4}{7} & - & - & - \\
 {\rm tr}\big[\Pi^{[1]} \Pi^{[3]} \big] & 0 & 0 & \frac{34}{15} & \frac{268}{315} & - & - & - \\
 {\rm tr}\big[\Pi^{[2]} \Pi^{[2]} \big] & 0 & 0 & \frac{2168}{735} & \frac{466}{735} & - & - & - \\
\hline
\end{array}$%
}
\caption{One-loop coefficients $s^{\rm 1L}_{b_0a_i;0}$ for $i=1,2,3$. Rows label the operators $b_0$, columns the operators $a_i$ and a dash indicates that the entry is not available at the truncation order considered.}
\label{tab:singlehardcoeff_a0b_q0}
\end{table}

\begin{table}[t]
\centering
\renewcommand{\arraystretch}{1.6}
\resizebox{\textwidth}{!}{%
$\begin{array}{c|ccccccc}
 & \multicolumn{7}{c}{s^{\rm 1L}_{b_1a_0;1}} \\
 b_1 \,\backslash\, a_0
 & {\rm tr}\big[ \Pi^{[1]} \big] & {\rm tr}\big[ \big( \Pi^{[1]} \big)^2 \big] & \Big( {\rm tr}\big[ \Pi^{[1]} \big] \Big)^2 & \Big( {\rm tr}\big[ \Pi^{[1]} \big] \Big)^3 & {\rm tr}\big[ \big( \Pi^{[1]} \big)^2 \big] {\rm tr}\big[ \Pi^{[1]} \big] & {\rm tr}\big[ \big( \Pi^{[1]} \big)^3 \big] & {\rm tr}\big[\Pi^{[1]} \Pi^{[2]} \big] \\
\hline
 \Psi & -\frac{13}{21} & -\frac{79}{2205} & \frac{101}{735} & -\frac{33301}{509355} & -\frac{1327}{169785} & -\frac{4184}{72765} & \frac{314}{4851} \\
\hline
 {\rm tr}\big[ \Pi^{[1]} \big]\Psi & \frac{13}{21} & \frac{79}{2205} & -\frac{101}{735} & \frac{33301}{509355} & \frac{1327}{169785} & \frac{4184}{72765} & -\frac{314}{4851} \\
\hline
 \Big( {\rm tr}\big[ \Pi^{[1]} \big] \Big)^2\Psi & 2 & 0 & \frac{47}{21} & \frac{253}{735} & \frac{158}{2205} & 0 & 0 \\
 {\rm tr}\big[ \big( \Pi^{[1]} \big)^2 \big] \Psi & \frac{2}{3} & -\frac{1}{21} & \frac{16}{21} & \frac{62}{735} & \frac{913}{2205} & \frac{56}{75} & -\frac{1066}{1575} \\
\hline
 {\Big( {\rm tr}\big[ \Pi^{[1]} \big] \Big)^3\Psi} & 0 & 0 & 3 & \frac{27}{7} & 0 & 0 & 0 \\
 {{\rm tr}\big[ \big( \Pi^{[1]} \big)^3 \big] \Psi} & 0 & 1 & 0 & 0 & \frac{8}{7} & \frac{3}{7} & 0 \\
 {{\rm tr}\big[ \big( \Pi^{[1]} \big)^2 \big] {\rm tr}\big[ \Pi^{[1]} \big]\Psi} & 0 & 1 & \frac{2}{3} & \frac{16}{21} & \frac{11}{7} & 0 & 0 \\
 {\rm tr}\big[\Pi^{[1]} \Pi^{[2]} \big]\Psi & 0 & \frac{4}{7} & \frac{13}{21} & \frac{3868}{6615} & \frac{2948}{2205} & \frac{544}{675} & -\frac{1301}{1575} \\
\hline
 {\Big( {\rm tr}\big[ \Pi^{[1]} \big] \Big)^4\Psi} & 0 & 0 & 0 & 4 & 0 & 0 & 0 \\
 {{\rm tr}\big[ \big( \Pi^{[1]} \big)^3 \big] {\rm tr}\big[ \Pi^{[1]} \big]\Psi} & 0 & 0 & 0 & 0 & 1 & 1 & 0 \\
 {{\rm tr}\big[ \big( \Pi^{[1]} \big)^2 \big] \Big({\rm tr}\big[ \Pi^{[1]} \big]\Big)^2\Psi} & 0 & 0 & 0 & \frac{2}{3} & 2 & 0 & 0 \\
 {\Big( {\rm tr}\big[ \big( \Pi^{[1]} \big)^2 \big] \Big)^2\Psi} & 0 & 0 & 0 & 0 & \frac{4}{3} & 0 & 0 \\
 {{\rm tr}\big[ \Pi^{[1]}\big] {\rm tr}\big[\Pi^{[1]} \Pi^{[2]} \big]\Psi} & 0 & 0 & 0 & \frac{13}{21} & \frac{4}{7} & 0 & 1 \\
 {{\rm tr}\big[\Pi^{[1]} \Pi^{[1]} \Pi^{[2]}\big]\Psi} & 0 & 0 & 0 & 0 & \frac{8}{21} & \frac{10}{21} & \frac{2}{3} \\
 {\rm tr}\big[\Pi^{[2]} \Pi^{[2]} \big]\Psi & 0 & 0 & 0 & \frac{80}{147} & \frac{32}{147} & 0 & \frac{20}{21} \\
 {\rm tr}\big[\Pi^{[1]} \Pi^{[3]} \big]\Psi & 0 & 0 & 0 & \frac{437}{945} & \frac{232}{315} & \frac{272}{675} & \frac{41}{225} \\
\hline
\end{array}$%
}
\caption{One-loop coefficients $s^{\rm 1L}_{b_1a_0;1}$. }
\label{tab:singlehardcoeff_a1b0_q1}
\end{table}
 
\begin{table}[t]
\centering
\renewcommand{\arraystretch}{1.6}
\resizebox{\textwidth}{!}{%
$\begin{array}{c|cccc|cccc}
 & \multicolumn{4}{c|}{s^{\rm 1L}_{b_1a_1;0}} & \multicolumn{4}{c}{s^{\rm 1L}_{b_1a_1;1}} \\
 b_1 \,\backslash\, a_1
 & \Psi & {\rm tr}\big[\Pi^{[1]}\big]\Psi & \Big({\rm tr}\big[\Pi^{[1]}\big]\Big)^2\Psi & {\rm tr}\big[\big(\Pi^{[1]}\big)^2\big]\Psi
 & \Psi & {\rm tr}\big[\Pi^{[1]}\big]\Psi & \Big({\rm tr}\big[\Pi^{[1]}\big]\Big)^2\Psi & {\rm tr}\big[\big(\Pi^{[1]}\big)^2\big]\Psi \\
\hline
\Psi
& 0 & 0 & 0 & 0 & -1 & -\frac{13}{21} & \frac{101}{735} & -\frac{79}{2205} \\
\hline
{\rm tr}\big[\Pi^{[1]}\big]\Psi
& 0 & 0 & 0 & 0 & 1 & \frac{13}{21} & -\frac{101}{735} & \frac{79}{2205} \\
\hline
\Big({\rm tr}\big[\Pi^{[1]}\big]\Big)^2\Psi
& 1 & \frac{68}{21} & \frac{2624}{735} & \frac{254}{2205} & 0 & 2 & \frac{47}{21} & 0 \\
{\rm tr}\big[\big(\Pi^{[1]}\big)^2\big]\Psi
& 1 & \frac{68}{21} & \frac{2624}{735} & \frac{254}{2205} & 0 & \frac{2}{3} & \frac{16}{21} & -\frac{1}{21} \\
\hline
{\Big({\rm tr}\big[\Pi^{[1]}\big]\Big)^3\Psi}
& 0 & 3 & \frac{68}{7} & 0 & 0 & 0 & 3 & 0 \\
{{\rm tr}\big[\big(\Pi^{[1]}\big)^3\big]\Psi}
& 0 & 1 & \frac{94}{35} & \frac{58}{35} & 0 & 0 & 0 & 1 \\
{{\rm tr}\big[\big(\Pi^{[1]}\big)^2\big]{\rm tr}\big[\Pi^{[1]}\big]\Psi}
& 0 & \frac{5}{3} & \frac{176}{35} & \frac{116}{105} & 0 & 0 & \frac{2}{3} & 1 \\
{\rm tr}\big[\Pi^{[1]}\Pi^{[2]}\big]\Psi
& 0 & \frac{41}{21} & \frac{268}{49} & \frac{76}{63} & 0 & 0 & \frac{13}{21} & \frac{4}{7} \\
\hline
{\Big({\rm tr}\big[\Pi^{[1]}\big]\Big)^4\Psi}
& 0 & 0 & 6 & 0 & - & - & - & - \\
{{\rm tr}\big[\big(\Pi^{[1]}\big)^3\big]{\rm tr}\big[\Pi^{[1]}\big]\Psi}
& 0 & 0 & 1 & 1 & - & - & - & - \\
{{\rm tr}\big[\big(\Pi^{[1]}\big)^2\big]\Big({\rm tr}\big[\Pi^{[1]}\big]\Big)^2\Psi}
& 0 & 0 & \frac{7}{3} & 1 & - & - & - & - \\
{\Big({\rm tr}\big[\big(\Pi^{[1]}\big)^2\big]\Big)^2\Psi}
& 0 & 0 & \frac{4}{15} & \frac{38}{15} & - & - & - & - \\
{{\rm tr}\big[\Pi^{[1]}\big]{\rm tr}\big[\Pi^{[1]}\Pi^{[2]}\big]\Psi}
& 0 & 0 & \frac{18}{7} & \frac{4}{7} & - & - & - & - \\
{{\rm tr}\big[\Pi^{[1]}\Pi^{[1]}\Pi^{[2]}\big]\Psi}
& 0 & 0 & \frac{17}{15} & \frac{103}{105} & - & - & - & - \\
{\rm tr}\big[\Pi^{[2]}\Pi^{[2]}\big]\Psi
& 0 & 0 & \frac{2168}{735} & \frac{466}{735} & - & - & - & - \\
{\rm tr}\big[\Pi^{[1]}\Pi^{[3]}\big]\Psi
& 0 & 0 & \frac{34}{15} & \frac{268}{315} & - & - & - & - \\
\hline
\end{array}$%
}
\caption{One-loop coefficients $s^{\rm 1L}_{b_1a_1;q}$ for $q=0,1$.}
\label{tab:singlehardcoeff_a1b1}
\end{table}
 
\begin{table}[t]
\centering
\renewcommand{\arraystretch}{1.6}
$\begin{array}{c|cc|cc|c}
 & \multicolumn{2}{c|}{s^{\rm 1L}_{b_1a_2;0}} & \multicolumn{2}{c|}{s^{\rm 1L}_{b_1a_2;1}} & s^{\rm 1L}_{b_1a_3;0} \\
 b_1 \,\backslash\, a_i
 & \Psi^2 & {\rm tr}\big[\Pi^{[1]}\big]\Psi^2
 & \Psi^2 & {\rm tr}\big[\Pi^{[1]}\big]\Psi^2
 & \Psi^3 \\
\hline
\Psi
& 0 & 0 & -1 & -\frac{13}{21} & 0 \\
\hline
{\rm tr}\big[\Pi^{[1]}\big]\Psi
& 0 & 0 & 1 & \frac{13}{21} & 0 \\
\hline
\Big({\rm tr}\big[\Pi^{[1]}\big]\Big)^2\Psi
& 1 & \frac{68}{21} & 0 & 2 & 1 \\
{\rm tr}\big[\big(\Pi^{[1]}\big)^2\big]\Psi
& 1 & \frac{68}{21} & 0 & \frac{2}{3} & 1 \\
\hline
{\Big({\rm tr}\big[\Pi^{[1]}\big]\Big)^3\Psi}
& 0 & 3 & - & - & - \\
{{\rm tr}\big[\big(\Pi^{[1]}\big)^3\big]\Psi}
& 0 & 1 & - & - & - \\
{{\rm tr}\big[\big(\Pi^{[1]}\big)^2\big]{\rm tr}\big[\Pi^{[1]}\big]\Psi}
& 0 & \frac{5}{3} & - & - & - \\
{\rm tr}\big[\Pi^{[1]}\Pi^{[2]}\big]\Psi
& 0 & \frac{41}{21} & - & - & - \\
\hline
\end{array}$
\caption{One-loop coefficients $s^{\rm 1L}_{b_1a_2;q}$ for $q=0,1$ and $s^{\rm 1L}_{b_1a_3;0}$. }
\label{tab:singlehardcoeff_a1b23}
\end{table}

\begin{table}[t]
\centering
\renewcommand{\arraystretch}{1.6}
\begin{adjustbox}{max width=\textwidth}
$\begin{array}{c|ccccccc}
 & \multicolumn{7}{c}{s^{\rm 1L}_{b_2a_0;2}} \\
 b_2 \,\backslash\, a_0
 & {\rm tr}\big[ \Pi^{[1]} \big] & {\rm tr}\big[ \big( \Pi^{[1]} \big)^2 \big] & \Big( {\rm tr}\big[ \Pi^{[1]} \big] \Big)^2 & \Big( {\rm tr}\big[ \Pi^{[1]} \big] \Big)^3 & {\rm tr}\big[ \big( \Pi^{[1]} \big)^2 \big] {\rm tr}\big[ \Pi^{[1]} \big] & {\rm tr}\big[ \big( \Pi^{[1]} \big)^3 \big] & {\rm tr}\big[\Pi^{[1]} \Pi^{[2]} \big] \\
\hline
 \Psi^2 & 0 & 0 & 0 & 0 & 0 & 0 & 0 \\
\hline
 {\rm tr}\big[ \Pi^{[1]} \big]\Psi^2 & 1 & 0 & 0 & 0 & 0 & 0 & 0 \\
\hline
 {\rm tr}\big[ \big( \Pi^{[1]} \big)^2 \big]\Psi^2 & 0 & 1 & 0 & 0 & 0 & 0 & 0 \\
 \Big( {\rm tr}\big[ \Pi^{[1]} \big] \Big)^2\Psi^2 & 0 & 0 & 1 & 0 & 0 & 0 & 0 \\
\hline
 \Big( {\rm tr}\big[ \Pi^{[1]} \big] \Big)^3\Psi^2 & 0 & 0 & 0 & 1 & 0 & 0 & 0 \\
 {\rm tr}\big[ \big( \Pi^{[1]} \big)^2 \big] {\rm tr}\big[ \Pi^{[1]} \big]\Psi^2 & 0 & 0 & 0 & 0 & 1 & 0 & 0 \\
 {\rm tr}\big[ \big( \Pi^{[1]} \big)^3 \big]\Psi^2 & 0 & 0 & 0 & 0 & 0 & 1 & 0 \\
 {\rm tr}\big[\Pi^{[1]} \Pi^{[2]} \big]\Psi^2 & 0 & 0 & 0 & 0 & 0 & 0 & 1 \\
\hline
\end{array}$
\end{adjustbox}
\caption{One-loop coefficients $s^{\rm 1L}_{b_2a_0;2}$.}
\label{tab:singlehardcoeff_a2b0_q2}
\end{table}

\begin{table}[t]
\centering
\renewcommand{\arraystretch}{1.6}
\begin{adjustbox}{max width=\textwidth}
$\begin{array}{c|cccc|cccc}
 & \multicolumn{4}{c|}{s^{\rm 1L}_{b_2a_1;1}} & \multicolumn{4}{c}{s^{\rm 1L}_{b_2a_1;2}} \\
 b_2 \,\backslash\, a_1
 & \Psi & {\rm tr}\big[\Pi^{[1]}\big]\Psi & \Big({\rm tr}\big[\Pi^{[1]}\big]\Big)^2\Psi & {\rm tr}\big[\big(\Pi^{[1]}\big)^2\big]\Psi
 & \Psi & {\rm tr}\big[\Pi^{[1]}\big]\Psi & \Big({\rm tr}\big[\Pi^{[1]}\big]\Big)^2\Psi & {\rm tr}\big[\big(\Pi^{[1]}\big)^2\big]\Psi \\
\hline
\Psi^2
& -2 & -\frac{26}{21} & \frac{202}{735} & -\frac{158}{2205} & 1 & 0 & 0 & 0 \\
\hline
{\rm tr}\big[\Pi^{[1]}\big]\Psi^2
& 2 & \frac{26}{21} & -\frac{202}{735} & \frac{158}{2205} & 0 & 1 & 0 & 0 \\
\hline
{\rm tr}\big[\big(\Pi^{[1]}\big)^2\big]\Psi^2
& 0 & \frac{4}{3} & \frac{32}{21} & -\frac{2}{21} & 0 & 0 & 0 & 1 \\
\Big({\rm tr}\big[\Pi^{[1]}\big]\Big)^2\Psi^2
& 0 & 4 & \frac{94}{21} & 0 & 0 & 0 & 1 & 0 \\
\hline
\Big({\rm tr}\big[\Pi^{[1]}\big]\Big)^3\Psi^2
& 0 & 0 & 6 & 0 & - & - & - & - \\
{\rm tr}\big[\big(\Pi^{[1]}\big)^2\big]{\rm tr}\big[\Pi^{[1]}\big]\Psi^2
& 0 & 0 & \frac{4}{3} & 2 & - & - & - & - \\
{\rm tr}\big[\big(\Pi^{[1]}\big)^3\big]\Psi^2
& 0 & 0 & 0 & 2 & - & - & - & - \\
{\rm tr}\big[\Pi^{[1]}\Pi^{[2]}\big]\Psi^2
& 0 & 0 & \frac{26}{21} & \frac{8}{7} & - & - & - & - \\
\hline
\end{array}$
\end{adjustbox}
\caption{One-loop coefficients $s^{\rm 1L}_{b_2a_1;q}$ for $q=1,2$. }
\label{tab:singlehardcoeff_a2b1}
\end{table}
 
\begin{table}[t]
\centering
\renewcommand{\arraystretch}{1.6}
\begin{adjustbox}{max width=\textwidth}
$\begin{array}{c|cc|cc|c}
 & \multicolumn{2}{c|}{s^{\rm 1L}_{b_2a_2;0}} & \multicolumn{2}{c|}{s^{\rm 1L}_{b_2a_2;1}} & s^{\rm 1L}_{b_2a_3;0} \\
 b_2 \,\backslash\, a_i
 & \Psi^2 & {\rm tr}\big[\Pi^{[1]}\big]\Psi^2
 & \Psi^2 & {\rm tr}\big[\Pi^{[1]}\big]\Psi^2
 & \Psi^3 \\
\hline
\Psi^2
& 0 & 0 & -2 & -\frac{26}{21} & 0 \\
\hline
{\rm tr}\big[\Pi^{[1]}\big]\Psi^2
& 0 & 0 & 2 & \frac{26}{21} & 0 \\
\hline
{\rm tr}\big[\big(\Pi^{[1]}\big)^2\big]\Psi^2
& 1 & \frac{68}{21} & 0 & \frac{4}{3} & 1 \\
\Big({\rm tr}\big[\Pi^{[1]}\big]\Big)^2\Psi^2
& 1 & \frac{68}{21} & 0 & 4 & 1 \\
\hline
\Big({\rm tr}\big[\Pi^{[1]}\big]\Big)^3\Psi^2
& 0 & 3 & - & - & - \\
{\rm tr}\big[\big(\Pi^{[1]}\big)^2\big]{\rm tr}\big[\Pi^{[1]}\big]\Psi^2
& 0 & \frac{5}{3} & - & - & - \\
{\rm tr}\big[\big(\Pi^{[1]}\big)^3\big]\Psi^2
& 0 & 1 & - & - & - \\
{\rm tr}\big[\Pi^{[1]}\Pi^{[2]}\big]\Psi^2
& 0 & \frac{41}{21} & - & - & - \\
\hline
\end{array}$
\end{adjustbox}
\caption{One-loop coefficients $s^{\rm 1L}_{b_2a_2;q}$ for $q=0,1$ and $s^{\rm 1L}_{b_2a_3;0}$. }
\label{tab:singlehardcoeff_a2b23}
\end{table}

\begin{table}[t]
\centering
\renewcommand{\arraystretch}{1.6}
\begin{adjustbox}{max width=\textwidth}
$\begin{array}{c|cccc|cc|c}
 & \multicolumn{4}{c|}{s^{\rm 1L}_{b_3a_1;2}} & \multicolumn{2}{c|}{s^{\rm 1L}_{b_3a_2;1}} & s^{\rm 1L}_{b_3a_3;0} \\
 b_3 \,\backslash\, a_i
 & \Psi & {\rm tr}\big[\Pi^{[1]}\big]\Psi & \Big({\rm tr}\big[\Pi^{[1]}\big]\Big)^2\Psi & {\rm tr}\big[\big(\Pi^{[1]}\big)^2\big]\Psi
 & \Psi^2 & {\rm tr}\big[\Pi^{[1]}\big]\Psi^2
 & \Psi^3 \\
\hline
\Psi^3
& 3 & 0 & 0 & 0 & -3 & -\frac{13}{7} & 0 \\
\hline
{\rm tr}\big[\Pi^{[1]}\big]\Psi^3
& 0 & 3 & 0 & 0 & 3 & \frac{13}{7} & 0 \\
\hline
{\rm tr}\big[\big(\Pi^{[1]}\big)^2\big]\Psi^3
& 0 & 0 & 0 & 3 & 0 & 2 & 1 \\
\Big({\rm tr}\big[\Pi^{[1]}\big]\Big)^2\Psi^3
& 0 & 0 & 3 & 0 & 0 & 6 & 1 \\
\hline
\end{array}$
\end{adjustbox}
\caption{One-loop coefficients $s^{\rm 1L}_{b_3a_1;2}$, $s^{\rm 1L}_{b_3a_2;1}$ and $s^{\rm 1L}_{b_3a_3;0}$.}
\label{tab:singlehardcoeff_a3b}
\end{table}

\begin{table}[t]
\centering
\renewcommand{\arraystretch}{0.8}
\begin{adjustbox}{max width=\textwidth}
$\begin{array}{c|cc}
 b & \hat{s}^{{\rm 2L},[22]}_{b\d;00} & \tilde{s}^{{\rm 2L},[22]}_{b\d;00} \\ \\
 \hline
 \\
 {\rm tr}\big[ \Pi^{[1]} \big] & 0 & 0 \\ \\
 \hline
 \\
 {\rm tr}\big[ \big( \Pi^{[1]} \big)^2 \big] & \frac{1724}{105} & \frac{752}{441} \\
 \Big( {\rm tr}\big[ \Pi^{[1]} \big] \Big)^2 & \frac{1724}{105} & \frac{752}{441} \\ \\
 \hline
 \\
 \Big( {\rm tr}\big[ \Pi^{[1]} \big] \Big)^3 & \frac{141478}{2205} & \frac{8}{7} \\
 {\rm tr}\big[ \big( \Pi^{[1]} \big)^2 \big] {\rm tr}\big[ \Pi^{[1]} \big] & \frac{5834}{147} & \frac{7160}{441} \\
 {\rm tr}\big[ \big( \Pi^{[1]} \big)^3 \big] & \frac{60526}{2205} & \frac{3496}{147} \\
 {\rm tr}\big[\Pi^{[1]} \Pi^{[2]} \big] & \frac{269914}{6615} & \frac{1480}{147} \\ \\
 \hline
 \\
 \Big( {\rm tr}\big[ \Pi^{[1]} \big] \Big)^4 & \frac{544}{7} & 0 \\
 {\rm tr}\big[ \big( \Pi^{[1]} \big)^3 \big] {\rm tr}\big[ \Pi^{[1]} \big] & \frac{164}{7} & \frac{50}{7} \\
 {\rm tr}\big[ \big( \Pi^{[1]} \big)^2 \big] \big({\rm tr}\big[ \Pi^{[1]} \big]\big)^2 & \frac{12704}{315} & \frac{40}{7} \\
 \Big( {\rm tr}\big[ \big( \Pi^{[1]} \big)^2 \big] \Big)^2 & \frac{1184}{45} & \frac{80}{21} \\
 {\rm tr}\big[\Pi^{[1]} \Pi^{[1]} \Pi^{[2]} \big] & \frac{32224}{1323} & \frac{7412}{441} \\
 {\rm tr}\big[ \Pi^{[1]} \big] {\rm tr}\big[\Pi^{[1]} \Pi^{[2]} \big] & \frac{36332}{945} & \frac{746}{49} \\
 {\rm tr}\big[\Pi^{[1]} \Pi^{[3]} \big] & \frac{25628}{735} & \frac{802}{147} \\
 {\rm tr}\big[\Pi^{[2]} \Pi^{[2]} \big] & \frac{55568}{1323} & \frac{464}{21} \\ \\
 \hline
 \\
 \Big( {\rm tr}\big[ \Pi^{[1]} \big] \Big)^5 & 30 & 0 \\
 {\rm tr}\big[ \big( \Pi^{[1]} \big)^3 \big] \left({\rm tr}\big[ \Pi^{[1]} \big]\right)^2 & \frac{22}{3} & 0 \\
 {\rm tr}\big[ \big( \Pi^{[1]} \big)^2 \big] \left({\rm tr}\big[ \Pi^{[1]} \big]\right)^3 & 14 & 0 \\
 {\rm tr}\big[ \big( \Pi^{[1]} \big)^3 \big]{\rm tr}\big[ \big( \Pi^{[1]} \big)^2 \big] & \frac{62}{15} & 0 \\
 {\rm tr}\big[ \Pi^{[1]} \big]\Big( {\rm tr}\big[ \big( \Pi^{[1]} \big)^2 \big] \Big)^2 & \frac{326}{45} & 0 \\
 \big({\rm tr}\big[ \Pi^{[1]}\big] \big)^2 {\rm tr}\big[\Pi^{[1]} \Pi^{[2]} \big] & \frac{94}{7} & \frac{20}{7} \\
 {\rm tr}\big[ \Pi^{[1]}  \Pi^{[1]} \big] {\rm tr}\big[\Pi^{[1]} \Pi^{[2]}\big] & \frac{346}{45} & \frac{20}{21} \\
 {\rm tr}\big[ \Pi^{[1]} \big] {\rm tr}\big[\Pi^{[1]} \Pi^{[1]} \Pi^{[2]}\big] & \frac{782}{105} & \frac{50}{21} \\
 {\rm tr}\big[\Pi^{[1]} \Pi^{[2]} \Pi^{[2]} \big] & \frac{394}{49} & \frac{820}{147} \\
 {\rm tr}\big[ \Pi^{[1]} \big] {\rm tr}\big[\Pi^{[2]} \Pi^{[2]} \big] & \frac{10114}{735} & \frac{1100}{147} \\
 {\rm tr}\big[\Pi^{[1]} \Pi^{[1]} \Pi^{[3]} \big] & \frac{1286}{189} & \frac{208}{63} \\
 {\rm tr}\big[ \Pi^{[1]} \big] {\rm tr}\big[\Pi^{[1]} \Pi^{[3]} \big] & \frac{10954}{945} & \frac{24}{7} \\
 {\rm tr}\big[\Pi^{[2]} \Pi^{[3]} \big] & \frac{17258}{1323} & \frac{982}{147} \\
 {\rm tr}\big[\Pi^{[1]} \Pi^{[4]} \big] & \frac{33403}{6615} & \frac{148}{441} \\
\end{array}$
\end{adjustbox}
\caption{Two-loop coefficients $\hat s^{\rm 2L}$ and $\tilde s^{\rm 2L}$ for the $[22]$ contribution with $\d;00$, which corresponds to the Gaussian case considered in \cite{Bakx:2025cvu}.}
\label{tab:twoloopcoeff_Gauss}
\end{table}

\begin{table}[t]
\centering
\renewcommand{\arraystretch}{1.6}
\begin{adjustbox}{max width=\textwidth}
$\begin{array}{c|cc|cc|cc}
b
& \hat s^{\rm 2L,[22]}_{b\Psi;00} & \tilde s^{\rm 2L,[22]}_{b\Psi;00}
& \hat s^{\rm 2L,[3]}_{b\delta;0} & \tilde s^{\rm 2L,[3]}_{b\delta;0}
& \hat s^{\rm 2L,[3]}_{b\Psi;0} & \tilde s^{\rm 2L,[3]}_{b\Psi;0} \\
\hline
{\rm tr}[\Pi^{[1]}] & 0 & 0 & 0 & 0 & 0 & 0 \\
\hline
{\rm tr}[(\Pi^{[1]})^2] & \frac{8126}{2205} & \frac{8}{21} & \frac{3139}{1323} & \frac{1964}{1323} & \frac{13}{21} & \frac{8}{21} \\
\big({\rm tr}[\Pi^{[1]}]\big)^2 & \frac{8126}{2205} & \frac{8}{21} & \frac{3139}{1323} & \frac{1964}{1323} & \frac{13}{21} & \frac{8}{21} \\
\hline
\big({\rm tr}[\Pi^{[1]}]\big)^3 & \frac{68}{7} & 0 & \frac{47}{7} & \frac{8}{7} & 1 & 0 \\
{\rm tr}[(\Pi^{[1]})^2]\,{\rm tr}[\Pi^{[1]}] & \frac{92}{15} & \frac{20}{7} & \frac{167}{63} & \frac{244}{63} & \frac{1}{3} & \frac{2}{3} \\
{\rm tr}[(\Pi^{[1]})^3] & \frac{152}{35} & \frac{30}{7} & \frac{13}{21} & \frac{110}{21} & 0 & 1 \\
{\rm tr}[\Pi^{[1]}\Pi^{[2]}] & \frac{2944}{441} & \frac{38}{21} & \frac{4315}{1323} & \frac{4190}{1323} & \frac{10}{21} & \frac{11}{21} \\
\hline
\big({\rm tr}[\Pi^{[1]}]\big)^4 & 6 & 0 & 4 & 0 & - & - \\
{\rm tr}[(\Pi^{[1]})^3]\,{\rm tr}[\Pi^{[1]}] & 2 & 0 & \frac{1}{3} & \frac{5}{3} & - & - \\
{\rm tr}[(\Pi^{[1]})^2]\big({\rm tr}[\Pi^{[1]}]\big)^2 & \frac{10}{3} & 0 & \frac{4}{3} & \frac{4}{3} & - & - \\
\big({\rm tr}[(\Pi^{[1]})^2]\big)^2 & \frac{14}{5} & 0 & \frac{4}{9} & \frac{8}{9} & - & - \\
{\rm tr}[\Pi^{[1]}\Pi^{[1]}\Pi^{[2]}] & \frac{74}{35} & \frac{10}{7} & \frac{20}{63} & \frac{124}{63} & - & - \\
{\rm tr}[\Pi^{[1]}]\,{\rm tr}[\Pi^{[1]}\Pi^{[2]}] & \frac{22}{7} & \frac{10}{7} & \frac{9}{7} & \frac{5}{3} & - & - \\
{\rm tr}[\Pi^{[1]}\Pi^{[3]}] & \frac{982}{315} & \frac{4}{21} & \frac{239}{189} & \frac{247}{189} & - & - \\
{\rm tr}[\Pi^{[2]}\Pi^{[2]}] & \frac{878}{245} & \frac{20}{7} & \frac{680}{441} & \frac{748}{441} & - & - \\
\hline
\end{array}$
\end{adjustbox}
\caption{Two-loop coefficients $\hat s^{\rm 2L}$ and $\tilde s^{\rm 2L}$ for the $[22]$ contribution with $\Psi;00$ and the $[3]$ contributions with $\delta;0$ and $\Psi;0$. Rows label the operators $b$ and a dash indicates that the entry is not available at the truncation order considered.}
\label{tab:twoloopcoeff_a_000}
\end{table}
 
\begin{table}[t]
\centering
\renewcommand{\arraystretch}{1.6}
\begin{adjustbox}{max width=\textwidth}
$\begin{array}{c|cc|cc|cc|cc|cc}
b
& \hat s^{\rm 2L,[22]}_{b\delta;10} & \tilde s^{\rm 2L,[22]}_{b\delta;10}
& \hat s^{\rm 2L,[22]}_{b\Psi;00} & \tilde s^{\rm 2L,[22]}_{b\Psi;00}
& \hat s^{\rm 2L,[22]}_{b\Psi;10} & \tilde s^{\rm 2L,[22]}_{b\Psi;10}
& \hat s^{\rm 2L,[3]}_{b\Psi;0} & \tilde s^{\rm 2L,[3]}_{b\Psi;0}
& \hat s^{\rm 2L,[3]}_{b\delta;1} & \tilde s^{\rm 2L,[3]}_{b\delta;1} \\
\hline
\Psi
& \frac{208}{2205} & \frac{-52}{147} & 0 & 0 & \frac{16}{315} & \frac{-4}{21} & 0 & 0 & \frac{23}{63} & \frac{-23}{63} \\
\hline
{\rm tr}[\Pi^{[1]}]\Psi
& \frac{-208}{2205} & \frac{52}{147} & 0 & 0 & \frac{-16}{315} & \frac{4}{21} & 0 & 0 & \frac{-23}{63} & \frac{23}{63} \\
\hline
\big({\rm tr}[\Pi^{[1]}]\big)^2\Psi
& \frac{56233}{6615} & \frac{8}{21} & \frac{8126}{2205} & \frac{8}{21} & \frac{47}{21} & 0 & \frac{13}{21} & \frac{8}{21} & \frac{8}{3} & \frac{4}{7} \\
{\rm tr}[(\Pi^{[1]})^2]\Psi
& \frac{37321}{6615} & \frac{544}{49} & \frac{8126}{2205} & \frac{8}{21} & \frac{31}{21} & \frac{20}{7} & \frac{13}{21} & \frac{8}{21} & \frac{-4}{21} & \frac{24}{7} \\
\hline
\big({\rm tr}[\Pi^{[1]}]\big)^3\Psi
& \frac{149}{7} & 0 & \frac{68}{7} & 0 & 3 & 0 & 1 & 0 & 3 & 0 \\
{\rm tr}[(\Pi^{[1]})^3]\Psi
& \frac{19}{3} & \frac{20}{7} & \frac{152}{35} & \frac{30}{7} & 1 & 0 & 0 & 1 & 0 & 1 \\
{\rm tr}[(\Pi^{[1]})^2]\,{\rm tr}[\Pi^{[1]}]\Psi
& \frac{697}{63} & \frac{20}{7} & \frac{92}{15} & \frac{20}{7} & \frac{5}{3} & 0 & \frac{1}{3} & \frac{2}{3} & \frac{2}{3} & 1 \\
{\rm tr}[\Pi^{[1]}\Pi^{[2]}]\Psi
& \frac{66091}{6615} & \frac{544}{49} & \frac{2944}{441} & \frac{38}{21} & \frac{11}{7} & \frac{10}{7} & \frac{10}{21} & \frac{11}{21} & \frac{5}{21} & \frac{12}{7} \\
\hline
\big({\rm tr}[\Pi^{[1]}]\big)^4\Psi
& 12 & 0 & 6 & 0 & - & - & - & - & - & - \\
{\rm tr}[(\Pi^{[1]})^3]\,{\rm tr}[\Pi^{[1]}]\Psi
& \frac{8}{3} & 0 & 2 & 0 & - & - & - & - & - & - \\
{\rm tr}[(\Pi^{[1]})^2]\big({\rm tr}[\Pi^{[1]}]\big)^2\Psi
& \frac{16}{3} & 0 & \frac{10}{3} & 0 & - & - & - & - & - & - \\
\big({\rm tr}[(\Pi^{[1]})^2]\big)^2\Psi
& \frac{20}{9} & 0 & \frac{14}{5} & 0 & - & - & - & - & - & - \\
{\rm tr}[\Pi^{[1]}]\,{\rm tr}[\Pi^{[1]}\Pi^{[2]}]\Psi
& \frac{36}{7} & \frac{10}{7} & \frac{22}{7} & \frac{10}{7} & - & - & - & - & - & - \\
{\rm tr}[\Pi^{[1]}\Pi^{[1]}\Pi^{[2]}]\Psi
& \frac{8}{3} & \frac{20}{21} & \frac{74}{35} & \frac{10}{7} & - & - & - & - & - & - \\
{\rm tr}[\Pi^{[2]}\Pi^{[2]}]\Psi
& \frac{748}{147} & \frac{680}{147} & \frac{878}{245} & \frac{20}{7} & - & - & - & - & - & - \\
{\rm tr}[\Pi^{[1]}\Pi^{[3]}]\Psi
& \frac{572}{135} & \frac{68}{21} & \frac{982}{315} & \frac{4}{21} & - & - & - & - & - & - \\
\hline
\end{array}$
\end{adjustbox}
\caption{Two-loop coefficients $\hat s^{\rm 2L}$ and $\tilde s^{\rm 2L}$ for operators $b$ containing one power of $\Psi$, for the $[22]$ contributions with $\delta;10$, $\Psi;00$ and $\Psi;10$, and the $[3]$ contributions with $\Psi;1$ and $\delta;1$. }
\label{tab:twoloopcoeff_aPsi}
\end{table}
 
\begin{table}[t]
\centering
\renewcommand{\arraystretch}{1.6}
\begin{adjustbox}{max width=\textwidth}
$\begin{array}{c|cc|cc|cc}
b
& \hat s^{\rm 2L,[22]}_{b\Psi;10} & \tilde s^{\rm 2L,[22]}_{b\Psi;10}
& \hat s^{\rm 2L,[22]}_{b\delta;11} & \tilde s^{\rm 2L,[22]}_{b\delta;11}
& \hat s^{\rm 2L,[22]}_{b\delta;20} & \tilde s^{\rm 2L,[22]}_{b\delta;20} \\
\hline
\Psi^2 & \frac{32}{315} & \frac{-8}{21} & \frac{184}{189} & 0 & 0 & 0 \\
\hline
{\rm tr}[\Pi^{[1]}]\Psi^2 & \frac{-32}{315} & \frac{8}{21} & \frac{-184}{189} & 0 & 0 & 0 \\
\hline
\big({\rm tr}[\Pi^{[1]}]\big)^2\Psi^2 & \frac{94}{21} & 0 & \frac{80}{7} & 0 & \frac{68}{21} & 0 \\
{\rm tr}[(\Pi^{[1]})^2]\Psi^2 & \frac{62}{21} & \frac{40}{7} & \frac{80}{21} & 0 & \frac{68}{21} & 0 \\
\hline
\big({\rm tr}[\Pi^{[1]}]\big)^3\Psi^2 & 6 & 0 & 12 & 0 & 3 & 0 \\
{\rm tr}[(\Pi^{[1]})^3]\Psi^2 & 2 & 0 & \frac{4}{3} & 0 & 1 & 0 \\
{\rm tr}[(\Pi^{[1]})^2]\,{\rm tr}[\Pi^{[1]}]\Psi^2 & \frac{10}{3} & 0 & 4 & 0 & \frac{5}{3} & 0 \\
{\rm tr}[\Pi^{[1]}\Pi^{[2]}]\Psi^2 & \frac{22}{7} & \frac{20}{7} & \frac{68}{21} & 0 & \frac{41}{21} & 0 \\
\hline
\end{array}$
\end{adjustbox}
\caption{Two-loop coefficients $\hat s^{\rm 2L,[22]}$ and $\tilde s^{\rm 2L,[22]}$ for operators $b$ containing $\Psi^2$, for the $[22]$ contributions with $\Psi;10$, $\delta;11$ and $\delta;20$. The coefficients for $\delta;02$ are identical to those for $\delta;20$ and are not shown. For $q=2$, all $\tilde s^{\rm 2L,[22]}$ vanish, so these coefficients are independent of $r$.}
\label{tab:twoloopcoeff_aPsi2}
\end{table}

\clearpage
\bibliographystyle{JHEP}
\bibliography{main}

@article{Bakx:2026rmd,
    author = "Bakx, Thomas and Garny, Mathias and Rubira, Henrique and Vlah, Zvonimir",
    title = "{Galaxy bias renormalization: Two-loop Power Spectrum, One-loop Trispectrum and Bispectrum}",
    eprint = "2606.31280",
    archivePrefix = "arXiv",
    primaryClass = "astro-ph.CO",
    month = "6",
    year = "2026"
}

@article{Bakx:2025cvu,
    author = "Bakx, Thomas and Garny, Mathias and Rubira, Henrique and Vlah, Zvonimir",
    title = "{Two-loop renormalization and running of galaxy bias}",
    eprint = "2507.13905",
    archivePrefix = "arXiv",
    primaryClass = "astro-ph.CO",
    reportNumber = "TUM-HEP-1566/25, RBI-ThPhys-2025-30",
    doi = "10.1088/1475-7516/2026/03/056",
    journal = "JCAP",
    volume = "03",
    pages = "056",
    year = "2026"
}

@ARTICLE{2015JCAP...10..032S,
       author = {{Schmidt}, Fabian and {Chisari}, Nora Elisa and {Dvorkin}, Cora},
        title = "{Imprint of inflation on galaxy shape correlations}",
      journal = {\jcap},
         year = 2015,
        month = oct,
       volume = {2015},
       number = {10},
        pages = {032-032},
          doi = {10.1088/1475-7516/2015/10/032},
archivePrefix = {arXiv},
       eprint = {1506.02671},
 primaryClass = {astro-ph.CO},
       adsurl = {https://ui.adsabs.harvard.edu/abs/2015JCAP...10..032S}
}

@ARTICLE{2018JCAP...09..003C,
       author = {{Cabass}, G. and {Pajer}, E. and {Schmidt}, F.},
        title = "{Imprints of Oscillatory Bispectra on Galaxy Clustering}",
      journal = {\jcap},
         year = 2018,
        month = sep,
       volume = {2018},
       number = {9},
          eid = {003},
        pages = {003},
          doi = {10.1088/1475-7516/2018/09/003},
archivePrefix = {arXiv},
       eprint = {1804.07295},
 primaryClass = {astro-ph.CO},
       adsurl = {https://ui.adsabs.harvard.edu/abs/2018JCAP...09..003C}
}

@ARTICLE{2008PhRvD..77l3514D,
       author = {{Dalal}, Neal and {Dor{\'e}}, Olivier and {Huterer}, Dragan and {Shirokov}, Alexander},
        title = "{Imprints of primordial non-Gaussianities on large-scale structure: Scale-dependent bias and abundance of virialized objects}",
      journal = {\prd},
         year = 2008,
        month = jun,
       volume = {77},
       number = {12},
          eid = {123514},
        pages = {123514},
          doi = {10.1103/PhysRevD.77.123514},
archivePrefix = {arXiv},
       eprint = {0710.4560},
 primaryClass = {astro-ph},
       adsurl = {https://ui.adsabs.harvard.edu/abs/2008PhRvD..77l3514D}
}

@ARTICLE{2008ApJ...677L..77M,
       author = {{Matarrese}, Sabino and {Verde}, Licia},
        title = "{The Effect of Primordial Non-Gaussianity on Halo Bias}",
      journal = {\apjl},
         year = 2008,
        month = apr,
       volume = {677},
       number = {2},
        pages = {L77},
          doi = {10.1086/587840},
archivePrefix = {arXiv},
       eprint = {0801.4826},
 primaryClass = {astro-ph},
       adsurl = {https://ui.adsabs.harvard.edu/abs/2008ApJ...677L..77M}
}

@article{PhysRevD.42.3936,
  title = {Nonlinear evolution of long-wavelength metric fluctuations in inflationary models},
  author = {Salopek, D. S. and Bond, J. R.},
  journal = {Phys. Rev. D},
  volume = {42},
  issue = {12},
  pages = {3936--3962},
  numpages = {0},
  year = {1990},
  month = {Dec},
  publisher = {American Physical Society},
  doi = {10.1103/PhysRevD.42.3936},
  url = {https://link.aps.org/doi/10.1103/PhysRevD.42.3936}
}

@ARTICLE{maldacena:2003,
       author = {{Maldacena}, Juan},
        title = "{Non-gaussian features of primordial fluctuations in single field inflationary models}",
      journal = {Journal of High Energy Physics},
         year = 2003,
        month = may,
       volume = {2003},
       number = {5},
          eid = {013},
        pages = {013},
          doi = {10.1088/1126-6708/2003/05/013},
archivePrefix = {arXiv},
       eprint = {astro-ph/0210603},
 primaryClass = {astro-ph},
       adsurl = {https://ui.adsabs.harvard.edu/abs/2003JHEP...05..013M}
}

@article{Yokoyama:2012az,
    author = "Yokoyama, Shuichiro and Matsubara, Takahiko",
    title = "{Scale-dependent bias with higher order primordial non-Gaussianity: Use of the Integrated Perturbation Theory}",
    eprint = "1210.2495",
    archivePrefix = "arXiv",
    primaryClass = "astro-ph.CO",
    reportNumber = "ICRR-REPORT-631-2012-20",
    doi = "10.1103/PhysRevD.87.023525",
    journal = "Phys. Rev. D",
    volume = "87",
    pages = "023525",
    year = "2013"
}

@ARTICLE{2013JCAP...05..001B,
       author = {{Baumann}, Daniel and {Ferraro}, Simone and {Green}, Daniel and {Smith}, Kendrick M.},
        title = "{Stochastic bias from non-Gaussian initial conditions}",
      journal = {\jcap},
         year = 2013,
        month = may,
       volume = {2013},
       number = {5},
          eid = {001},
        pages = {001},
          doi = {10.1088/1475-7516/2013/05/001},
archivePrefix = {arXiv},
       eprint = {1209.2173},
 primaryClass = {astro-ph.CO},
       adsurl = {https://ui.adsabs.harvard.edu/abs/2013JCAP...05..001B}
}

@article{Philcox:2026myi,
    author = "Philcox, Oliver H. E.",
    title = "{The Redshift-Space Galaxy Power Spectrum at Two-Loop Order}",
    eprint = "2610.13867",
    archivePrefix = "arXiv",
    primaryClass = "astro-ph.CO",
    month = "10",
    year = "2026"
}

@ARTICLE{MSZ,
   author = {{Mirbabayi}, M. and {Schmidt}, F. and {Zaldarriaga}, M.},
    title = "{Biased tracers and time evolution}",
  journal = {\jcap},
archivePrefix = "arXiv",
   eprint = {1412.5169},
     year = 2015,
    month = jul,
   volume = 7,
      eid = {030},
    pages = {30},
      doi = {10.1088/1475-7516/2015/07/030},
   adsurl = {http://adsabs.harvard.edu/abs/2015JCAP...07..030M}
}

@article{Nikolis:2024kbx,
    author = "Nikolis, Charalampos and Rubira, Henrique and Schmidt, Fabian",
    title = "{The renormalization group for large-scale structure: primordial non-Gaussianities}",
    eprint = "2405.21002",
    archivePrefix = "arXiv",
    primaryClass = "astro-ph.CO",
    doi = "10.1088/1475-7516/2024/08/017",
    journal = "JCAP",
    volume = "08",
    pages = "017",
    year = "2024"
}

@article{Assassi2014,
    author = "Assassi, Valentin and Baumann, Daniel and Green, Daniel and Zaldarriaga, Matias",
    title = "{Renormalized Halo Bias}",
    eprint = "1402.5916",
    archivePrefix = "arXiv",
    primaryClass = "astro-ph.CO",
    doi = "10.1088/1475-7516/2014/08/056",
    journal = "JCAP",
    volume = "08",
    pages = "056",
    year = "2014"
}

@article{Gleyzes:2016tdh,
    author = "Gleyzes, J{\'e}r{\^o}me and de Putter, Roland and Green, Daniel and Dor{\'e}, Olivier",
    title = "{Biasing and the search for primordial non-Gaussianity beyond the local type}",
    eprint = "1612.06366",
    archivePrefix = "arXiv",
    primaryClass = "astro-ph.CO",
    doi = "10.1088/1475-7516/2017/04/002",
    journal = "JCAP",
    volume = "04",
    pages = "002",
    year = "2017"
}

@article{MoradinezhadDizgah:2017szk,
    author = "Moradinezhad Dizgah, Azadeh and Dvorkin, Cora",
    title = "{Scale-Dependent Galaxy Bias from Massive Particles with Spin during Inflation}",
    eprint = "1708.06473",
    archivePrefix = "arXiv",
    primaryClass = "astro-ph.CO",
    doi = "10.1088/1475-7516/2018/01/010",
    journal = "JCAP",
    volume = "01",
    pages = "010",
    year = "2018"
}

@article{Giannantonio:2009ak,
    author = "Giannantonio, Tommaso and Porciani, Cristiano",
    title = "{Structure formation from non-Gaussian initial conditions: multivariate biasing, statistics, and comparison with N-body simulations}",
    eprint = "0911.0017",
    archivePrefix = "arXiv",
    primaryClass = "astro-ph.CO",
    doi = "10.1103/PhysRevD.81.063530",
    journal = "Phys. Rev. D",
    volume = "81",
    pages = "063530",
    year = "2010"
}

@article{McDonald:2008sc,
    author = "McDonald, Patrick",
    title = "{Primordial non-Gaussianity: large-scale structure signature in the perturbative bias model}",
    eprint = "0806.1061",
    archivePrefix = "arXiv",
    primaryClass = "astro-ph",
    doi = "10.1103/PhysRevD.78.123519",
    journal = "Phys. Rev. D",
    volume = "78",
    pages = "123519",
    year = "2008"
}

@article{Desjacques:2016bnm,
    author = "Desjacques, Vincent and Jeong, Donghui and Schmidt, Fabian",
    title = "{Large-Scale Galaxy Bias}",
    eprint = "1611.09787",
    archivePrefix = "arXiv",
    primaryClass = "astro-ph.CO",
    doi = "10.1016/j.physrep.2017.12.002",
    journal = "Phys. Rept.",
    volume = "733",
    pages = "1--193",
    year = "2018"
}

@article{Rubira:2025rqo,
    author = "Rubira, Henrique and Schmidt, Fabian",
    title = "{Non-Gaussian Galaxy Stochasticity and the Noise-Field Formulation}",
    eprint = "2511.05484",
    archivePrefix = "arXiv",
    primaryClass = "astro-ph.CO",
    journal = "JCAP",
    volume = "089",
    pages = "1125",
    year = "2026"
}

@article{Rubira:2023vzw,
    author = "Rubira, Henrique and Schmidt, Fabian",
    title = "{Galaxy bias renormalization group}",
    eprint = "2307.15031",
    archivePrefix = "arXiv",
    primaryClass = "astro-ph.CO",
    reportNumber = "TUM-HEP- 1464/23",
    doi = "10.1088/1475-7516/2024/01/031",
    journal = "JCAP",
    volume = "01",
    pages = "031",
    year = "2024"
}

@article{Schmidt:2010gw,
    author = "Schmidt, Fabian and Kamionkowski, Marc",
    title = "{Halo Clustering with Non-Local Non-Gaussianity}",
    eprint = "1008.0638",
    archivePrefix = "arXiv",
    primaryClass = "astro-ph.CO",
    doi = "10.1103/PhysRevD.82.103002",
    journal = "Phys. Rev. D",
    volume = "82",
    pages = "103002",
    year = "2010"
}

@article{Assassi:2015fma,
    author = "Assassi, Valentin and Baumann, Daniel and Schmidt, Fabian",
    title = "{Galaxy Bias and Primordial Non-Gaussianity}",
    eprint = "1510.03723",
    archivePrefix = "arXiv",
    primaryClass = "astro-ph.CO",
    doi = "10.1088/1475-7516/2015/12/043",
    journal = "JCAP",
    volume = "12",
    pages = "043",
    year = "2015"
}

@article{Huang:2026zvi,
    author = "Huang, Hsiang-Ming and Rubira, Henrique and Schmidt, Fabian",
    title = "{Connecting field-level and n-point function measurements of galaxy bias}",
    eprint = "2611.14456",
    archivePrefix = "arXiv",
    primaryClass = "astro-ph.CO",
    month = "11",
    year = "2026"
}

@article{Assassi:2015EFT,
	doi = {10.1088/1475-7516/2015/11/024},
  
	url = {https://doi.org/10.1088%2F1475-7516%2F2015%2F11%2F024},
  
	year = 2015,
	month = {nov},
  
	publisher = {{IOP} Publishing},
  
	volume = {2015},
  
	number = {11},
  
	pages = {024--024},
  
	author = {Valentin Assassi and Daniel Baumann and Enrico Pajer and Yvette Welling and Drian van der Woude},
  
	title = {Effective theory of large-scale structure with primordial non-Gaussianity},
  
	journal = {Journal of Cosmology and Astroparticle Physics}
}

@article{Komatsu:2001rj,
    author = "Komatsu, Eiichiro and Spergel, David N.",
    title = "{Acoustic signatures in the primary microwave background bispectrum}",
    eprint = "astro-ph/0005036",
    archivePrefix = "arXiv",
    doi = "10.1103/PhysRevD.63.063002",
    journal = "Phys. Rev. D",
    volume = "63",
    pages = "063002",
    year = "2001"
}

@article{Cheung:2007st,
    author = "Cheung, Clifford and Creminelli, Paolo and Fitzpatrick, A. Liam and Kaplan, Jared and Senatore, Leonardo",
    title = "{The Effective Field Theory of Inflation}",
    eprint = "0709.0293",
    archivePrefix = "arXiv",
    primaryClass = "hep-th",
    reportNumber = "IC-2007-032",
    doi = "10.1088/1126-6708/2008/03/014",
    journal = "JHEP",
    volume = "03",
    pages = "014",
    year = "2008"
}

@article{Chen_2010,
   title={Quasi-single field inflation and non-Gaussianities},
   volume={2010},
   ISSN={1475-7516},
   url={http://dx.doi.org/10.1088/1475-7516/2010/04/027},
   DOI={10.1088/1475-7516/2010/04/027},
   number={04},
   journal={Journal of Cosmology and Astroparticle Physics},
   publisher={IOP Publishing},
   author={Chen, Xingang and Wang, Yi},
   year={2010},
   month=apr, pages={027–027} }

@misc{arkanihamed2015cosmological,
      title={Cosmological Collider Physics}, 
      author={Nima Arkani-Hamed and Juan Maldacena},
      year={2015},
      eprint={1503.08043},
      archivePrefix={arXiv},
      primaryClass={hep-th}
}

@article{Lee_2016,
   title={Non-Gaussianity as a particle detector},
   volume={2016},
   ISSN={1029-8479},
   url={http://dx.doi.org/10.1007/JHEP12(2016)040},
   DOI={10.1007/jhep12(2016)040},
   number={12},
   journal={Journal of High Energy Physics},
   publisher={Springer Science and Business Media LLC},
   author={Lee, Hayden and Baumann, Daniel and Pimentel, Guilherme L.},
   year={2016},
   month=dec }

@article{Rubira:2024tea,
    author = "Rubira, Henrique and Schmidt, Fabian",
    title = "{The renormalization group for large-scale structure: origin of galaxy stochasticity}",
    eprint = "2404.16929",
    archivePrefix = "arXiv",
    primaryClass = "astro-ph.CO",
    doi = "10.1088/1475-7516/2024/10/092",
    journal = "JCAP",
    volume = "10",
    pages = "092",
    year = "2024"
}

@article{arkanihamed2019cosmological,
    author = "Arkani-Hamed, Nima and Baumann, Daniel and Lee, Hayden and Pimentel, Guilherme L.",
    title = "{The Cosmological Bootstrap: Inflationary Correlators from Symmetries and Singularities}",
    eprint = "1811.00024",
    archivePrefix = "arXiv",
    primaryClass = "hep-th",
    doi = "10.1007/JHEP04(2020)105",
    journal = "JHEP",
    volume = "04",
    pages = "105",
    year = "2020"
}

@article{baumann2022snowmass,
    author = "Baumann, Daniel and Green, Daniel and Joyce, Austin and Pajer, Enrico and Pimentel, Guilherme L. and Sleight, Charlotte and Taronna, Massimo",
    title = "{Snowmass White Paper: The Cosmological Bootstrap}",
    eprint = "2203.08121",
    archivePrefix = "arXiv",
    primaryClass = "hep-th",
    doi = "10.21468/SciPostPhysCommRep.1",
    journal = "SciPost Phys. Comm. Rep.",
    volume = "2024",
    pages = "1",
    year = "2024"
}

@article{Guth:1980zm,
    author = "Guth, Alan H.",
    editor = "Fang, Li-Zhi and Ruffini, R.",
    title = "{The Inflationary Universe: A Possible Solution to the Horizon and Flatness Problems}",
    reportNumber = "SLAC-PUB-2576",
    doi = "10.1103/PhysRevD.23.347",
    journal = "Phys. Rev. D",
    volume = "23",
    pages = "347--356",
    year = "1981"
}

@article{Linde:1981mu,
    author = "Linde, Andrei D.",
    editor = "Fang, Li-Zhi and Ruffini, R.",
    title = "{A New Inflationary Universe Scenario: A Possible Solution of the Horizon, Flatness, Homogeneity, Isotropy and Primordial Monopole Problems}",
    reportNumber = "LEBEDEV-81-229",
    doi = "10.1016/0370-2693(82)91219-9",
    journal = "Phys. Lett. B",
    volume = "108",
    pages = "389--393",
    year = "1982"
}

@article{Guth:1982ec,
    author = "Guth, Alan H. and Pi, S. Y.",
    title = "{Fluctuations in the New Inflationary Universe}",
    doi = "10.1103/PhysRevLett.49.1110",
    journal = "Phys. Rev. Lett.",
    volume = "49",
    pages = "1110--1113",
    year = "1982"
}

@article{McDonald:2006mx,
    author = "McDonald, Patrick",
    title = "{Clustering of dark matter tracers: Renormalizing the bias parameters}",
    eprint = "astro-ph/0609413",
    archivePrefix = "arXiv",
    doi = "10.1103/PhysRevD.74.129901",
    journal = "Phys. Rev. D",
    volume = "74",
    pages = "103512",
    year = "2006",
    note = "[Erratum: Phys.Rev.D 74, 129901 (2006)]"
}

@incollection{baldauf:2020bsd,
	title = {Effective {Field} {Theory} of {Large}-{Scale} {Structure}},
	isbn = {978-0-19-885574-3},
	url = {https://doi.org/10.1093/oso/9780198855743.003.0007},
	doi = {10.1093/oso/9780198855743.003.0007},
	booktitle = {Effective {Field} {Theory} in {Particle} {Physics} and {Cosmology}: {Lecture} {Notes} of the {Les} {Houches} {Summer} {School}: {Volume} 108, {July} 2017},
	publisher = {Oxford University Press},
	author = {Baldauf, Tobias},
	editor = {Davidson, Sacha and Gambino, Paolo and Laine, Mikko and Neubert, Matthias and Salomon, Christophe},
	month = apr,
	year = {2020},
}

@article{Joyce_2015,
   title={Multiple soft limits of cosmological correlation functions},
   volume={2015},
   ISSN={1475-7516},
   url={http://dx.doi.org/10.1088/1475-7516/2015/01/012},
   DOI={10.1088/1475-7516/2015/01/012},
   number={01},
   journal={Journal of Cosmology and Astroparticle Physics},
   publisher={IOP Publishing},
   author={Joyce, Austin and Khoury, Justin and Simonović, Marko},
   year={2015},
   month=jan, pages={012–012} }

@article{Mirbabayi_2016,
   title={Effective theory of squeezed correlation functions},
   volume={2016},
   ISSN={1475-7516},
   url={http://dx.doi.org/10.1088/1475-7516/2016/03/056},
   DOI={10.1088/1475-7516/2016/03/056},
   number={03},
   journal={Journal of Cosmology and Astroparticle Physics},
   publisher={IOP Publishing},
   author={Mirbabayi, Mehrdad and Simonović, Marko},
   year={2016},
   month=mar, pages={056–056} }

@article{Assassi_2012,
   title={On soft limits of inflationary correlation functions},
   volume={2012},
   ISSN={1475-7516},
   url={http://dx.doi.org/10.1088/1475-7516/2012/11/047},
   DOI={10.1088/1475-7516/2012/11/047},
   number={11},
   journal={Journal of Cosmology and Astroparticle Physics},
   publisher={IOP Publishing},
   author={Assassi, Valentin and Baumann, Daniel and Green, Daniel},
   year={2012},
   month=Nov, pages={047–047} }

@article{Baumann_2012,
   title={Signature of supersymmetry from the early universe},
   volume={85},
   ISSN={1550-2368},
   url={http://dx.doi.org/10.1103/PhysRevD.85.103520},
   DOI={10.1103/physrevd.85.103520},
   number={10},
   journal={Physical Review D},
   publisher={American Physical Society (APS)},
   author={Baumann, Daniel and Green, Daniel},
   year={2012},
   month=May }

@article{Paolo_Creminelli_2004,
   title={A single-field consistency relation for the three-point function},
   volume={2004},
   ISSN={1475-7516},
   url={http://dx.doi.org/10.1088/1475-7516/2004/10/006},
   DOI={10.1088/1475-7516/2004/10/006},
   number={10},
   journal={Journal of Cosmology and Astroparticle Physics},
   publisher={IOP Publishing},
   author={Paolo Creminelli and Matias Zaldarriaga},
   year={2004},
   month=Oct, pages={006–006} }

@article{Werth:2024aui,
    author = "Werth, Denis and Pinol, Lucas and Renaux-Petel, S{\'e}bastien",
    title = "{CosmoFlow: Python Package for Cosmological Correlators}",
    eprint = "2402.03693",
    archivePrefix = "arXiv",
    primaryClass = "astro-ph.CO",
    doi = "10.1088/1361-6382/ad6740",
    journal = "Class. Quant. Grav.",
    volume = "41",
    number = "17",
    pages = "175015",
    year = "2024"
}

@article{Beneke:1997zp,
    author = "Beneke, M. and Smirnov, Vladimir A.",
    title = "{Asymptotic expansion of Feynman integrals near threshold}",
    eprint = "hep-ph/9711391",
    archivePrefix = "arXiv",
    reportNumber = "CERN-TH-97-315",
    doi = "10.1016/S0550-3213(98)00138-2",
    journal = "Nucl. Phys. B",
    volume = "522",
    pages = "321--344",
    year = "1998"
}

@article{Azyzy:2025zif,
    author = "Azyzy, Sha and Jamieson, Drew and Komatsu, Eiichiro and Kurita, Toshiki",
    title = "{Nonlinear evolution of the matter trispectrum with primordial parity violation}",
    eprint = "2510.06164",
    archivePrefix = "arXiv",
    primaryClass = "astro-ph.CO",
    doi = "10.1088/1475-7516/2026/06/054",
    journal = "JCAP",
    volume = "06",
    pages = "054",
    year = "2026"
}

@article{Kurita:2025hmp,
    author = "Kurita, Toshiki and Jamieson, Drew and Komatsu, Eiichiro and Schmidt, Fabian",
    title = "{Parity violation in galaxy shapes: Primordial non-Gaussianity}",
    eprint = "2509.08787",
    archivePrefix = "arXiv",
    primaryClass = "astro-ph.CO",
    doi = "10.1103/fxh6-hpmk",
    journal = "Phys. Rev. D",
    volume = "113",
    number = "6",
    pages = "063557",
    year = "2026"
}

@article{Schmidt:2013nsa,
    author = "Schmidt, Fabian",
    title = "{Non-Gaussian Halo Bias Beyond the Squeezed Limit}",
    eprint = "1304.1817",
    archivePrefix = "arXiv",
    primaryClass = "astro-ph.CO",
    doi = "10.1103/PhysRevD.87.123518",
    journal = "Phys. Rev. D",
    volume = "87",
    number = "12",
    pages = "123518",
    year = "2013"
}

@article{PhysRevD.82.043531,
  title = {Non-Gaussianity and large-scale structure in a two-field inflationary model},
  author = {Tseliakhovich, Dmitriy and Hirata, Christopher and Slosar, An{\ifmmode \check{z}\else \v{z}\fi{}}e},
  journal = {Phys. Rev. D},
  volume = {82},
  issue = {4},
  pages = {043531},
  numpages = {9},
  year = {2010},
  month = {Aug},
  publisher = {American Physical Society},
  doi = {10.1103/PhysRevD.82.043531},
  url = {https://link.aps.org/doi/10.1103/PhysRevD.82.043531}
}

@article{PhysRevLett.107.191301,
  title = {Universal Bound on $N$-Point Correlations from Inflation},
  author = {Smith, Kendrick M. and LoVerde, Marilena and Zaldarriaga, Matias},
  journal = {Phys. Rev. Lett.},
  volume = {107},
  issue = {19},
  pages = {191301},
  numpages = {5},
  year = {2011},
  month = {Nov},
  publisher = {American Physical Society},
  doi = {10.1103/PhysRevLett.107.191301},
  url = {https://link.aps.org/doi/10.1103/PhysRevLett.107.191301}
}

@article{K-Smith,
doi = {10.1088/1475-7516/2012/03/032},
url = {https://doi.org/10.1088/1475-7516/2012/03/032},
year = {2012},
month = {mar},
publisher = {},
volume = {2012},
number = {03},
pages = {032},
author = {Kendrick M. Smith and Simone Ferraro and Marilena LoVerde},
title = {Halo clustering and gNL-type primordial non-gaussianity},
journal = {Journal of Cosmology and Astroparticle Physics}
}

\end{document}